\documentclass[useAMS,usenatbib,onecolumn]{mn2e}
\usepackage{graphicx}

\title[The double-power approach]{The double-power approach to spherically symmetric astrophysical systems}
\author[M. Lingam and P.H. Nguyen]{Manasvi Lingam$^{1}$\thanks{E-mail:
manasvi@physics.utexas.edu} and Phuc H. Nguyen$^{2}$\thanks{E-mail:
phn229@physics.utexas.edu}\\
$^{1}$Institute for Fusion Studies, The University of Texas, Austin, TX 78712, USA\\
$^{2}$Center for Relativity and Texas Cosmology Center, The University of Texas, Austin, TX 78712, USA}
\begin{document}

\date{}

\pagerange{\pageref{firstpage}--\pageref{lastpage}} \pubyear{2013}

\maketitle

\label{firstpage}

\begin{abstract}
In this paper, we present two simple approaches for deriving anisotropic distribution functions for a wide range of spherical models. The first method involves multiplying and dividing a basic augmented density with polynomials in $r$ and constructing more complex augmented densities in the process, from which we obtain the double-power distribution functions. This procedure is applied to a specific case of the Veltmann models that is known to closely approximate the Navarro--Frenk--White (NFW) profile, and also to the Plummer and Hernquist profiles (in the appendix). The second part of the paper is concerned with obtaining hypervirial distribution functions, i.e. distribution functions that satisfy the local virial theorem, for several well-known models. In order to construct the hypervirial augmented densities and the corresponding distribution functions, we start with an appropriate ansatz for the former and proceed to determine the coefficients appearing in that ansatz by expanding the potential--density pair as a series, around $r=0$ and $r=\infty$. By doing so, we obtain hypervirial distribution functions, valid in these two limits, that can generate the potential--density pairs of these models to an arbitrarily high degree of accuracy. This procedure is explicitly carried out for the H\'enon isochrone, Jaffe, Dehnen and NFW models and the accuracy of this procedure is established. Finally, we derive some universal properties for these hypervirial distribution functions, involving the asymptotic behaviour of the anisotropy parameter and its relation to the density slope in this regime. In particular, we show that the cusp slope--central anisotropy inequality is saturated.
\end{abstract}


\section{Introduction}\label{SectIntro}
Although the notion of spherical symmetry is an ideal one, it is known that many systems such as globular clusters, galactic bulges and dark matter haloes can be modelled as being roughly spherically symmetric. In order to model the distribution of matter in such systems in a self-consistent manner, one solves the Vlasov--Poisson system, and the equilibrium solutions are known to be functions of the integrals of motion via Jeans's theorem. The usual method in the literature involves the `$\rho$ to $f$' approach. Here, one starts off with a given potential--density pair and certain assumptions about the velocity structure, and proceeds to find the distribution function \citep{pl11,ed16, ly62, os79, ja83, b200, bi87, he90, pe08a} via an appropriate integral transform. A less common method is the `$f$ to $\rho$' approach which postulates a certain functional form for the distribution function, and solving the Poisson equation. In general, solving this non-linear differential equation for analytical solutions is very difficult. The self-consistent approach continues to be widely used, especially in the context of dark matter haloes \citep{fo13,dev14,me14,siu14}.

The equilibrium distribution functions in a spherically symmetric setting depend on the two integrals of motion, namely the magnitude of the angular momentum $L$ and the energy $E$. Methods for generating two-integral distribution functions have been widely studied in the literature \citep{b123,ji07a,ji07b,ji07c,ras10} and originated with \cite{b21}, who noticed that whenever the density profile could be written as a sum of monomials of $R$ (in cylindrical coordinates) and $\Phi$ in the axisymmetric setting, it was possible to construct a distribution function expressible as a sum of monomials in angular momentum and energy. This procedure was generalized by different authors to spherical and axial symmetry \citep{m63,pt70,k76,jf85,dj86,r88,e93,e94,ji07a,ji07c,pe08a,pe08b,pe08c}. Relativistic treatments along the same lines have also been undertaken by several authors, see e.g. \citet{ag11,ram12,ng13} and \citet{li13}. 

The approach that we make use of in this paper is akin to the ones mentioned above. However, it must be mentioned that the powers of the energy and angular momentum thus obtained in spherical symmetry differ from those obtained in an axisymmetric setting. Moreover, the decomposition into powers of $r$ and $\Phi$ in the spherical case is non-unique, as opposed to the axisymmetric one. Owing to this non-uniqueness, one can construct a wide range of dynamic properties such as the velocity dispersions and anisotropy parameters. In addition, finding such a decomposition for an arbitrary potential--density pair is rather difficult, with no well-defined algorithm. The method of double-power distribution functions was employed by \citet{li13}, who used it to generate two-term double-power distribution functions, and it was shown that a wide class of potential--density pairs could thereby be obtained. The double-power approach introduced in this paper is an example of a hybrid approach since it starts off with a known functional form of the distribution function, which is a sum of anisotropic polytropes, but also assumes a priori knowledge of the potential--density pair. Once the augmented density has been found, one could also obtain the distribution function through a different approach which would involve performing a fractional derivative inversion \citep{an12}. The two approaches are equivalent and the method that we use in this paper is exactly the same as the use of fractional derivatives.

The standard method of constructing distribution functions from a given density profile $\rho{(r)}$ is via integral transforms. The simplest such transform is given by Eddington's formula, which is used for constructing isotropic distribution functions:
\begin{equation}
f(\mathcal{E}) = \frac{1}{\sqrt{8}\pi^{2}} \frac{d}{d\mathcal{E}} \int_{0}^{\mathcal{E}} \frac{d\Psi}{\sqrt{\mathcal{E}-\Psi}} \frac{d\rho}{d\Psi},
\end{equation}
where $\Psi=-\Phi$ is the negative of the Newtonian potential, $\mathcal{E}=-E$ is the binding energy and $\rho(\Psi)$ is found by substituting $r(\Psi)$ into $\rho(r)$. Even in this case, the integral is often non-trivial. As a result, the distribution functions and their properties such as the velocity dispersions and anisotropy parameters cannot be easily reduced to analytical expressions. The situation is worsened for the case of anisotropic distribution functions, which depend on both energy and angular momentum. In this case, the fundamental integral relation yields a function $\rho{(r,\Psi)}$ (for the isotropic case, it is only a function of $\Psi$) which we will refer to as the augmented density, and an analogue of Eddington's formula exists but is substantially more complicated \citep{dj87}. First, we perform a double-integral transform to construct a function $f'{(\mathcal{E},L)}$:
\begin{equation} \label{AnInvScheme1}
f'{(\mathcal{E},L)} = \frac{1}{\sqrt{\pi}} \frac{d^{2}}{d\mathcal{E}^{2}} \left[\sqrt{\mathcal{E}} \int_{0}^{1} \frac{du}{\sqrt{1-u}} \int_{0}^{1-u} \frac{dv}{v\sqrt{1-v-u}} \rho{\left(\mathcal{E}u,\frac{L/\sqrt{2\mathcal{E}}}{\sqrt{v}}\right)} \right].
\end{equation}
Then, we analytically continue the second argument to the complex domain, and the distribution function is found by taking the imaginary part of the result:
\begin{equation} \label{AnInvScheme2}
f{(\mathcal{E},L)} = -\frac{1}{2^{3/2}\pi^{5/2}} \mathrm{Im} [f'(\mathcal{E},L)].
\end{equation}
In addition, a further disadvantage of this approach stems from the fact that the above inversion scheme, as pointed out by \citet{dj87} is numerically unstable.

\citet{ea05} derived an elegant family of models, which possessed hyperviriality, i.e. they satisfied the virial theorem locally. This aspect has important (positive) consequences for stability, as the authors point out, because it constrains the number of high-velocity stars that can escape and give rise to instability. The importance of the local virial theorem, i.e. hyperviriality, in self-gravitating systems has also been studied by other authors \citep{ig06,so06,so08,hen09}. Subsequently, \citet{an05} went on to derive a two-component family of hypervirial distribution functions, which were presented as candidates for modelling dark matter haloes. The family of models that are described by the two different hypervirial families include a wide range of commonly used models such as the Plummer (Eddington in 1916 discovered the hypervirial nature of its distribution function) and Hernquist models, but there are several other commonly used potential--density pairs which do not have a known hypervirial distribution function such as the H\'enon isochrone, Jaffe, Dehnen and NFW models. 

\citet{ae06} derived a very general theorem relating two global properties, the cusp slope $\left(\gamma_{0}\right)$ and the central anisotropy $\left(\beta_{0}\right)$ through the relation  $\gamma_{0}\geq2\beta_{0}$. Other work that connects the cusp slope to global quantities includes \citet{ha04}, which relates it to the Tsallis entropic index. \citet{ae06} proved that this holds true for distribution functions with constant anisotropy, and generalized this to systems with non-constant anisotropy via the series expansion. This theorem was generalized by \citet{ci10a,ci10b,va11}, who studied the global equivalent, termed the global density-slope anisotropy inequality (GDSAI), for separable augmented densities, and established the conditions under which it held true.  We emphasize that some of the augmented densities constructed through the double-power approach in this paper are not separable, although they can be expressed as a sum of separable augmented densities. 

The outline of the paper is as follows. In Section \ref{SectI}, we introduce the double-power approach, which involves constructing an augmented density that is expressible as a sum of monomials in $\Psi$ and $r$. By using this augmented density, we construct the corresponding distribution functions and associated properties, such as velocity dispersions, and perform an asymptotic analysis to deduce some of its general features. This procedure is explicitly carried out for the $m=1/2$ Veltmann model, and some of the simpler distribution functions are calculated. This procedure is also extended to the Plummer and Hernquist models in Appendix \ref{AppA}. The second part of the paper is concerned with investigating the behaviour of hypervirial distribution functions which satisfy the virial theorem locally. In Section \ref{SectII}, we present a generic method of constructing augmented densities that are hypervirial for a wide range of physical models from a basic ansatz. We make use of this procedure to find the corresponding hypervirial augmented densities and the distribution functions for well-known models such as the H\'enon isochrone, NFW, Jaffe and Dehnen models. In Section \ref{SectIII}, we take the generic hypervirial augmented density and derive universal properties, which include an algebraic relation between the density slope and the anisotropy parameter in the inner regime.

\section{Distribution functions for the Veltmann models}\label{SectI}
In this section, we develop the formalism and illustrate it by deriving various families of distribution functions for the degenerate Veltmann isochrone models \citep{ve79a,ve79b}. Even though most distribution functions discussed in this section are found in the literature, a few have never been explicitly written down to the authors' knowledge. The degenerate Veltmann isochrone models comprise of the following one-parameter family of potential--density pairs:
\begin{equation}\label{phiVeltmann}
\Phi=-\frac{GM}{a}\left(1+\left(\frac{r}{a}\right)^{m}\right)^{-1/m},
\end{equation}
\begin{equation}\label{rhoVeltmann}
\rho(r)=\frac{\Phi_{0}(1+m)a^{-m}}{4\pi G}r^{m-2}\left[1+\left(\frac{r}{a}\right)^{m}\right]^{-(1/m)-2},
\end{equation}
where the parameter $m$ is positive. These models constitute the limiting case of the more general family described by Veltmann, and we shall henceforth abbreviate them as the Veltmann models. Two of the best known spherical models of astrophysics are included in the family: the Plummer model (for $m=2$) and the Hernquist model (for $m=1$), and the case $m=\frac{1}{2}$ is a good approximation to the NFW profile commonly used to model dark matter haloes \citep{zh96}. We will find it convenient to define $\Psi=-\Phi$ and $\Phi_{0}=\frac{GM}{a}=-\Phi{(r=0)}$. 

A distribution function of the anisotropic polytropic form for this potential--density pair was derived in \citet{ea05} and \citet{ng13}:
\begin{equation}\label{hypervirialDF}
f({\mathcal{E}},L) = CL^{m-2} {\mathcal{E}}^{(3m+1)/2},
\end{equation}
where $C$ is an unimportant normalization factor, and it is implicitly understood that the distribution function vanishes whenever $\mathcal{E} = -E < 0$. As discussed in \citet{ea05}, this distribution function possesses the remarkable property that the virial theorem holds locally (i.e. hyperviriality). In this section, we proceed to find distribution functions of the more general form:
\begin{equation}\label{doublepowerDF}
f{({\mathcal{E}},L)} = \sum_{i} C_{i} L^{2\alpha_{i}} {\mathcal{E}}^{\beta_{i}},
\end{equation}
where we restrict $C_{i}$ to be positive constants, so that the distribution function is manifestly nonnegative. We will variously refer to a distribution function of the above form as a multi component (anisotropic) polytrope, or, following \citet{bi87}, as a double-power distribution function. Obviously, the distribution function (\ref{hypervirialDF}) is a special case consisting of one component. The distribution function (\ref{doublepowerDF}) gives rise to the Poisson equation:
\begin{equation}\label{mastereq}
-\nabla^{2}\Psi = \sum_{i} C_{i} 2^{\alpha_{i}+(7/2)}\pi^{5/2}G\frac{\Gamma{(\alpha_{i}+1)}\Gamma{(\beta_{i}+1)}}{\Gamma{(\alpha_{i}+\beta_{i}+\frac{5}{2})}}r^{2\alpha_{i}} \Psi^{\alpha_{i}+\beta_{i}+(3/2)} = 4\pi G \rho{(r,\Psi)},
\end{equation}
through the evaluation of integrals of the form
\begin{equation} \label{centralintegral}
\int_{0}^{\pi}\sin^{2\alpha+1}{\eta}d\eta = \sqrt{\pi}\frac{\Gamma{(1+\alpha)}}{\Gamma{(\frac{3}{2}+\alpha)}},
\end{equation}
\begin{equation}
\int_{0}^{\sqrt{2\Psi}} v^{2\alpha+2}\left(-\frac{v^2}{2}+\Psi\right)^{\beta} dv = 2^{\alpha+(1/2)}\Psi^{\alpha+\beta+(3/2)}\frac{\Gamma{(\frac{3}{2}+\alpha)}\Gamma{(1+\alpha)}}{\Gamma{(\alpha+\beta+\frac{5}{2})}},
\end{equation}
where the first integral above converges for $\alpha > -1$ and the second converges for $\alpha > -\frac{3}{2}$ and $\beta > -1$. It follows that the density is finite provided $\alpha_{i} > -1$ and $\beta_{i} > -1$ for all $i$. Our task now will be to come up with some augmented density $\rho(\Psi,r)$ for the Veltmann potential--density pair, expand the augmented density as a series in $r$ and $\Psi$ to recast it in the form of equation (\ref{mastereq}), and deduce the powers and coefficients appearing in equation (\ref{doublepowerDF}). It is well known that solving for the distribution function for a given potential--density pair is a degenerate problem, and there are infinitely many ways to come up with an augmented density.
Once an augmented density is known, it is also possible to calculate the velocity dispersions without having to resort to finding the explicit form of the distribution function. Such a procedure is discussed in \citet{dj87}, and the corresponding expressions are given by
\begin{equation} \label{sigmar2}
\sigma_r^2 \equiv \left\langle v_{r}^{2}\right\rangle  = \frac{1}{\rho(\Psi,r)} \int_0^\Psi{\rho(\Psi',r)}d\Psi',
\end{equation}
\begin{equation} \label{sigmat2}
\sigma_\theta^2 \equiv \left\langle v_{\theta}^{2}\right\rangle = \frac{1}{\rho(\Psi,r)} \int_{0}^{\Psi} \partial_{r^2}\left[r^2\rho(\Psi',r)\right] d\Psi',
\end{equation}
and it must be noted that $\sigma_\theta^2 = \sigma_\phi^2$. This emerges from the fact that, when the expectation values of $v_\theta^2$ and $v_\phi^2$ are computed, they result in integrals of the form $\int_0^{2\pi} \sin^2\chi d\chi$ and  $\int_0^{2\pi} \cos^2\chi d\chi$, which are exactly equal to each other. The expectation values of $v_r$, $v_\theta$ and $v_\phi$ are all zero as well, which arises from the fact that $\int_0^{2\pi} \sin\chi d\chi = \int_0^{2\pi} \cos\chi d\chi = 0$. Hence, the relations $\left\langle v_{s}^{2}\right\rangle \equiv \sigma_s^2$ hold true for $s = r,\theta,\phi$. The anisotropy parameter $\beta$ is
\begin{equation} \label{betafromv}
\beta = 1 - \frac{\sigma_\theta^2}{\sigma_r^2}.
\end{equation}

\subsection{Distribution functions with finitely many components}
In this subsection, we will introduce an augmented density ansatz that comprises of a finite number of components. We perform an asymptotic analysis to obtain the limiting values of the anisotropy parameter. In addition, we shall explicitly set down some expressions for the distribution functions and velocity structure profiles for the NFW-like profile ($m=1/2$) discussed by \citet{zh96}.

\subsubsection{The augmented density ansatz and asymptotic structure}
We first focus on the case where the augmented density consists of a finite number of components. From equation (\label{phiVeltmann}), we obtain the relation
\begin{equation}\label{Veltmanneq}
1+x^{m}=\left(\frac{\Psi}{\Phi_{0}}\right)^{-m},
\end{equation}
where $x = r/a$. This immediately leads to the one-component augmented density which corresponds to the distribution function (\ref{hypervirialDF})
\begin{equation}\label{rho1term}
\rho\left(\Psi,x\right)=\frac{m+1}{\Phi_{0}^{2m}4\pi G a^{2}}x^{m-2}\Psi^{2m+1}.
\end{equation}
Next, we multiply and divide the right-hand side of equation (\ref{rho1term}) with $[1+x^{m}]^q$ for any natural number $q$, and cast the augmented density as
\begin{equation}\label{rhoq}
\rho\left(\Psi,x\right)=\frac{\Phi_{0}^{-(2+q)m}(1+m)a^{-2}}{4\pi G}x^{m-2}\Psi^{(q+2)m+1}\left[1+x^{m}\right]^{q}.
\end{equation}
Since $q$ is a natural number, we may use the binomial theorem to obtain the finite sum
\begin{equation}\label{rhoqbinom}
\rho\left(\Psi,x\right)=\frac{\Phi_{0}^{-(2+q)m}(1+m)a^{-2}}{4\pi G}x^{m-2}\Psi^{(q+2)m+1}\sum_{k=0}^{q}\left(\begin{array}{c}
q\\
k
\end{array}\right)x^{mk}.
\end{equation}
It is worth noting that $q$ need not necessarily be an integer. Since this section is concerned with a distribution function that has a finite number of double-power components, an integral value of $q$ was chosen to satisfy this criterion.
Each of the $x$ factors in equation (\ref{rhoqbinom}) can be multiplied and divided themselves by some positive integral power of $1+x^{m}$ as undertaken earlier. In general, one can choose different powers for different terms. Let us say that the kth term is multiplied and divided with $[1+x^{m}]^{n_{k}}$ for some positive integer $n_{k}$. We then have
\begin{equation}
\rho\left(\Psi,x\right)=\frac{\Phi_{0}^{-(2+q)m}(1+m)a^{-2}}{4\pi G}x^{m-2}\Psi^{(q+2)m+1}\sum_{k=0}^{q}\left(\begin{array}{c}
q\\
k
\end{array}\right)x^{mk}\left[\frac{1+x^{m}}{1+x^{m}}\right]^{n_{k}}.
\end{equation}
By using the binomial theorem again, the augmented density becomes
\begin{equation}\label{masterrho1}
\rho\left(\Psi,x\right)=\frac{\Phi_{0}^{-(2+q)m}(1+m)a^{-2}}{4\pi G}\sum_{k=0}^{q}\sum_{l=0}^{n_{k}}\left(\begin{array}{c}
q\\
k
\end{array}\right)\left(\begin{array}{c}
n_{k}\\
l
\end{array}\right)\left(\Phi_{0}\right)^{-mn_{k}}x^{m(l+k+1)-2} \Psi^{m\left(n_{k}+q+2\right)+1}.
\end{equation}
We will perform an asymptotic analysis of the above augmented density ansatz to obtain the limiting values of the anisotropy parameter, which is important in investigating the GDSAI. Most of the work involving the GDSAI \citep{ae06,ci10a,ci10b,va11} was undertaken for separable distribution functions, but it must be noted that this ansatz is not separable. One can use equations (\ref{sigmar2})--(\ref{betafromv}) to compute the velocity dispersion profiles and the anisotropy parameter for all values of $r$. The resulting function is a function of $r$ and $\Psi$, of which the latter itself is a function of $r$.

Now, in the limit $r\rightarrow 0$ it is worth noting that only the lowest powers of $r$ dominate the other terms. In this limit, $\Psi \rightarrow \frac{GM}{a}$, which is a constant, which simplifies matters a great deal. By considering the \emph{smallest} powers of $r$ in the numerator and the denominator of the velocity structure profiles, it is found that
\begin{equation}\label{sigmarorigin}
\sigma_{r}^{2}\left(r=0\right)=\frac{\Phi_{0}}{m\left(n_{0}+q+2\right)+2},
\end{equation}
\begin{equation}\label{sigmatorigin}
\sigma_{\phi}^{2}\left(r=0\right)=\frac{m\Phi_{0}}{2m\left(n_{0}+q+2\right)+4},
\end{equation}
\begin{equation}
\beta(0) = 1-\frac{m}{2}.
\end{equation}
The remarkable property of the above expression is that the anisotropy parameter is completely independent of the choice of the $n_{k}$ and $q$ under the limit $r\rightarrow0$. In fact, for any distribution function of the form $\sum_{i}L^{2\alpha_{i}}f_{i}{(\mathcal{E})}$, it is relatively straightforward to argue that as $r \rightarrow 0$, $-\beta \rightarrow \mathrm{min}(\alpha_{i})$. This implies that the total number of terms in our distribution function is irrelevant when it comes to determining the central value of the anisotropy parameter. On the other hand, the leading power for the density profile at small $r$ is $\rho \approx r^{-(2-m)}$; thus, for $m<2$, we have an inner cusp $\gamma_0 = 2-m$. From this, we can conclude that the cusp slope-central anisotropy inequality is saturated, and one has $\gamma_0 = 2\beta_0$.

We can perform a similar asymptotic analysis in the limit $r \rightarrow \infty$. However, in this limit, one must note that $\Psi$ falls off as $1/r$. Thus, one must be more careful in evaluating the anisotropy parameter in this region because $\Psi$ also has an $r$-dependence. As we are interested in $r \rightarrow \infty$, we must take into consideration the highest powers of $r$ (inclusive of the contribution from $\Psi$) in the denominator and the numerator of the velocity structure profiles, since they dominate the expression in this limit. The velocity structure profiles take on the form
\begin{equation}
\sigma_{r}^{2}\left(r\rightarrow\infty\right) \sim \frac{\Phi_{0}}{m\left(n_{q}+q+2\right)+2}\frac{1}{x},
\end{equation}
\begin{equation}
\sigma_{\phi}^{2}\left(r\rightarrow\infty\right) \sim \frac{\Phi_{0}m(q+n_{q}+1)}{m\left(n_{q}+q+2\right)+2}\frac{1}{x},
\end{equation}
\begin{equation} \label{betatheorem001}
\beta\left(r\rightarrow\infty\right)=1-\frac{m\left(q+n_{q}+1\right)}{2}.
\end{equation}
This is again a fairly simple expression that is independent of most of the free parameters inherent in the ansatz. As before, for a distribution function of the form $\sum_{i} L^{2\alpha_{i}}f_{i}{(\mathcal{E})}$, it is straightforward to argue that as $r \rightarrow \infty$, one has $-\beta \rightarrow \mathrm{max}{(\alpha_{i})}$.
The above expression can also be rewritten as
\begin{equation} \label{betatheorem002}
\beta_{\infty}=\beta_{0}-\frac{m}{2}\left(q+n_{q}\right).
\end{equation}
Since we know that $q$ and $n_q$ are both positive, it immediately follows that $\beta_\infty \leq \beta_0$ for all values of $m$. This can also be established in the following manner. We have earlier stated that $\beta_0 = - \mathrm{min}\left(\alpha_i\right)$ and that $\beta_\infty = - \mathrm{max}\left(\alpha_i\right)$. From these two conditions, it is evident that $\beta_0 \geq \beta_\infty$.

\subsubsection{The NFW-like models}\label{NFWlikemodels}
The cases of the Plummer and Hernquist model have been extensively analysed in the literature (see, for example, \citet{dj87,ba02,bvh07}). We will supplement these analyses by studying in some detail the distribution functions for the NFW-like profile ($m=1/2$) discussed by \citet{zh96}. 

\begin{figure*}
$$
\begin{array}{ccc}
 \includegraphics[width=6.8cm]{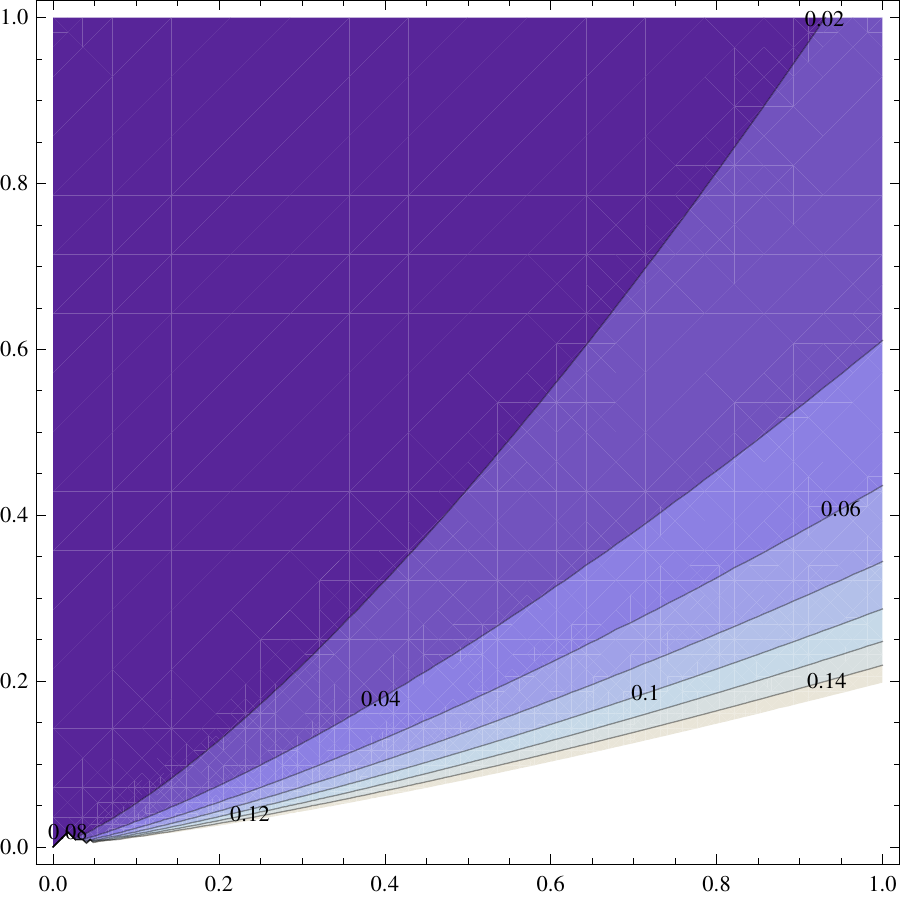} & \includegraphics[width=6.8cm]{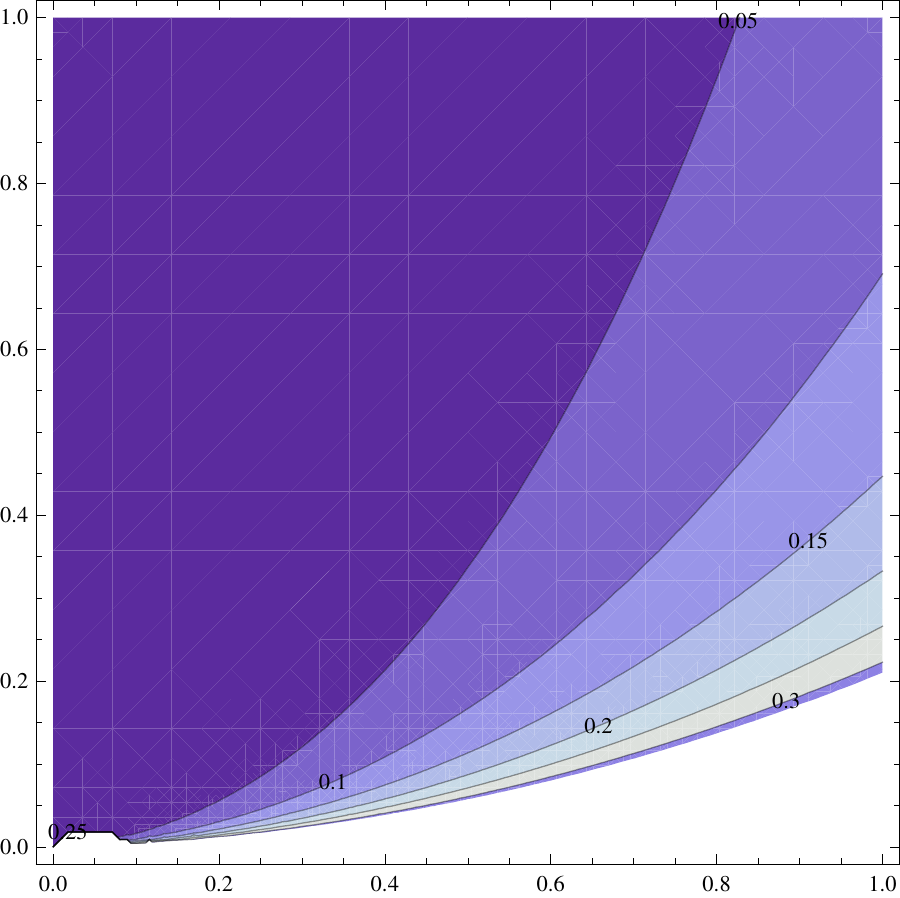} \\
 \includegraphics[width=6.8cm]{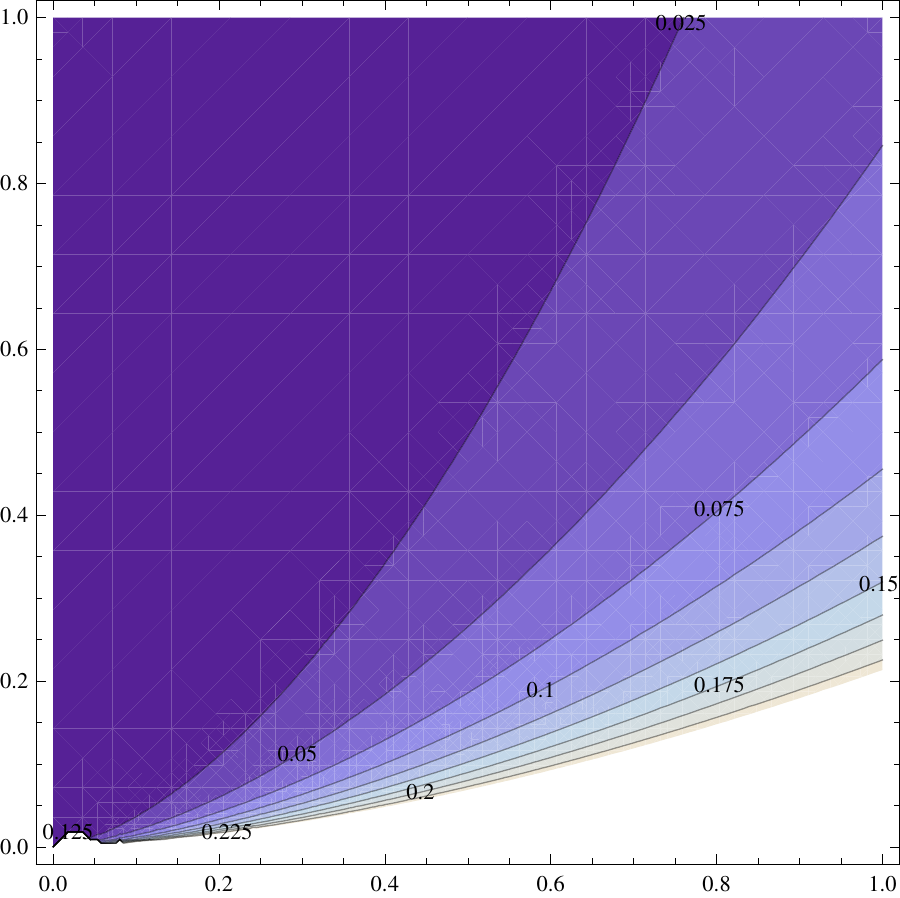} & \includegraphics[width=6.8cm]{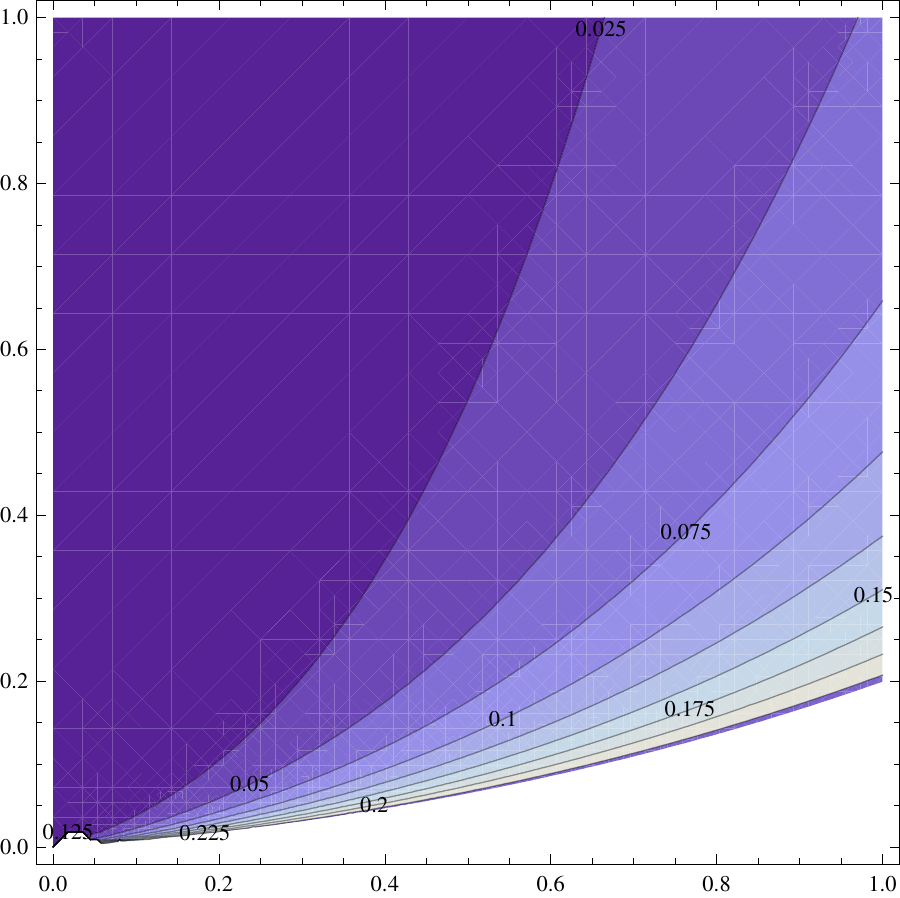}
\end{array}
$$
\caption{(color figures online) Contour plots of the NFW-like distribution functions. The horizontal direction is $-E/\Phi_0$ and the vertical direction is $L/(a\sqrt{\Phi_0})$. Top-left panel: the distribution function of equation (\ref{NFW2}). Top-right panel: the distribution function of equation (\ref{NFW3a}). Bottom-left panel: the distribution function of equation (\ref{NFW3b}). Bottom-right panel: the distribution function of equation (\ref{NFW3c}).}
\label{contourNFWlike}
\end{figure*}

For $q=0$, we have the one-component augmented density
\begin{equation} \label{NFW1}
\rho(x,\Psi) \propto \Psi^2 x^{-3/2},
\end{equation}
which gives rise to the hypervirial distribution function (\ref{HypervirialDFNFW}). Next, consider the two-component model ($q=1$ and $n_{0}=n_{1}=0$). The augmented density is
\begin{equation} \label{NFW2}
\rho{(x,\Psi)} \propto \Psi^{5/2} x^{-3/2} \left(1 + \sqrt{x}\right).
\end{equation}
The corresponding distribution function is given by
\begin{equation}
f({\mathcal{E}},L) = \frac{45 a}{64 G^{5/2} M^{3/2} \pi^{2} 2^{3/4} \Gamma(1/4) \Gamma(11/4)} L^{-3/2} {\mathcal{E}}^{7/4} + \frac{15 \sqrt{a}}{32 \pi^{3} G^{5/2} M^{3/2}} L^{-1} {\mathcal{E}}^{3/2}.
\end{equation}
There are three ways to construct three-component distribution functions for the NFW-like model. The first ($q=2$ and $n_{0}=n_{1}=n_{2}=0$) is found from the augmented density
\begin{equation} \label{NFW3a}
\rho{(x,\Psi)} \propto \Psi^3 x^{-3/2} \left(1+\sqrt{x}\right)^2,
\end{equation}
and the distribution function is given by
\begin{equation}
f({\mathcal{E}},L) = \frac{9 a^{3/2}} {2^{3/4} 4 \pi^{5/2} G^3 M^2 \Gamma{(1/4)}\Gamma{(13/4)}} L^{-3/2} {\mathcal{E}}^{9/4} + \frac{9 a}{8 \pi^{3} G^3 M^2} L^{-1} {\mathcal{E}}^{2} + \frac{9 \sqrt{a}}{2^{1/4} 8 \pi^{5/2} G^3 M^2 \Gamma{(3/4)}\Gamma{(11/4)}} L^{-1/2} {\mathcal{E}}^{7/4}.
\end{equation}
The second distribution function is obtained by setting $q=1$, $n_{0}=1$ and $n_{1}=0$. The augmented density is
\begin{equation} \label{NFW3b}
\rho{(x,\Psi)} \propto \Psi^{5/2} x^{-3/2} \left(\sqrt{\frac{\Psi}{\Phi_{0}}} \left(1+\sqrt{x}\right) + \sqrt{x}\right),
\end{equation}
and the corresponding distribution function is
\begin{equation}
f({\mathcal{E}},L) = \frac{9 a^{3/2}} {2^{3/4} 4 \pi^{5/2} G^3 M^2 \Gamma{(1/4)}\Gamma{(13/4)}} L^{-3/2} {\mathcal{E}}^{9/4} + \frac{9 a}{16 \pi^{3} G^3 M^2} L^{-1} {\mathcal{E}}^{2} + \frac{15 \sqrt{a}}{32 \pi^{3} G^{5/2} M^{3/2}} L^{-1} {\mathcal{E}}^{3/2}.
\end{equation}
For the third distribution function, the parameters are $q=1$, $n_{0}=0$ and $n_{1}=1$ and the augmented density takes on the form
\begin{equation} \label{NFW3c}
\rho{(x,\Psi)} \propto \Psi^{5/2} x^{-3/2} \left(1 + \sqrt{\frac{\Psi}{\Phi_{0}}} \left(x+\sqrt{x}\right) \right).
\end{equation}
The corresponding distribution function is given by
\begin{equation}
f({\mathcal{E}},L) = \frac{45 a} {2^{3/4} 64 \pi^{2} G^{5/2} M^{3/2} \Gamma(1/4) \Gamma(11/4)} L^{-3/2} {\mathcal{E}}^{7/4} + \frac{9 a}{16 \pi^{3} G^3 M^2} L^{-1} {\mathcal{E}}^{2} + \frac{9 \sqrt{a}}{2^{1/4} 8 \pi^{5/2} G^{3} M^{2} \Gamma(3/4) \Gamma(11/4)} L^{-1/2} {\mathcal{E}}^{7/4}.
\end{equation}
In Fig. \ref{contourNFWlike}, the contour plots for some of these distribution functions have been plotted as functions of the dimensionless binding energy and dimensionless angular momentum. 
We will now proceed to compute the velocity dispersion profiles and the anisotropy parameter from the distribution functions obtained above for the $m=1/2$ Veltmann model.
\begin{figure*}
$$
\begin{array}{ccc}
 \includegraphics[width=5.2cm]{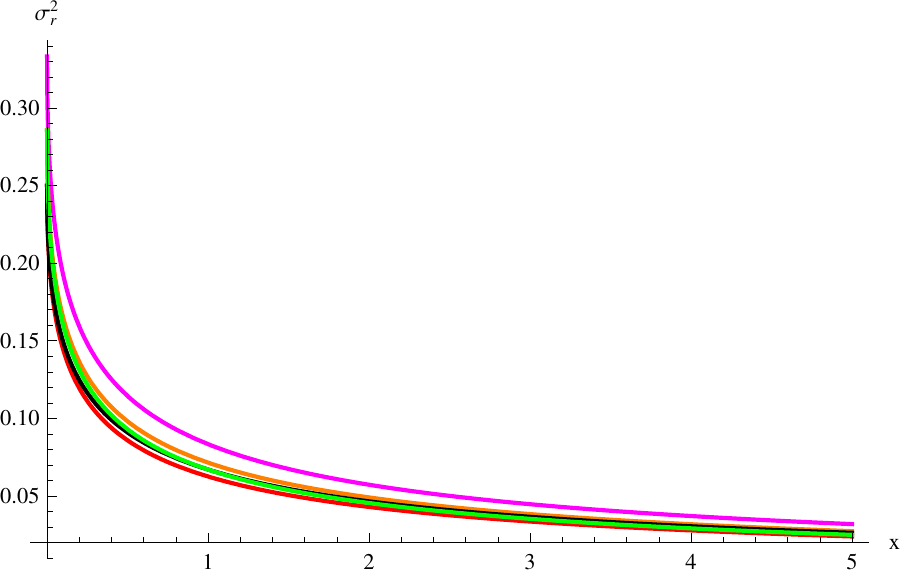} & \includegraphics[width=5.2cm]{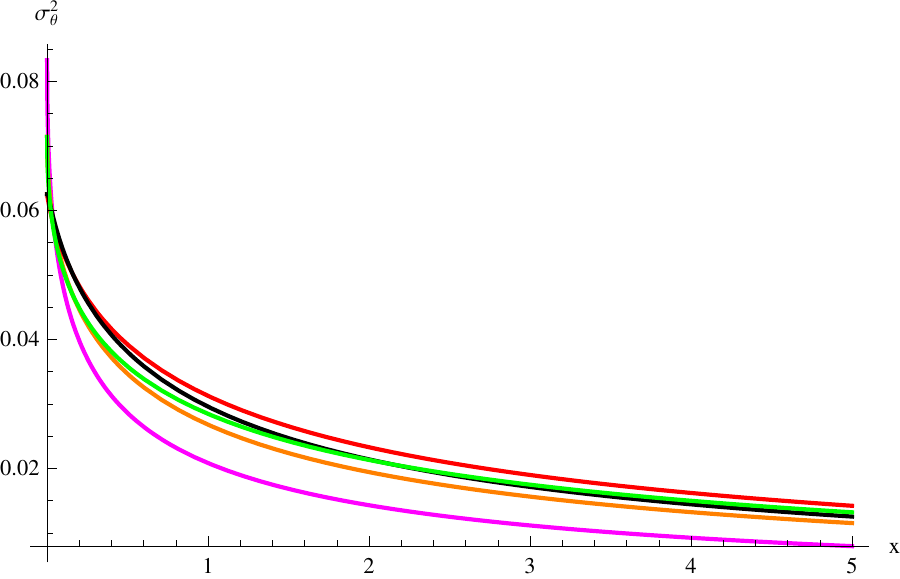} & \includegraphics[width=5.2cm]{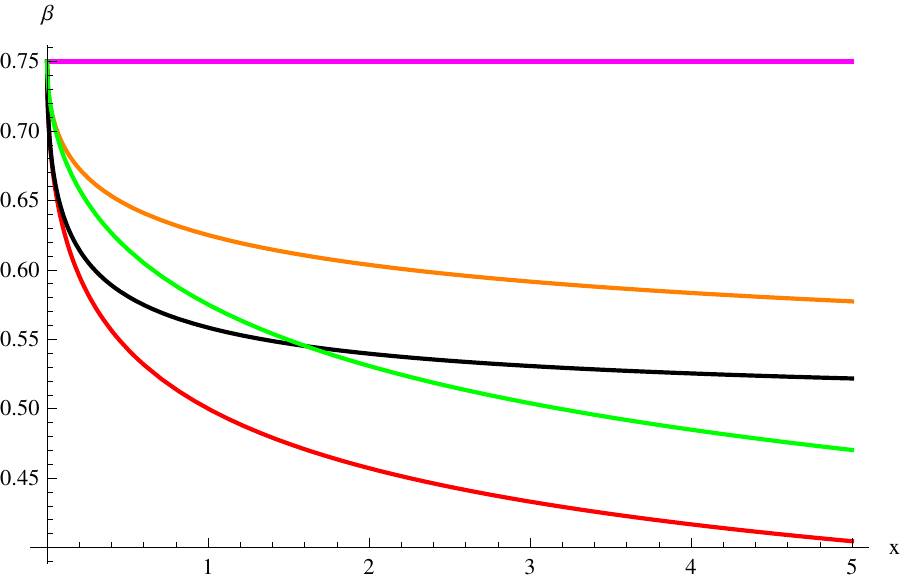}\\
 \quad\quad(a) & \quad\quad(b) & \quad\quad(c)\\
\end{array}
$$
\caption{(color figures online) The velocity dispersion tensor and the anisotropy parameter for multi-component NFW-like models. Left-hand panel: $\sigma^{2}_{r}$ as a function of $x$. Central panel: $\sigma^{2}_{\theta}$ as a function of $x$. Right-hand panel: $\beta$ as a function of $x$. For all panels: we plotted the one-component model (magenta), the two-component model (orange) and the three-component models (red, black and green, in the order in which they are discussed in Section \ref{NFWlikemodels}). Note: the velocity dispersions have been converted to dimensionless units before plotting.}
\label{fig3}
\end{figure*}

The simplest of all the models is the one with the hypervirial distribution function. It takes on the form given by equation (\ref{NFW1}), and the velocity dispersion profiles are as follows:
\begin{equation}
\sigma^2_r = 4 \sigma^2_\phi = \frac{1}{3} \frac{\Phi_{0}}{(1+\sqrt{x})^2},
\end{equation}
\begin{equation}
\beta = \frac{3}{4}.
\end{equation}
The two-component distribution function corresponds to a density that has the form (\ref{NFW2}). For this model, the velocity dispersions and anisotropy parameter are
\begin{equation}
\sigma^2_r = \frac{2}{7} \frac{\Phi_{0}}{(1+\sqrt{x})^2},
\end{equation}
\begin{equation}
\sigma^2_\phi = \frac{\Phi_{0}}{14} \frac{1+2\sqrt{x}}{(1+\sqrt{x})^3},
\end{equation}
\begin{equation}
\beta = \frac{3+2\sqrt{x}}{4\left(1+\sqrt{x}\right)}.
\end{equation}
We move on to the three-term distribution functions. The first of these, represented by the relation (\ref{NFW3a}), has the following velocity dispersion profiles and anisotropy parameter:
\begin{equation}
\sigma^2_r = \frac{1}{4} \frac{\Phi_{0}}{(1+\sqrt{x})^2},
\end{equation}
\begin{equation}
\sigma^2_\phi = \frac{\Phi_{0}}{16}\frac{\left(1+4\sqrt{x}+3x\right)}{\left(1+\sqrt{x}\right)^{4}},
\end{equation}
\begin{equation}
\beta = \frac{1}{4} \frac{3+4\sqrt{x}+x}{\left(1+\sqrt{x}\right)^{2}}.
\end{equation}
The next three-term distribution function is given by equation (\ref{NFW3b}). For this model, the corresponding values are
\begin{equation}
\sigma^2_r = \frac{\Phi_{0}}{28} \frac{7+8x+15\sqrt{x}}{\left(1+\sqrt{x}\right)^{4}},
\end{equation}
\begin{equation}
\sigma^2_\phi =  \frac{\Phi_{0}}{112} \frac{7+16x+30\sqrt{x}}{\left(1+\sqrt{x}\right)^{4}},
\end{equation}
\begin{equation}
\beta = \frac{21+16x+30\sqrt{x}}{28+32x+60\sqrt{x}}.
\end{equation}
For the last of the three-term NFW models, the density can be expressed as in equation (\ref{NFW3c}). For this model, one proceeds to calculate the velocity dispersions and the anisotropy parameter, which are given by
\begin{equation}
\sigma^2_r = \frac{\Phi_{0}}{28} \frac{8+7x+15\sqrt{x}}{\left(1+\sqrt{x}\right)^{4}},
\end{equation}
\begin{equation}
\sigma^2_\phi =  \frac{\Phi_{0}}{112} \frac{8+21x+22\sqrt{x}}{\left(1+\sqrt{x}\right)^{4}},
\end{equation}
\begin{equation}
\beta = \frac{24+7x+38\sqrt{x}}{32+28x+60\sqrt{x}}.
\end{equation}
The radial and tangential velocity dispersions, as well as the anisotropy parameter, for the distribution functions for the NFW-like model discussed in this section have been plotted in Fig. \ref{fig3}. This class of models have in common the fact that they are tangentially anisotropic in the outer regions. Also, we note that the models constructed here are stable to radial perturbations by the Doremus--Feix--Baumann theorem since all of them are decreasing functions of the energy.

\subsection{Constant anisotropy distribution functions}
Distribution functions which give rise to constant anisotropy parameters have been explored in numerous papers, and in particular, such distribution functions have been written down for the Veltmann models in, for instance, \citet{ea05}, \citet{an06}, \citet{ea06}, and \citet{an13}. In this section, we briefly re-derive those results from our approach and we provide explicit expressions for the case $m=1/2$. To start, we multiply and divide the augmented density (\ref{rho1term}) by some power of $x^{m}$:
\begin{equation}
\rho\left(\Psi,x\right)=\frac{\Phi_{0}(1+m)}{4\pi Ga^{2}}x^{m-2}\left(\frac{\Psi}{\Phi_{0}}\right)^{2m+1}x^{-ms}x^{ms}.
\end{equation}
where $s$ is positive. Using equation (\ref{Veltmanneq}), we obtain the augmented density
\begin{equation}\label{rho1classII}
\rho\left(\Psi,x\right)=\frac{\Phi_{0}(1+m)}{4\pi Ga^{2}}x^{m\left(s+1\right)-2}\left(\frac{\Psi}{\Phi_{0}}\right)^{m(2+s)+1}\left[1-\left(\frac{\Psi}{\Phi_{0}}\right)^{m}\right]^{-s}.
\end{equation}
The case $s=0$ is trivial, since in this case we recover equation $(\ref{rho1term})$. Also, it is easily seen that the parameter $s$ is related to the constant value of the anisotropy parameter by
\begin{equation}\label{betaclassII}
\beta_{c} = 1 - \frac{m}{2}(s+1).
\end{equation}
The right-hand side can now be expanded if $s\geq0$, using a generalized binomial theorem:
\begin{equation} \label{rho1classIIexpand}
\rho\left(\Psi,r\right)=\frac{\Phi_{0}(1+m)}{4\pi Ga^{2}}x^{m\left(s+1\right)-2}\left(\frac{\Psi}{\Phi_{0}}\right)^{m(2+s)+1}\sum_{k=0}^{\infty}\left(\begin{array}{c}
s+k-1\\
k
\end{array}\right)\left(\frac{\Psi}{\Phi_{0}}\right)^{mk}.
\end{equation}
The series converges since $\Psi \leq \Phi_{0}$ everywhere, for all Veltmann models. From this expansion, we can use our recipe to construct a distribution function in a series form and, if possible, sum up the series to obtain a closed form expression. For the $m=1/2$ model, we find the following family of distribution functions:
\begin{eqnarray}\label{classIINFW}
f({\mathcal{E}},L) &\propto& \frac{\left(\sqrt{2}a\Phi_{0}\right)^{-s/2}}{\Gamma{\left(\frac{s+1}{4}\right)}\Gamma\left(\frac{9+s}{4}\right)\Gamma\left(\frac{11+s}{4}\right)}
\Bigg[s\sqrt{\frac{\mathcal{E}}{\Phi_0}}\,\Gamma\left(\frac{7+s}{2}\right)\Gamma\left(\frac{9+s}{4}\right) {}_{3}F_{2}\left(\frac{1}{2}+\frac{s}{2},1+\frac{s}{2},\frac{7}{2}+\frac{s}{2};\frac{3}{2},\frac{11}{4}+\frac{s}{4};\frac{\mathcal{E}}{\Phi_0}\right) \\ \nonumber
&+&  \Gamma\left(\frac{6+s}{2}\right)\Gamma\left(\frac{11+s}{4}\right) {}_{3}F_{2}\left(\frac{1}{2}+\frac{s}{2},3+\frac{s}{2},\frac{s}{2};\frac{1}{2},\frac{9}{4}+\frac{s}{4};\frac{\mathcal{E}}{\Phi_0}\right)\Bigg]\, \left(\frac{\mathcal{E}}{\Phi_0}\right)^{(s+5)/4} L^{(s-3)/2}.
\end{eqnarray}
The choice $s=0$ reduces to the hypervirial distribution function discussed in \citet{ng13} and \citet{ea05}:
\begin{equation}\label{HypervirialDFNFW}
f({\mathcal{E}},L) \propto \left(\frac{\mathcal{E}}{\Phi_0}\right)^{5/4}L^{-3/2}.
\end{equation}
The choice $s=1$ gives the model
\begin{equation}
f({\mathcal{E}},L) \propto L^{-1} \left(1-\frac{\mathcal{E}}{\Phi_{0}}\right)^{-2} \left[\frac{5}{2}\left(\frac{\mathcal{E}}{\Phi_{0}}\right)^{3/2} - \frac{3}{2}\left(\frac{\mathcal{E}}{\Phi_{0}}\right)^{5/2}+3\left(\frac{\mathcal{E}}{\Phi_{0}}\right)^{2} -2\left(\frac{\mathcal{E}}{\Phi_{0}}\right)^{3}\right].
\end{equation}
Other values of $s$ where the distribution function can be written in terms of elementary functions are $s=5,9$ and $13$.

\section{Hypervirial distribution functions for an arbitrary potential--density pair} \label{SectII}
In this section, we derive the generic form of an augmented density which has the property of hyperviriality, and then apply the recipe above to construct hypervirial distribution functions for a few well-known potential--density pairs. We start with the statement of hyperviriality, which is that
\begin{equation}\label{hyperviriality}
\sigma_r^2 + \sigma_t^2 =\frac{1}{2}\Psi.
\end{equation}
Using the formulae to compute the velocity dispersion from the augmented density, this can be written as
\begin{equation}
\frac{1}{\rho}\intop_{0}^{\Psi}\rho\left(r,\Psi'\right)d\Psi'+2\frac{1}{\rho}\intop_{0}^{\Psi}\partial_{r^{2}}\left[r^{2}\rho\left(r,\Psi'\right)\right]d\Psi'=\frac{1}{2}\Psi.
\end{equation}
Differentiating both sides with respect to $\Psi$, we obtain the governing partial differential equation (PDE) for the augmented density:
\begin{equation}
\rho\left(r,\Psi\right)+4\partial_{r^{2}}\left[r^{2}\rho\left(r,\Psi\right)\right]=\Psi\frac{\partial\rho}{\partial\Psi}.
\end{equation}
On further simplification, we can cast the above equation in the form
\begin{equation} \label{hypergov1}
5\rho\left(r,\Psi\right)+2r\frac{\partial\rho\left(r,\Psi\right)}{\partial r}=\Psi\frac{\partial\rho}{\partial\Psi}.
\end{equation}
Note that the PDE is linear, so it can be solved using the series method. It is easily seen that any augmented density of the following form is a solution:
\begin{equation}
\rho = \sum_{k} C_{k} r^{p_{k}-2} \Psi^{2p_{k}+1},
\end{equation}
where the coefficients $C_{k}$ and the powers $p_{k}$ are freely specifiable. Note that this can be rewritten as a series in $\sqrt{r}\Psi$:
\begin{equation}\label{hypervirialrho}
\rho = r^{-5/2} \sum_{k}C_{k}(\sqrt{r}\Psi)^{2p_{k}+1}.
\end{equation}
This suggests a procedure to construct a hypervirial distribution function given an arbitrary potential--density pair. If we can express $r^{5/2}\rho$ as a function of $\sqrt{r}\Psi$:
\begin{equation}
r^{5/2}\rho = g(\sqrt{r}\Psi),
\end{equation}
then, through the Taylor series for $g$, we can compute the coefficients $C_{k}$ and powers $p_{k}$ for the augmented density, and use the double-power recipe to compute the distribution function in series form (the recipe only works if all $p_{k}$ are positive, or equivalently, if all powers in the Taylor series for $g$ are larger than 1). At this point, we point out a caveat: in general the function $g$ cannot be defined at all radii. This is because to construct $g$, we need to invert the function $\sqrt{r}\Psi$ as a function of $r$, but for all the models considered below $\sqrt{r}\Psi{(r)}$ is never a monotonic function. Instead, this function always has one maximum. As a result, the function $g$ has two branches forming a loop, one branch for the inner region and another branch for the outer region. The best we can do, then, is to compute hypervirial distribution functions that approximate either the inner region or the outer region well, but not both.\\
Note that the Veltmann models are quite special with respect to hyperviriality: indeed, for this family of potential--density pairs, the two branches coincide and we have a well-defined $g$ at all radii: $r^{5/2}\rho \propto (\sqrt{r}\Psi)^{2m+1}$, and we recover the usual, one-term hypervirial family. In the remainder of the section, we put this formalism to use to compute hypervirial distribution functions for some of the most commonly used profiles in astrophysics: the H\'enon isochrone model, the NFW profile, the Jaffe and another particular case of the Dehnen models. This can also be used to construct specific cases of the potential--density pair introduced by \citet{ta09} and whose relevance was further explored by \citet{w32} and \citet{li13}.\\
We will work with a truncated augmented density that comprises of a finite number of terms. The aim of this multi term augmented density, generated from a suitable double-power distribution function, is to model potential--density pairs of known models to an arbitrary degree of accuracy. Depending on the level of desired accuracy, we can choose the order of truncation and adjust the accuracy to an arbitrarily high order simply by including a sufficiently high number of terms. The coefficients in the augmented density are found by expanding the functions $\sqrt{r} \Psi(r)$ and $r^{5/2} \rho(r)$ to obtain series in the two asymptotic limits, around $r \rightarrow 0$ and $r \rightarrow \infty$, and matching the powers to find the coefficients. It is important to note that some of these terms may be negative, and this could very well lead to an invalid distribution function in these two regimes. Hence, it is necessary to establish the positivity and convergence of the augmented density ansatz defined by equation (\ref{hypervirialrho}) and the corresponding distribution function. When studying each model explicitly, we shall address these issues and establish the validity of the distribution function over an appropriate (finite) range in the two asymptotic limits.
We are also interested in determining the relative error, which measures the deviation of the augmented density from the exact one. The relative error is defined to be
\begin{equation}
\Delta = \left | \frac{\rho\left(\Psi(r),r\right) - \rho (r)}{\rho(r)} \right|.
\end{equation}
By evaluating $\Delta$ in the two limits, one can determine the accuracy of the augmented density (and the distribution function) in modelling the exact density.  Since the paper relies on expanding the potential--density pair as a series in a given asymptotic regime, the plots that we present of several quantities such as the velocity structure profiles are plotted only in the ranges in which these expressions are valid, i.e. for $x \ll 1$ and $x \gg 1$, respectively.

Finally, a few remarks regarding this approach are in order. It is not an approach that is always guaranteed to work since the matching of coefficients in equation (\ref{hypervirialrho}) may lead to internal inconsistencies. In some other cases, it may turn out that there is no series expansion of the potential--density pair. For these reasons, it was found that profiles such as the Einasto profile and the singular isothermal sphere could not be modelled using this approach, owing to the former and latter reasons, respectively.

\subsection{The hypervirial H\'enon isochrone model}
\begin{figure*}
$$
\begin{array}{ccc}
 \includegraphics[width=6.8cm]{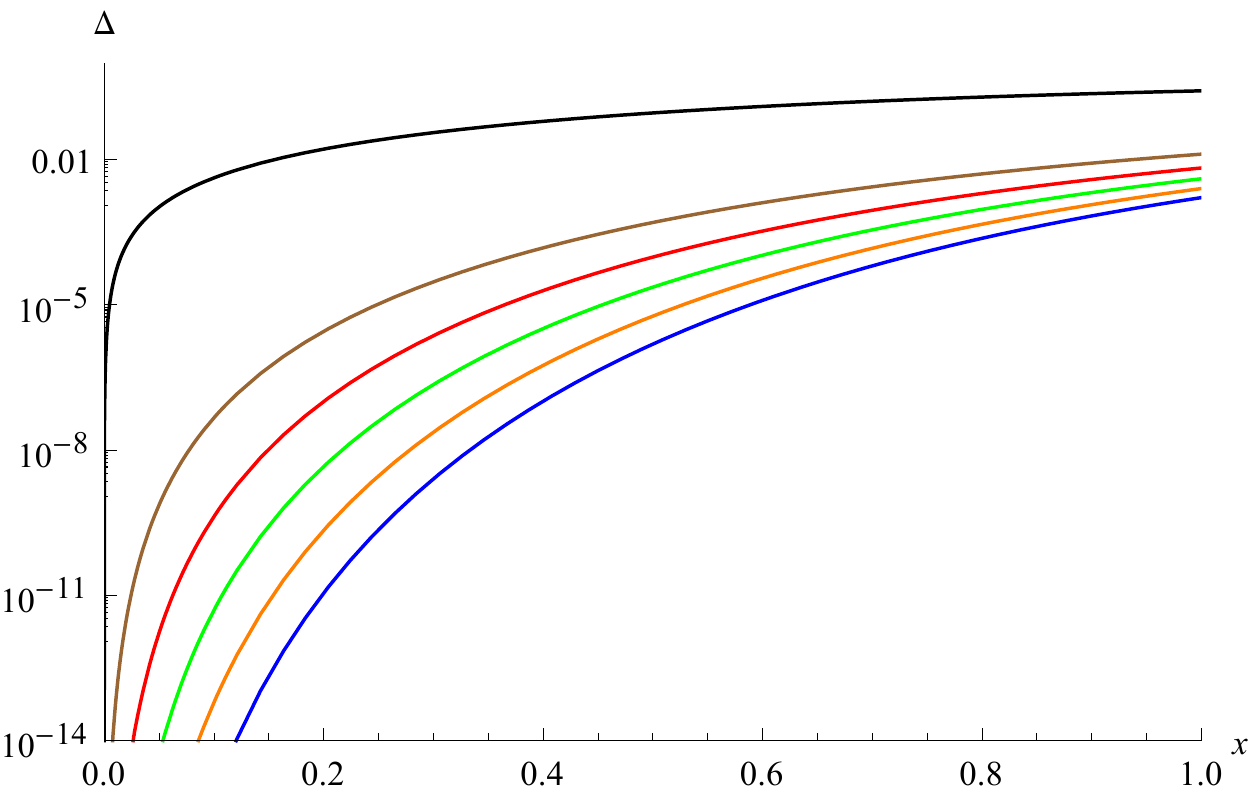} & \includegraphics[width=6.8cm]{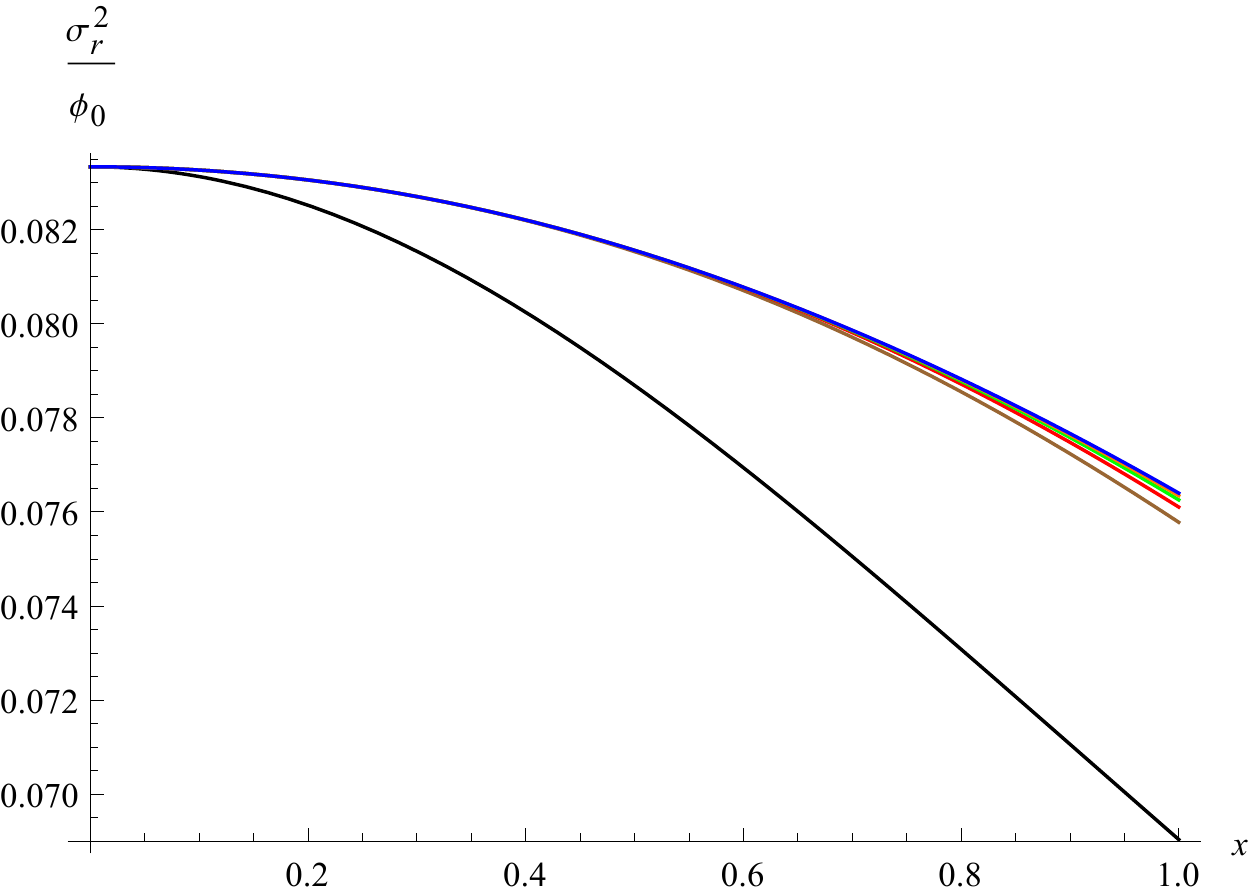} \\
 \includegraphics[width=6.8cm]{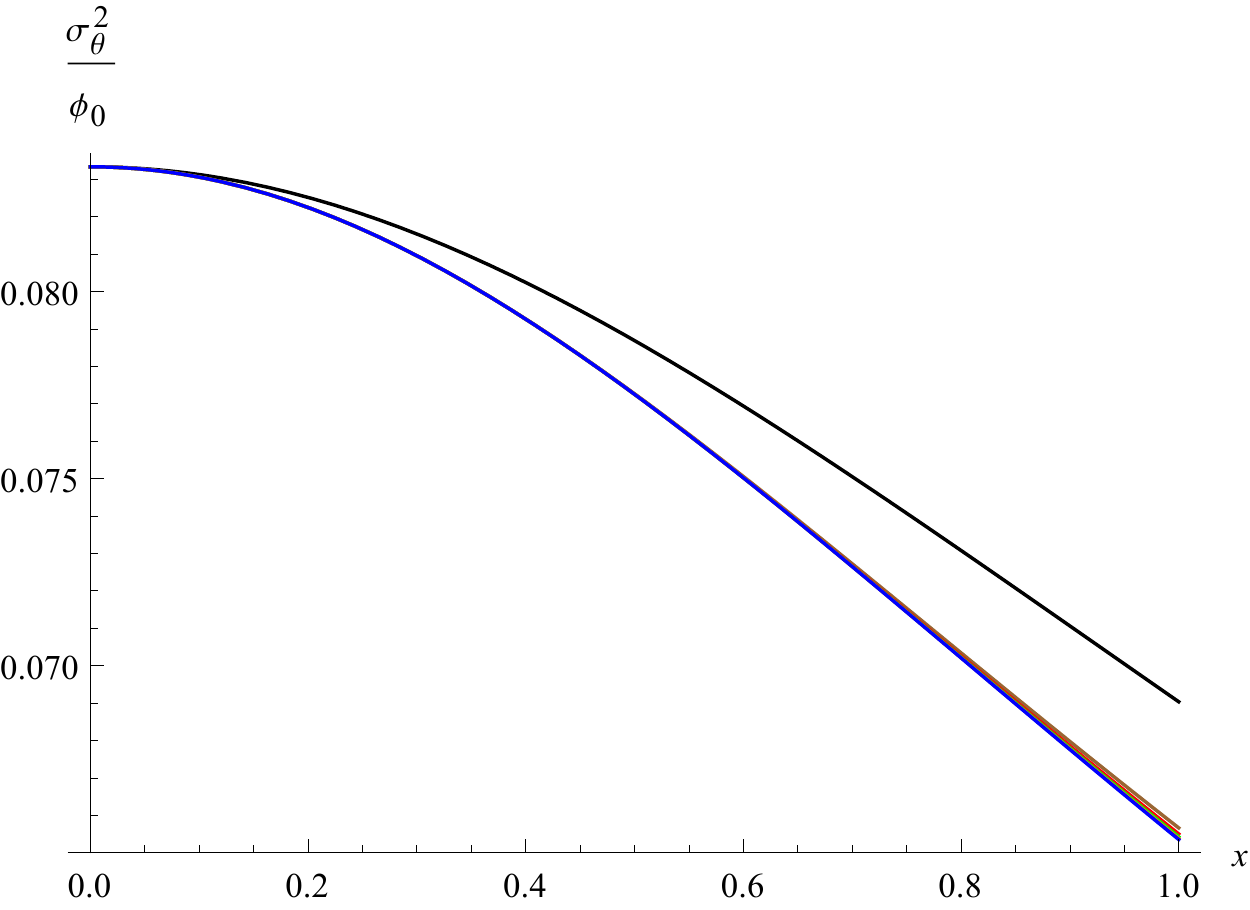} & \includegraphics[width=6.8cm]{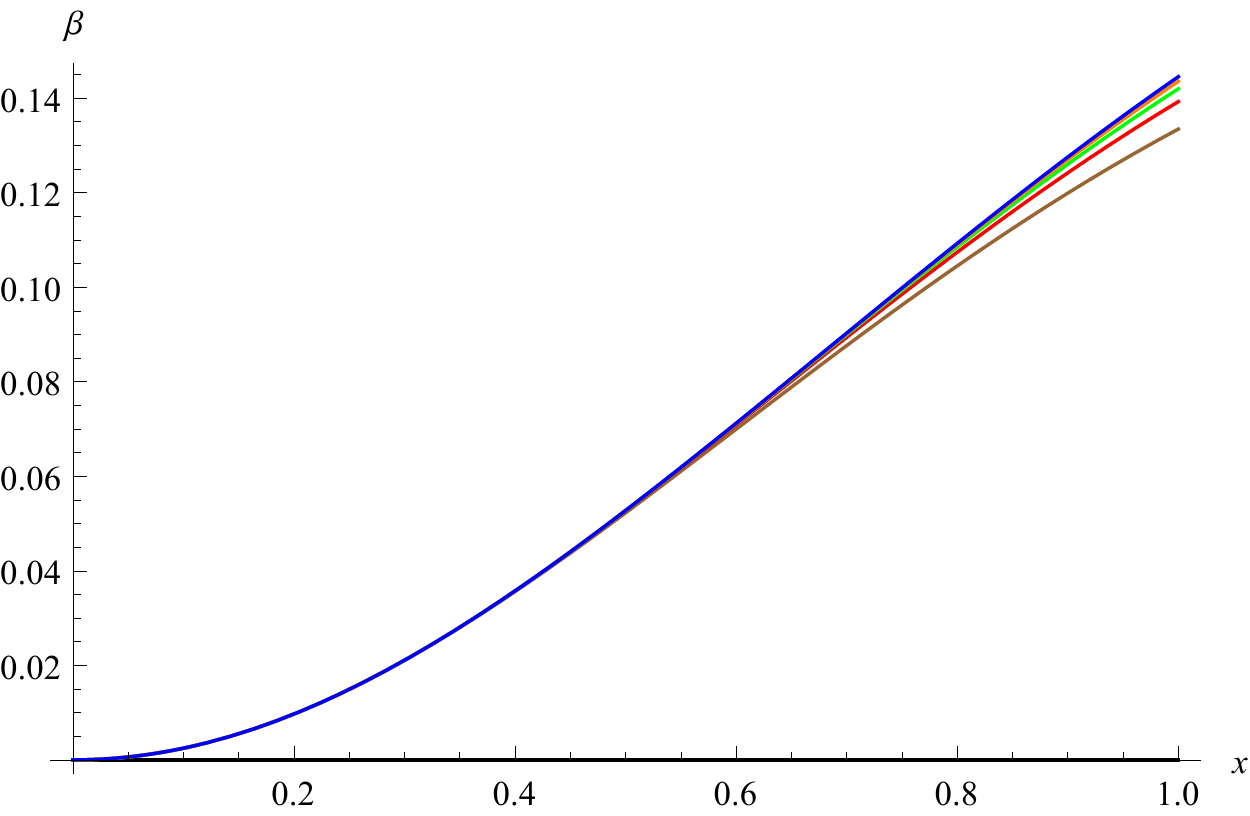}
\end{array}
$$
\caption{(color figures online) Plots for the hypervirial isochrone model near the origin. Top-left panel: $\Delta$ versus $x$. Top-right panel: the radial velocity dispersion versus $x$. Bottom-left panel: the tangential velocity dispersion versus $x$. Bottom-right panel: the anisotropy parameter $\beta$ versus $x$. For all panels: the black, brown, red, green, orange and blue curves correspond to the distribution function with 1, 2, 3, 4, 5 and 6 terms, respectively.}
\label{fig2}
\end{figure*}

The potential--density pair for the H\'enon isochrone model \citep{hen59} is:
\begin{equation}\label{Psiisochrone}
\Psi = \frac{GM}{a+r_{*}},
\end{equation}
\begin{equation}
\rho = \frac{Ma(a+2r_{*})}{4\pi r_{*}^{3}(a+r_{*})^{2}},
\end{equation}
where $r_{*}=\sqrt{r^{2}+a^{2}}$. From the first equation above, we have:
\begin{equation}
y = \frac{\sqrt{x}}{1+\sqrt{1+x^2}},
\end{equation}
where we defined the dimensionless quantities $y = \sqrt{ra}\Psi/GM$ and $x = r/a$. As claimed above, $y(x)$ is not monotonic, and $g$ is multi valued. We will work with the inner branch of $g$. Inverting $y(x)$, we find three roots, which, when substituted into $r^{5/2}\rho$ give the following expansions:
\begin{equation}
r^{5/2}\rho \propto 2y^{3}-9y^{5} + \cdots
\end{equation}
\begin{equation}
r^{5/2}\rho \propto 24y^{5} - 160y^{9} + \cdots
\end{equation}
\begin{equation}
r^{5/2}\rho \propto 2y^{3} + 3y^{5} + 6y^{7} + \cdots
\end{equation}
In order to tell which of the three expansions is the correct one for the inner branch, we consider the leading terms. The leading terms in the first and third expansions give the Hernquist distribution function, and therefore give rise to infinite density at the origin. On the other hand, the second expansion starts with the Plummer distribution function, and therefore has a finite central density (all subleading terms are distribution functions for shell-like models, whose central density vanishes). Since the isochrone model has a core in the centre rather than a cusp, the second expansion is correct.\\
The augmented density is
\begin{equation}\label{rhoisochronesmallr}
\rho{(r,\Psi)} = \frac{6a^{2}}{\pi G^{5}M^{4}}\Psi^{5}-\frac{40a^{4}}{\pi G^{9}M^{8}}r^{2}\Psi^{9} -\frac{1152a^{8}}{\pi G^{17}M^{16}}r^{6}\Psi^{17}
- \frac{16896 a^{10}}{\pi G^{21} M^{20}}r^8 \Psi^{21} - \frac{319488a^{12}}{\pi G^{25}M^{24}}r^{10}\Psi^{25} - \frac{6389760 a^{14}}{\pi G^{29}M^{28}}r^{12}\Psi^{29}.
\end{equation}
Using the recipe from Section \ref{SectI}, the corresponding distribution function is found to be
\begin{eqnarray}\label{fisochronesmallr}
f({\mathcal{E}},L) &=& \frac{192\sqrt{2}a^2}{7\pi^{3} G^5 M^4} {\mathcal{E}}^{7/2} - \frac{245760\sqrt{2} a^4}{143 \pi^{3} G^9 M^8} L^2 {\mathcal{E}}^{13/2} - \frac{4831838208 a^8}{2185 \pi^{3} G^{17} M^{16}} L^6 {\mathcal{E}}^{25/2} - \frac{3968549781504\sqrt{2}a^{10}}{20677 \pi^{3} G^{21} M^{20}} L^8 {\mathcal{E}}^{31/2} \nonumber \\ 
&-& \frac{686095255732224\sqrt{2}a^{12}}{33263 \pi^{3} G^{25} M^{24}} L^{10} {\mathcal{E}}^{37/2} - \frac{4610560118520545280\sqrt{2}a^{14}}{2022161\pi^{3}G^{29}M^{28}}L^{12}\mathcal{E}^{43/2}.
\end{eqnarray}
At this point, a remark concerning the convergence of the series above is in order. While it is clear that the series (\ref{rhoisochronesmallr}) converges to the function $g$, it is important to keep in mind that convergence only happens within a finite radius around $\sqrt{r}\Psi=0$ (i.e. around the origin). Ideally, we would like to compute the radius of convergence but such a task is inordinately difficult, since it involves deriving a formula for all the coefficients in the series. The same issue applies to the series (\ref{fisochronesmallr}). The latter clearly converges within a sufficiently small neighbourhood of the origin (since $L \rightarrow 0$ as $r \rightarrow 0$ while $\mathcal{E}$ is bounded by $GM/a$ at all radii), and it is also clearly positive definite sufficiently close to the origin, since the dominant term is positive. For large enough values of $L$, the series (\ref{fisochronesmallr}) becomes negative, but it is unclear whether this happens within the radius of convergence or outside this radius \footnote{To illustrate this point, consider the expansion of the exponential function $\mathrm{e}^{-x} \approx 1-x$. The expansion to first order becomes negative for $x > 1$, but this happens outside of the radius of convergence, and the actual function is always positive.}. We will therefore content ourselves with working in the small $r$ limit. \\
To illustrate the fact that the result above does not approximate the outer region well, note that while the Plummer model and the isochrone model have the same inner cusp ($r^{0}$), the outer falloff of the isochrone model ($r^{-4}$) is less steep than that of its hypervirial approximation ($r^{-5}$). Finally, for the isochrone model near the origin, the value of $\Delta$ to leading order is
\begin{equation}
\Delta = \frac{4097}{49152}x^{14}.
\end{equation} 
We see that the approximation is excellent. The velocity dispersion tensor is
\begin{equation}
\sigma_r^2 = \frac{\Psi}{2} \frac{-1+4x^{2}(\frac{\Psi}{\phi_{0}})^{4}+64x^{6}(\frac{\Psi}{\phi_{0}})^{12}+768x^{8}(\frac{\Psi}{\phi_{0}})^{16}+12288x^{10}(\frac{\Psi}{\phi_{0}})^{20}+212992x^{12}(\frac{\Psi}{\phi_{0}})^{24}}{-3+20x^{2}(\frac{\Psi}{\phi_{0}})^{4}+576x^{6}\Psi^{12}+8448x^{8}(\frac{\Psi}{\phi_{0}})^{16}+159744x^{10}(\frac{\Psi}{\phi_{0}})^{20}+3194880x^{12}(\frac{\Psi}{\phi_{0}})^{24}},
\end{equation}
\begin{equation}
\sigma_\theta^2 = \frac{\Psi}{2}\frac{-1+8x^{2}(\frac{\Psi}{\phi_{0}})^{4}+256x^{6}(\frac{\Psi}{\phi_{0}})^{12}+3840x^{8}(\frac{\Psi}{\phi_{0}})^{16}+73728x^{10}(\frac{\Psi}{\phi_{0}})^{20}+1490944x^{12}(\frac{\Psi}{\phi_{0}})^{24}}{-3+20x^{2}(\frac{\Psi}{\phi_{0}})^{4}+576x^{6}(\frac{\Psi}{\phi_{0}})^{12}+8448x^{8}(\frac{\Psi}{\phi_{0}})^{16}+159744x^{10}(\frac{\Psi}{\phi_{0}})^{20}+3194880x^{12}(\frac{\Psi}{\phi_{0}})^{24}},
\end{equation}
and with the following anisotropy parameter
\begin{equation}
\beta = -4x^{2}\left(\frac{\Psi}{\phi_{0}}\right)^{4}\frac{1+48x^{4}(\frac{\Psi}{\phi_{0}})^{8}+768x^{6}(\frac{\Psi}{\phi_{0}})^{12}+15360x^{8}(\frac{\Psi}{\phi_{0}})^{16}+319488x^{10}(\frac{\Psi}{\phi_{0}})^{20}}{-1+4x^{2}(\frac{\Psi}{\phi_{0}})^{4}+64x^{6}(\frac{\Psi}{\phi_{0}})^{12}+768x^{8}(\frac{\Psi}{\phi_{0}})^{16}+12288x^{10}(\frac{\Psi}{\phi_{0}})^{20}+212992x^{12}(\frac{\Psi}{\phi_{0}})^{24}}.
\end{equation}
The velocity dispersion, the anisotropy parameter and $\Delta$ are plotted in Fig. \ref{fig2} to various orders of approximation. As can be seen from this figure, the convergence in the velocity structure profiles is excellent as long as we remain close to the origin. Furthermore, we also see that incorporating successive terms leads to a boost in the accuracy of these models, particularly near $r \rightarrow 0$. 

\begin{figure*}
$$
\begin{array}{ccc}
 \includegraphics[width=6.8cm]{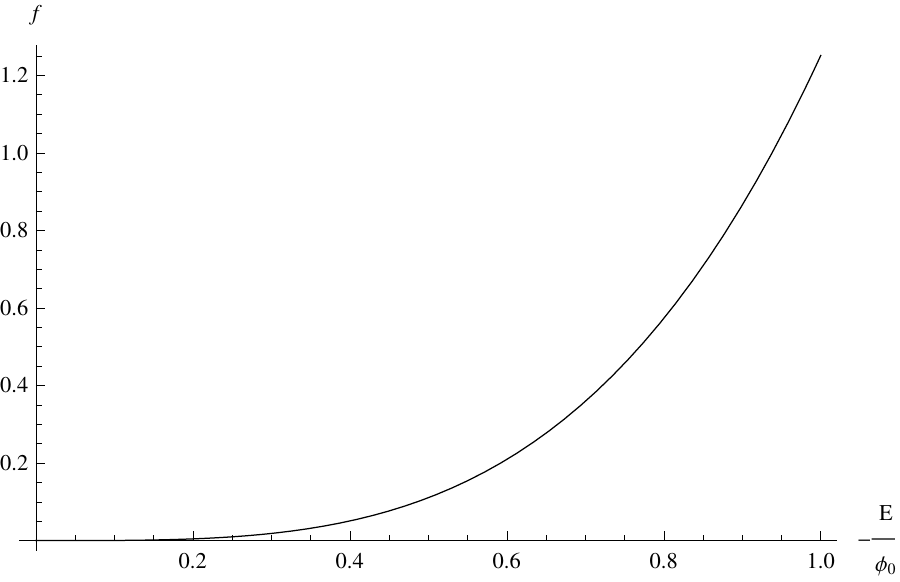} & \includegraphics[width=6.8cm]{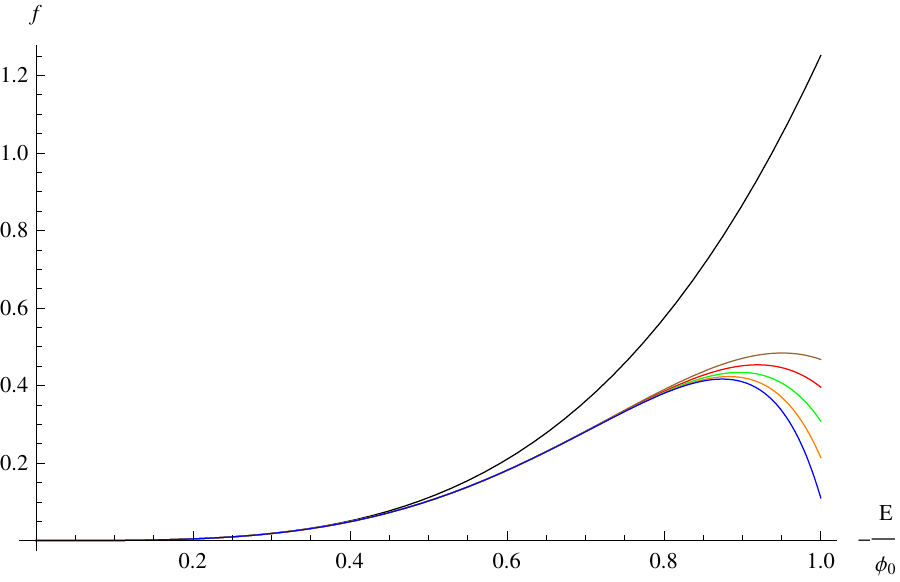} \\
 \includegraphics[width=6.8cm]{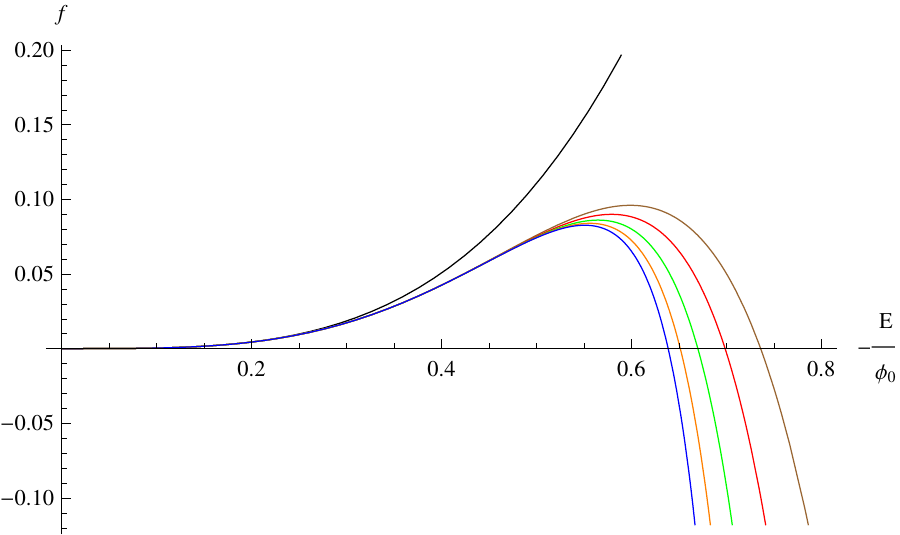} & \includegraphics[width=6.8cm]{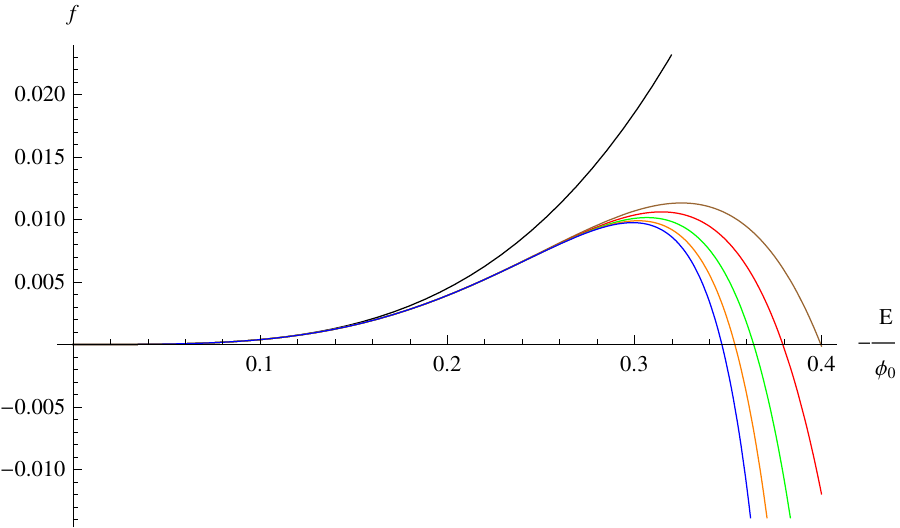}
\end{array}
$$
\caption{(color figures online) Plots of the distribution function as a function of the energy, for different fixed values of angular momentum, for the hypervirial isochrone model near the origin. Top left-hand panel: $L=0$. Top right-hand panel: $L=0.1 a\sqrt{\Phi_0}$. Bottom-left panel: $L=0.2 a\sqrt{\Phi_0}$. Bottom-right panel: $L=0.5 a\sqrt{\Phi_0}$. For all panels: the black, brown, red, green, orange and blue curves correspond to the distribution function with 1, 2, 3, 4, 5 and 6 terms, respectively.}
\label{fig3dot1}
\end{figure*}

We also plot the dimensionless distribution function as a function of the dimensionless binding energy $-\frac{E}{\Phi_0}$, for differing fixed values of the angular momentum in Fig. \ref{fig3dot1}. We see that as $L$ increases, the radius of convergence gets smaller (for $L=0.1 a\sqrt{\Phi_0}$ it is about 0.8, for $L=0.2 a\sqrt{\Phi_0}$ it is about 0.5, and for $L=0.5 a\sqrt{\Phi_0}$ it is about 0.3) but there is indeed a non-zero radius of convergence. A second crucial point is that the plots go negative, but this happens at each time outside of the radius of convergence. This provides compelling evidence that the exact distribution function may be positive definite. A final remark is due regarding the chosen values of $L$ -- we are working with functions of $r$ and $v$, but the latter is bounded from above. Since the former is very small, it follows that the values of $L$ will also turn out to be accordingly small.\\
Next, we consider the outer branch. By repeating the procedure and following a similar line of reasoning, we find the following augmented density and distribution function:
\begin{eqnarray}\label{rhoisochronelarger}
\rho &=& \left(\frac{a}{2\pi G^{3}M^{2}}\right)r^{-1}\Psi^{3} + \left(\frac{3 a^2}{4\pi G^{5}M^{4}}\right)\Psi^{5} + \left(\frac{3a^3}{2\pi G^{7}M^{6}}\right)r\Psi^{7} + \left(\frac{15 a^{4}}{4\pi G^{9}M^{8}}\right)r^{2}\Psi^{9} \\ \nonumber
&+& \left(\frac{177a^5}{16\pi G^{11}M^{10}}\right)r^{3}\Psi^{11} + \left(\frac{147a^{6}}{4\pi G^{13}M^{12}}\right)r^{4}\Psi^{13},
\end{eqnarray}
\begin{eqnarray}\label{fisochronelarger}
f(\mathcal{E},L) &=& \frac{3a}{4\pi^{3}G^{3}M^{2}}\frac{\mathcal{E}^{2}}{L} + \frac{24\sqrt{2}a^{2}}{7\pi^{3}G^{5}M^{4}}\mathcal{E}^{7/2}+\frac{63a^{3}}{2\pi^{3}G^{7}M^{6}}L\mathcal{E}^{5}+\frac{23040\sqrt{2}a^{4}}{143\pi^{3}G^{9}M^{8}}L^{2}\mathcal{E}^{13/2} \\ \nonumber
&+& \frac{29205 a^{5}}{16\pi^{3}G^{11}M^{10}}L^{3}\mathcal{E}^{8} + \frac{3612672\sqrt{2}a^{6}}{323\pi^{3}G^{13}M^{12}}L^{4}\mathcal{E}^{19/2}.
\end{eqnarray}
The dominant term in equation (\ref{fisochronelarger}) is the Hernquist model. This is expected, since the isochrone and the Hernquist profiles do have a density that goes as $r^{-4}$ in the limit $r \rightarrow \infty$. 

Again, a remark concerning the convergence of equations (\ref{rhoisochronelarger}) and (\ref{fisochronelarger}) is in order. While it is clear that the former converges sufficiently far from the origin, the convergence of the latter is more subtle. This follows from the fact that, as $r \rightarrow \infty$, $L$ is not bounded from above but the upper-bound for $\mathcal{E}$ tends to zero, because of the cutoff imposed on the velocity $v < \sqrt{2\Psi}$. To investigate what happens in the large $r$ limit, we first factor out the dominant term in equation (\ref{fisochronelarger}). We then have an expression of the form:
\begin{equation}
f(\mathcal{E},L) = K_{1}\frac{\mathcal{E}^2}{L}\left[1 + K_{2}L\mathcal{E}^{3/2} + K_{3}L^{2}\mathcal{E}^{3} + \cdots \right],
\end{equation}
where $K_{1} > 0$. Converting the factor inside the square bracket to $(r,v)$ variables, and writing $v^{2} = 2\alpha\Psi$ for some $0<\alpha<1$, the above becomes:
\begin{equation} \label{asymptoticiso}
f(\mathcal{E},L) = K_{1}\frac{\mathcal{E}^2}{L}\left[1 + K_{2}\sqrt{2\alpha}(1-\alpha)^{3/2}|\sin{\eta}|r\Psi^{2} + K_{3}2\alpha(1-\alpha)^{3}\sin^{2}{\eta}r^{2}\Psi^{4} + \cdots \right].
\end{equation}
Finally, substituting the isochrone potential (\ref{Psiisochrone}) in the above, and expanding around infinity, we find $r\Psi^{2} \approx r^{-1}$, $r^{2}\Psi^{4} \approx r^{-2}$, etc. Thus, we have established that the terms in the expansion above become smaller and smaller, and the series converges (and is clearly positive).\\

\begin{figure*}
$$
\begin{array}{ccc}
 \includegraphics[width=6.8cm]{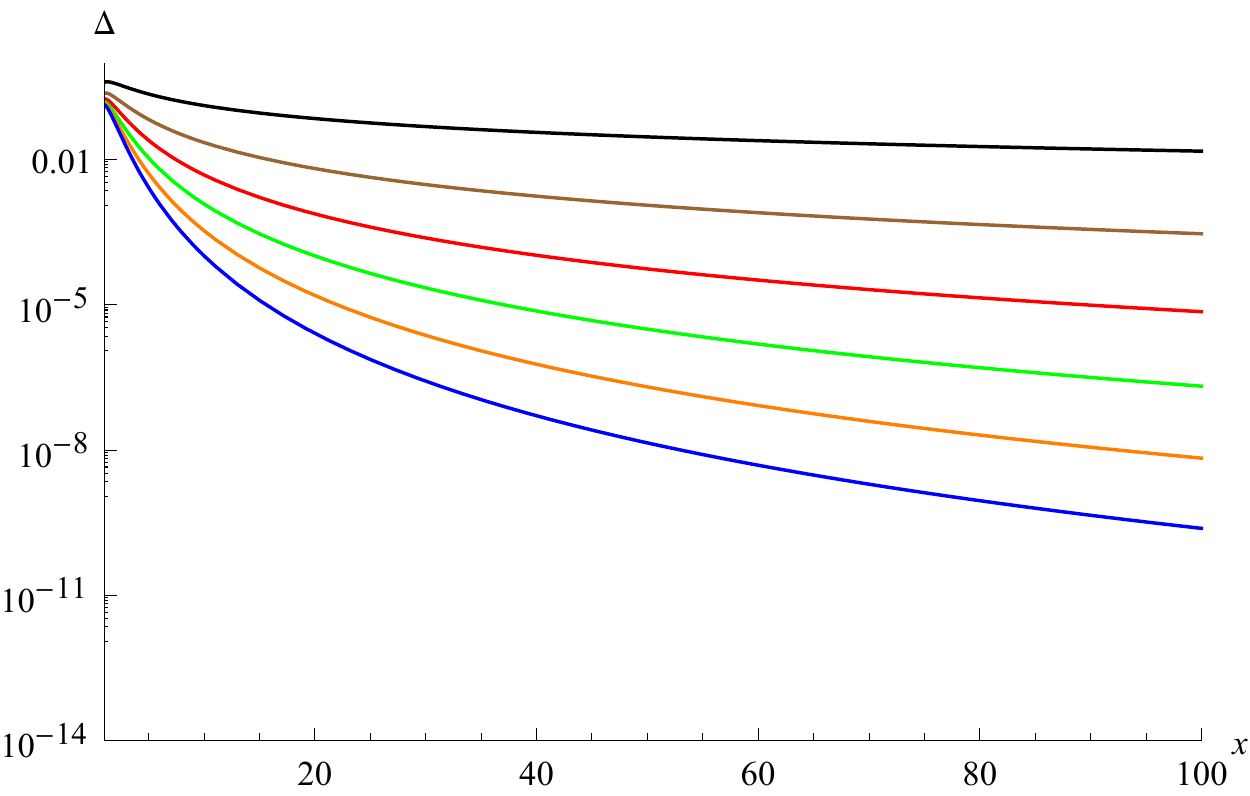} & \includegraphics[width=6.8cm]{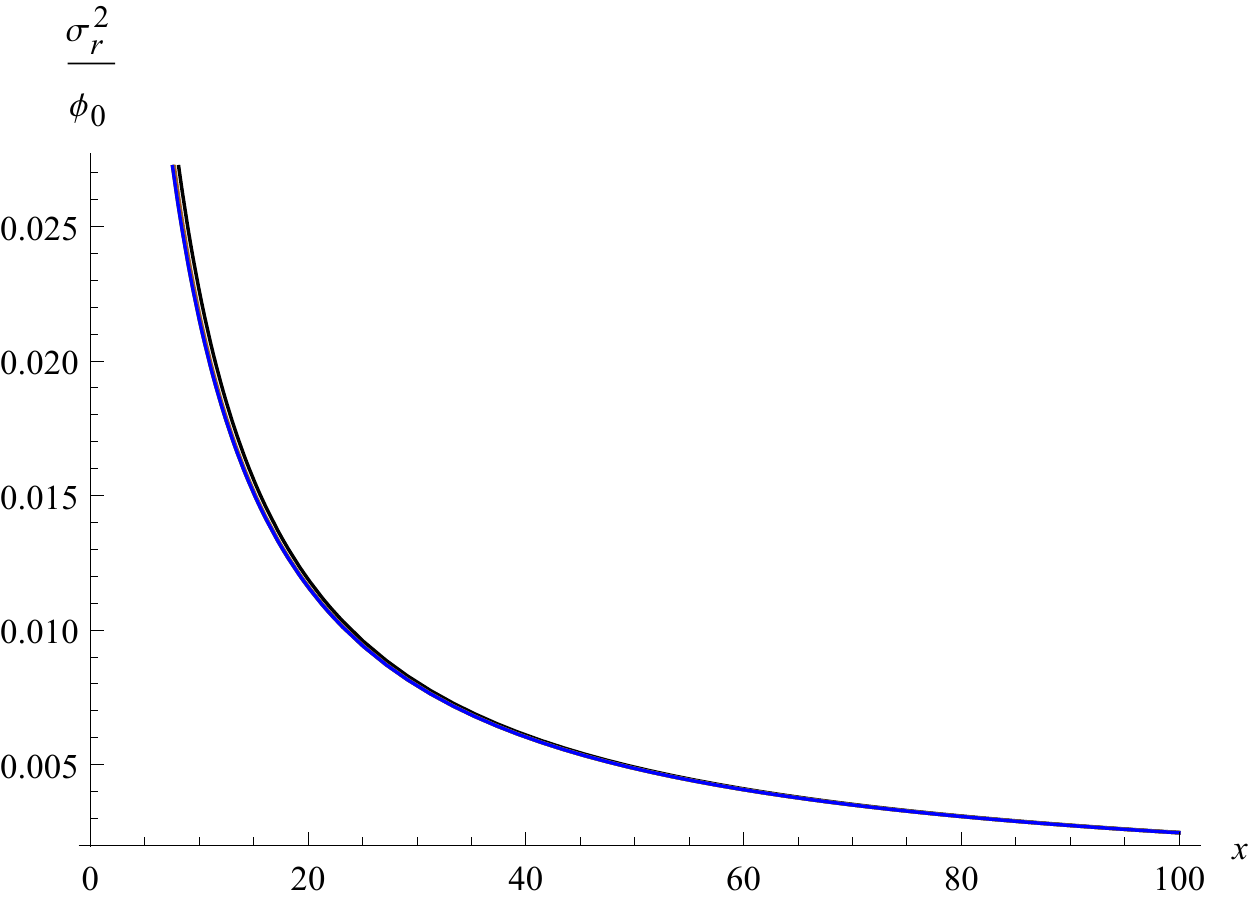} \\
 \includegraphics[width=6.8cm]{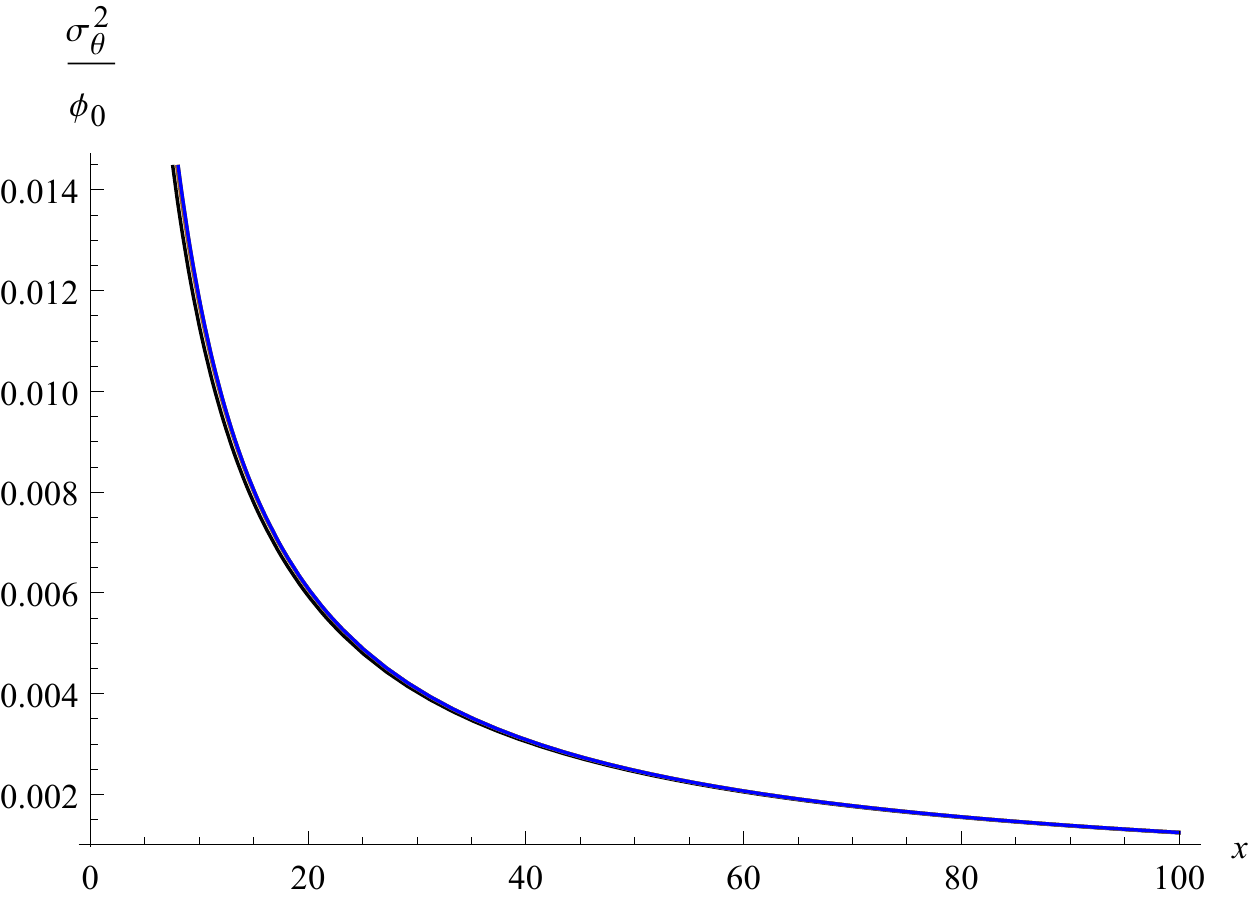} & \includegraphics[width=6.8cm]{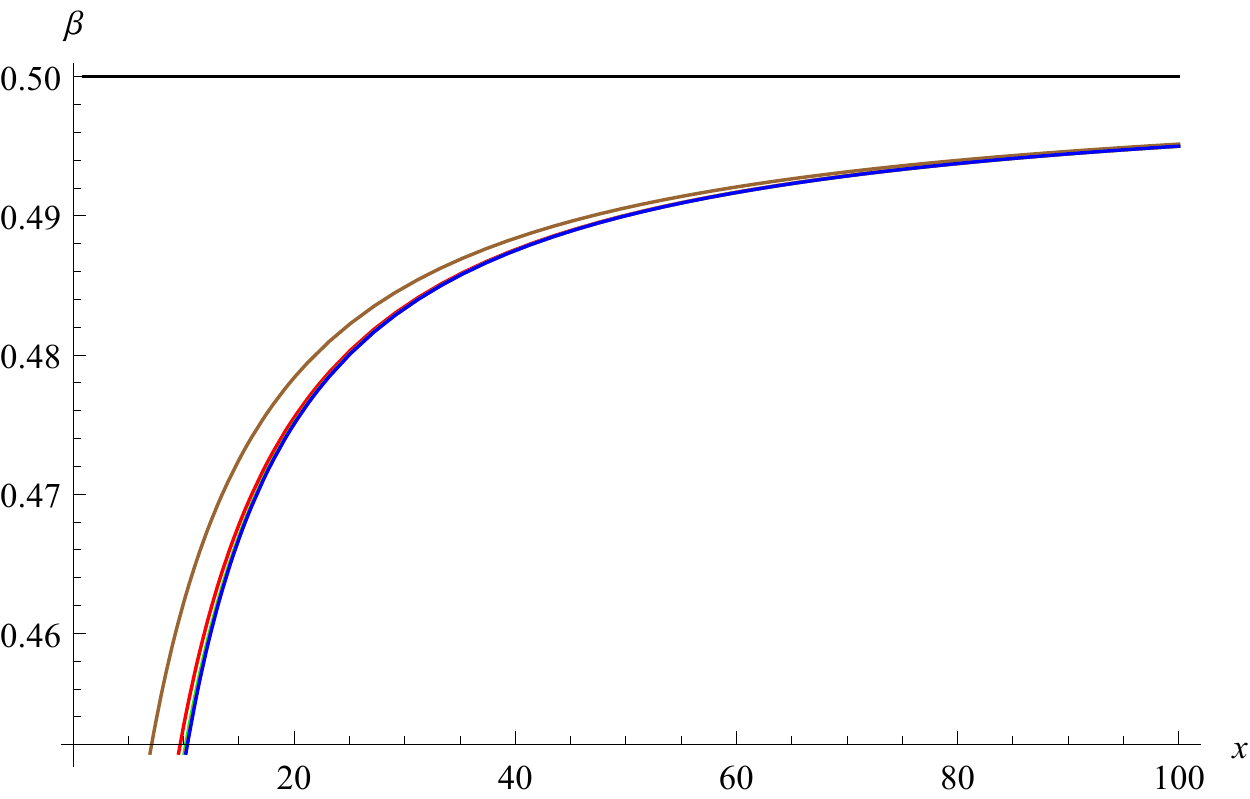}
\end{array}
$$
\caption{(color figures online) Plots for the hypervirial isochrone model near infinity. Top-left panel: $\Delta$ versus $x$. Top-right panel: the radial velocity dispersion versus $x$. Bottom-left panel: the tangential velocity dispersion versus $x$. Bottom-right panel: the anisotropy parameter $\beta$ versus $x$. For all panels: the black, brown, red, green, orange and blue curves correspond to the distribution function with 1, 2, 3, 4, 5 and 6 terms, respectively.}
\label{fig3dot5}
\end{figure*}

The property $\lim_{r \rightarrow \infty} r\Psi^{2} = r^{-1}$ is actually very general: it is common to all models of finite total mass. To see this, notice that for such a model the mass contained inside a radius $r$ tends to a constant as $r \rightarrow \infty$, and is the total mass. Thus, $\Psi \approx r^{-1}$ in this limit, being just the potential of a point mass, and the property follows.\\
The value of $\Delta$ is
\begin{equation}
\Delta = \frac{1063}{4}x^{-6}.
\end{equation}
The velocity dispersion profiles are given by
\begin{equation}
\sigma_{r}^{2} = \frac{\Psi}{4}\frac{8+8x(\frac{\Psi}{\phi_{0}})^{2}+12x^{2}(\frac{\Psi}{\phi_{0}})^{4}+24x^{3}(\frac{\Psi}{\phi_{0}})^{6}+59x^{4}(\frac{\Psi}{\phi_{0}})^{8}+168x^{5}(\frac{\Psi}{\phi_{0}})^{10}}{8+12x(\frac{\Psi}{\phi_{0}})^{2}+24x^{2}(\frac{\Psi}{\phi_{0}})^{4}+60x^{3}(\frac{\Psi}{\phi_{0}})^{6}+177x^{4}(\frac{\Psi}{\phi_{0}})^{8}+588x^{5}(\frac{\Psi}{\phi_{0}})^{10}},
\end{equation}
\begin{equation}
\sigma_{\theta}^{2} = \frac{\Psi}{8}\frac{8+16x(\frac{\Psi}{\phi_{0}})^{2}+36x^{2}(\frac{\Psi}{\phi_{0}})^{4}+96x^{3}(\frac{\Psi}{\phi_{0}})^{6}+295x^{4}(\frac{\Psi}{\phi_{0}})^{8}+1008x^{5}(\frac{\Psi}{\phi_{0}})^{10}}{8+12x(\frac{\Psi}{\phi_{0}})^{2}+24x^{2}(\frac{\Psi}{\phi_{0}})^{4}+60x^{3}(\frac{\Psi}{\phi_{0}})^{6}+177x^{4}(\frac{\Psi}{\phi_{0}})^{8}+588x^{5}(\frac{\Psi}{\phi_{0}})^{10}}.
\end{equation}
The anisotropy parameter is
\begin{equation}
\beta = \frac{8-12x^{2}(\frac{\Psi}{\phi_{0}})^{4}-48x^{3}(\frac{\Psi}{\phi_{0}})^{6}-177x^{4}(\frac{\Psi}{\phi_{0}})^{8}-672x^{5}(\frac{\Psi}{\phi_{0}})^{10}}{2(8+8x(\frac{\Psi}{\phi_{0}})^{2}+12x^{2}(\frac{\Psi}{\phi_{0}})^{4}+24x^{3}(\frac{\Psi}{\phi_{0}})^{6}+59x^{4}(\frac{\Psi}{\phi_{0}})^{8}+168x^{5}(\frac{\Psi}{\phi_{0}})^{10})}.
\end{equation}
The velocity dispersion, anisotropy parameter and $\Delta$ for the outer branch of the isochrone model are plotted in Fig. \ref{fig3dot5}. As before, we have plotted the quantities as a function of $x$ and including successively higher order terms to illustrate the convergence of these models, and the increase in accuracy.

\begin{figure*}
$$
\begin{array}{ccc}
 \includegraphics[width=6.8cm]{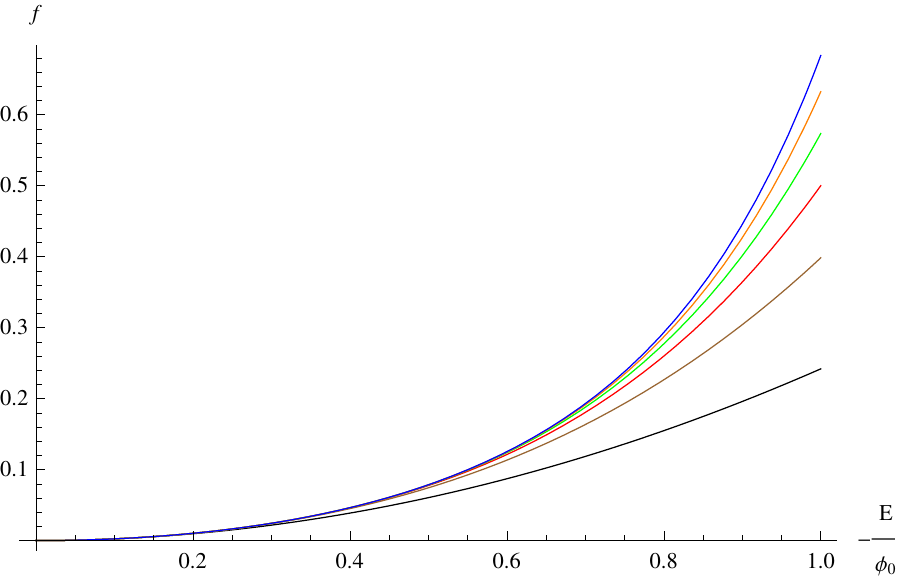} & \includegraphics[width=6.8cm]{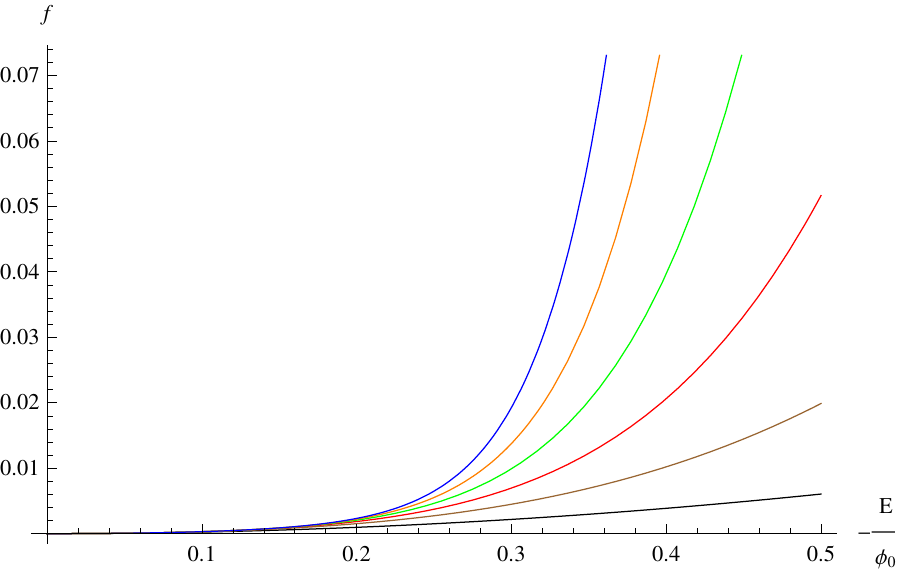} \\
 \includegraphics[width=6.8cm]{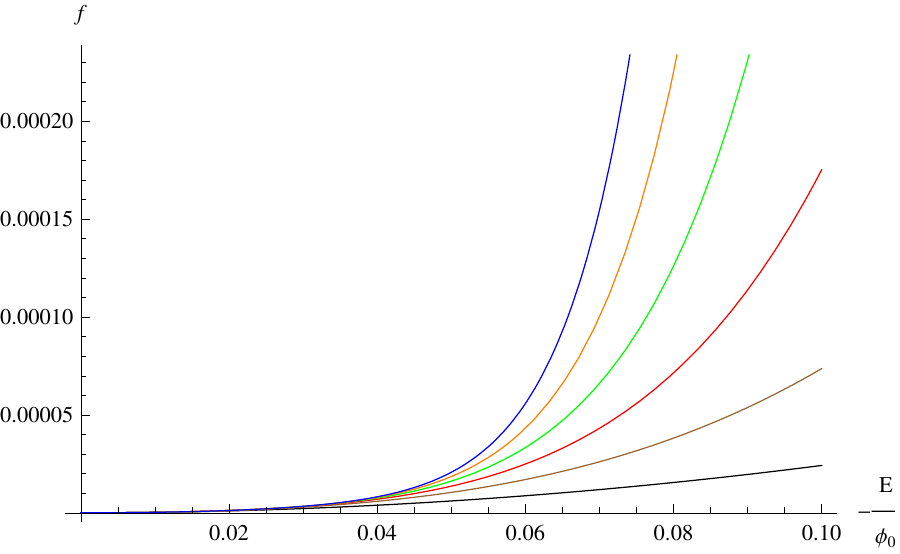} & \includegraphics[width=6.8cm]{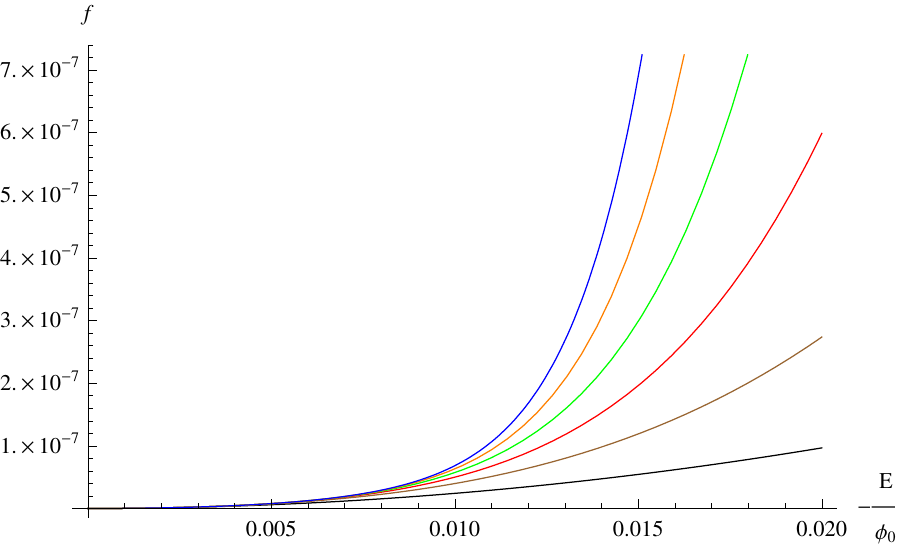}
\end{array}
$$
\caption{(color figures online) Plots of the distribution function as a function of the energy, for different fixed values of angular momentum, for the hypervirial isochrone model far from the origin. Top-left panel: $L=0.1 a\sqrt{\Phi_0}$. Top-right panel: $L=a\sqrt{\Phi_0}$. Bottom-left panel: $L=10 a\sqrt{\Phi_0}$. Bottom-right panel: $L=10a\sqrt{\Phi_0}$. For all panels: the black, brown, red, green, orange and blue curves correspond to the distribution function with 1, 2, 3, 4, 5 and 6 terms, respectively.}
\label{fig5dot5}
\end{figure*}

In Fig. \ref{fig5dot5}, the distribution functions have been plotted as functions of the dimensionless binding energy $-\frac{E}{\Phi_0}$ for differing fixed values of the angular momentum $L$. It must be noted that $L$ can both be small or large, even though $r$ is large. This is because of the fact that one can choose $v$ to be as small as one wishes, which has motivated our choices of $L$. In all the figures, we see that the distribution function remains identically positive, and that there exists a finite radius of convergence. It is also seen that this radius of convergence diminishes as one moves to higher values of $L$. Once again, this is an expected result because a larger value of $L$ implies that $r$ becomes larger, and that $\alpha$ from equation (\ref{asymptoticiso}) is chosen to be as high as possible. But, these choices result in simultaneously reducing the dimensionless binding energy, because $\Psi$ falls off with $r$ and $1-\alpha$ becomes smaller as well. For these reasons, one sees a successively decreasing radius of convergence as one increase the value of $L$.

\subsection{The hypervirial NFW model}
The NFW profile remains one of the most commonly used profiles in modelling dark matter haloes, and was first introduced by \citet{nfw96}. The potential--density pair for the NFW profile is
\begin{equation}
\Psi = \frac{GM}{a}\frac{\ln{(1+x)}}{x},
\end{equation}
\begin{equation}
\rho = \frac{M}{4\pi a^3}\frac{1}{x(1+x)^2}.
\end{equation}
It should be noted that the parameter $M$ this time is not the total mass (which is infinite). Technically, distribution functions do not make sense for models with infinite mass since such a distribution function is not normalizable. Nevertheless, we can truncate the system at some radius, which we can interpret as the scale radius $a$ and the mass contained within this radius can be interpreted as the mass $M$. This will introduce a discontinuity in the distribution function, but such a discontinuity will be `small' if the truncation happens at a very large radius.\\

This time, we cannot analytically obtain $x$ as a function of $\sqrt{x}\Psi$, so we will proceed differently. Expanding the potential--density pair as a Taylor series, we have
\begin{equation}\label{NFWaugd}
x^{5/2}\rho = \frac{M}{4\pi a^3} \left(x^{3/2} - 2x^{5/2} + 3x^{7/2} - 4x^{9/2} + 5x^{11/2} - 6x^{13/2} + \cdots \right),
\end{equation}
\begin{equation}\label{NFWaugr}
\sqrt{x}\Psi = \frac{GM}{a} \left(\sqrt{x} - \frac{x^{3/2}}{2} + \frac{x^{5/2}}{3} - \frac{x^{7/2}}{4} + \frac{x^{9/2}}{5} + \cdots \right).
\end{equation}
By inspection, it is easy to see that only the coefficients $C_{k}$ with odd $k$ in equation (\ref{hypervirialrho}) are non-zero. Using the expansions (\ref{NFWaugd}) and (\ref{NFWaugr}), and collecting like powers, we can solve for the coefficients. In the end, we find the following augmented density and distribution function by truncating them to include the first five non-zero terms. The augmented density is
\begin{equation} \label{NFWaugdens}
\rho = \frac{a}{4\pi G^3 M^2} r^{-1} \Psi^{3} - \frac{a^2}{8\pi G^5 M^4} \Psi^{5} - \frac{a^4}{96\pi G^9 M^8} r^2 \Psi^{9}- \frac{3a^5}{320\pi G^{11} M^{10}} r^3 \Psi^{11} - \frac{29 a^{6}}{2880 \pi G^{13} M^{12}} r^4 \Psi^{13} - \frac{349a^{7}}{30240\pi G^{15}M^{14}}r^{5}\Psi^{15}.
\end{equation}
Using the recipe of Section \ref{SectI}, the corresponding distribution function is found to be
\begin{eqnarray} \label{NFWDF}
f({\mathcal{E}},L) &=& \frac{3 a}{8 \pi^{3} G^3 M^2} L^{-1} {\mathcal{E}}^2 - \frac{4\sqrt{2} a^2}{7 \pi^{3} G^5 M^4} {\mathcal{E}}^{7/2} - \frac{64\sqrt{2} a^4}{143 \pi^{3} G^9 M^8} L^2 {\mathcal{E}}^{13/2} \\ \nonumber
&-& \frac{99 a^5}{64 \pi^{3} G^{11} M^{10}} L^3 {\mathcal{E}}^{8} - \frac{14848 \sqrt{2} a^6}{4845 \pi^{3} G^{13} M^{12}} L^4 {\mathcal{E}}^{19/2} - \frac{4537a^{7}}{360\pi^{3}G^{15}M^{14}}L^{5}\mathcal{E}^{11}.
\end{eqnarray}
Again, the convergence of the series (\ref{NFWaugdens}) and (\ref{NFWDF}) sufficiently close to $r=0$ is clear (by the same reasoning as for the isochrone model). Note that the leading term is the Hernquist model; this is not surprising since the NFW profile has a cusp whose slope is exactly equal to that of the Hernquist profile. The quantity $\Delta$, to lowest order, takes on the form
\begin{equation} \label{NFWDELTA}
\Delta = \frac{2237}{40320}x^7.
\end{equation}
We proceed to compute the velocity dispersions from the six-term distribution function. They are found to be
\begin{equation}
\sigma_r^2 = \frac{\Psi}{8} \frac{-30240+10080x(\Psi/\Phi_0)^2+504x^3(\Psi/\Phi_0)^6+378x^4(\Psi/\Phi_0)^8+348x^5(\Psi/\Phi_0)^{10}+349x^6(\Psi/\Phi_0)^{12}}{-15120+7560x(\Psi/\Phi_0)^2+630x^3(\Psi/\Phi_0)^6+567x^4(\Psi/\Phi_0)^8+609x^5(\Psi/\Phi_0)^{10}+698x^6(\Psi/\Phi_0)^{12}},
\end{equation}
\begin{equation}
\sigma_\theta^2 = \frac{\Psi}{16} \frac{-30240+20160x(\Psi/\Phi_0)^2+2016x^3(\Psi/\Phi_0)^6+1890x^4(\Psi/\Phi_0)^8+2088x^5(\Psi/\Phi_0)^{10}+2443x^6(\Psi/\Phi_0)^{12}}{-15120+7560x(\Psi/\Phi_0)^2+630x^3(\Psi/\Phi_0)^6+567x^4(\Psi/\Phi_0)^8+609x^5(\Psi/\Phi_0)^{10}+698x^6(\Psi/\Phi_0)^{12}},
\end{equation}
and the corresponding anisotropy parameter is
\begin{equation}
\beta = \frac{-\left(30240+1008x^3(\Psi/\Phi_0)^6+1134x^4(\Psi/\Phi_0)^8+1392x^5(\Psi/\Phi_0)^{10}+1745x^6(\Psi/\Phi_0)^{12}\right)}{-60480+20160x(\Psi/\Phi_0)^2+1008x^3(\Psi/\Phi_0)^6+756x^4(\Psi/\Phi_0)^8+696x^5(\Psi/\Phi_0)^{10}+698x^6(\Psi/\Phi_0)^{12}}.
\end{equation}
\begin{figure*}
$$
\begin{array}{ccc}
 \includegraphics[width=6.8cm]{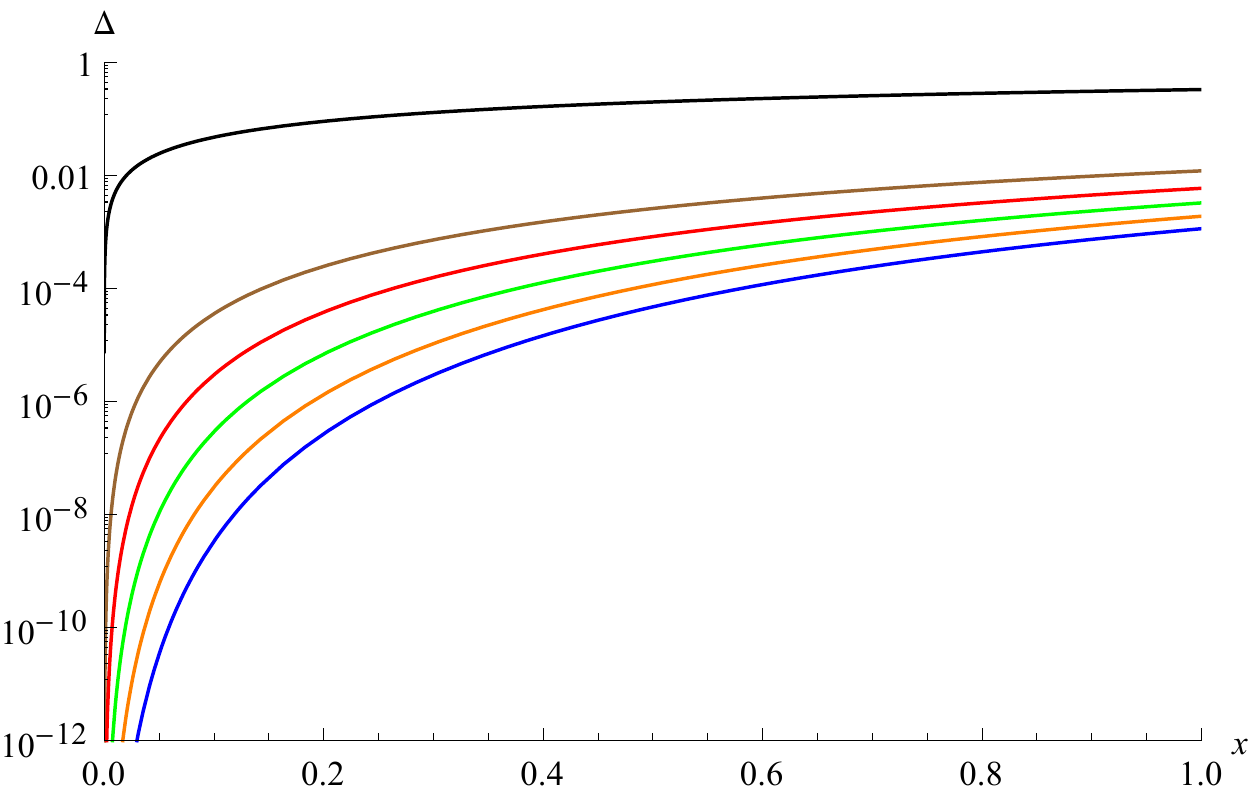} & \includegraphics[width=6.8cm]{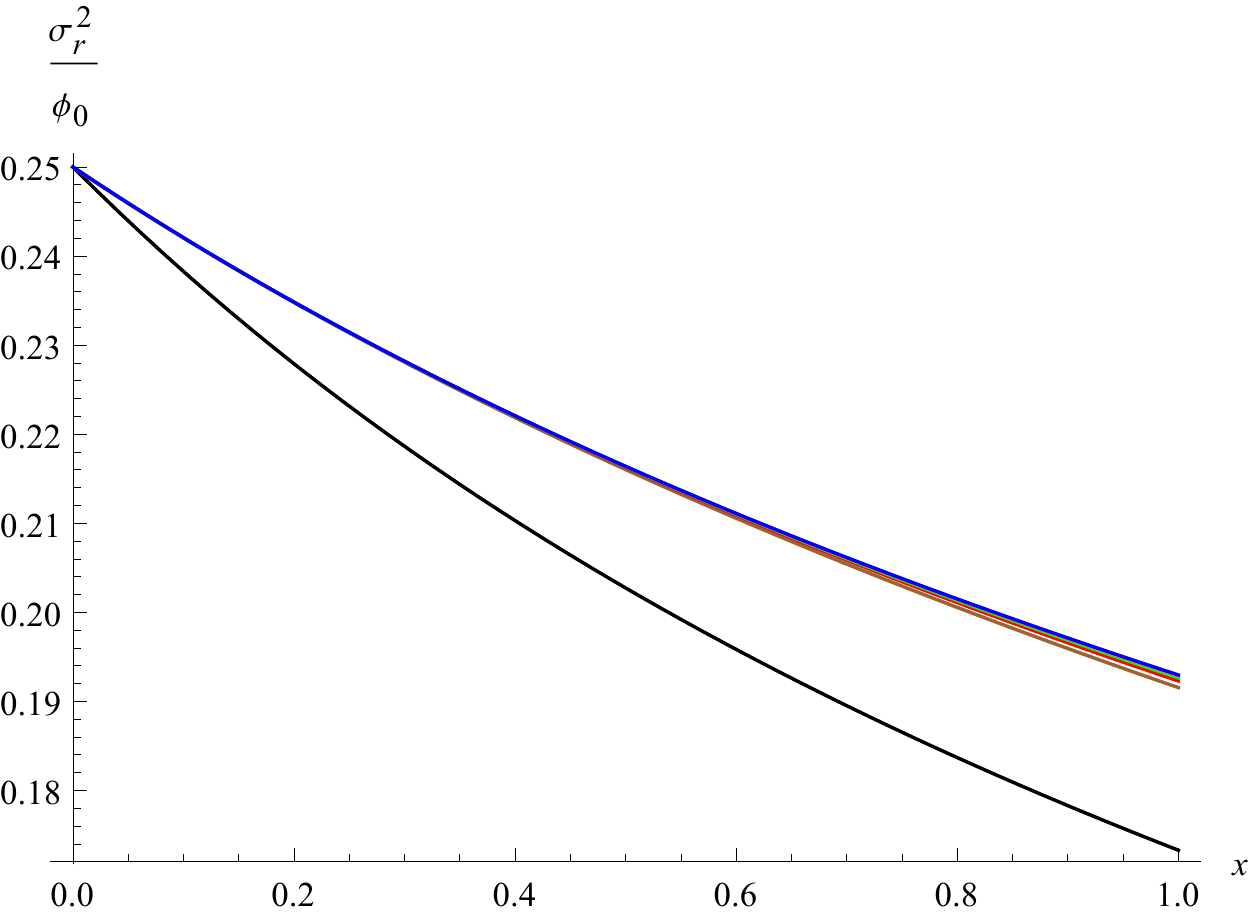} \\
 \includegraphics[width=6.8cm]{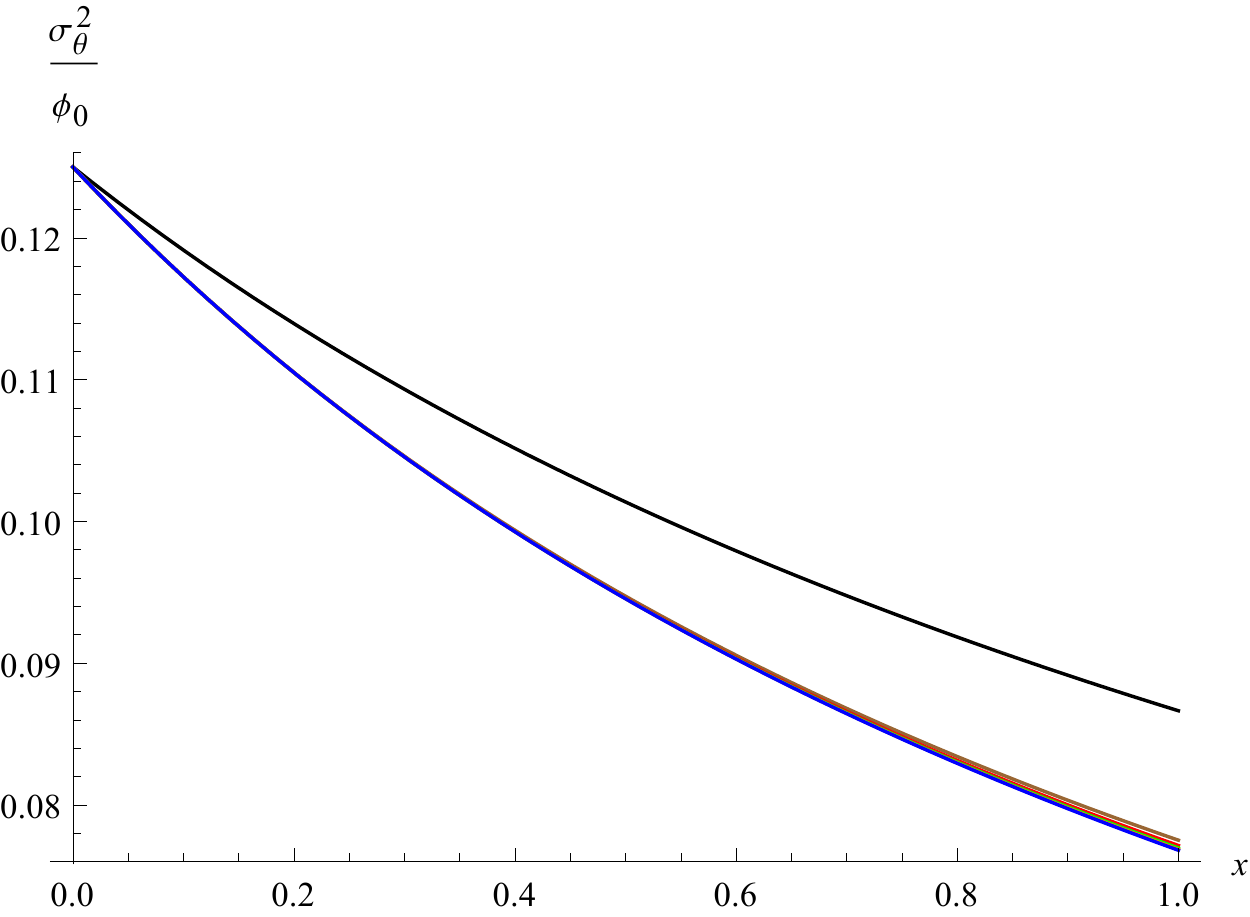} & \includegraphics[width=6.8cm]{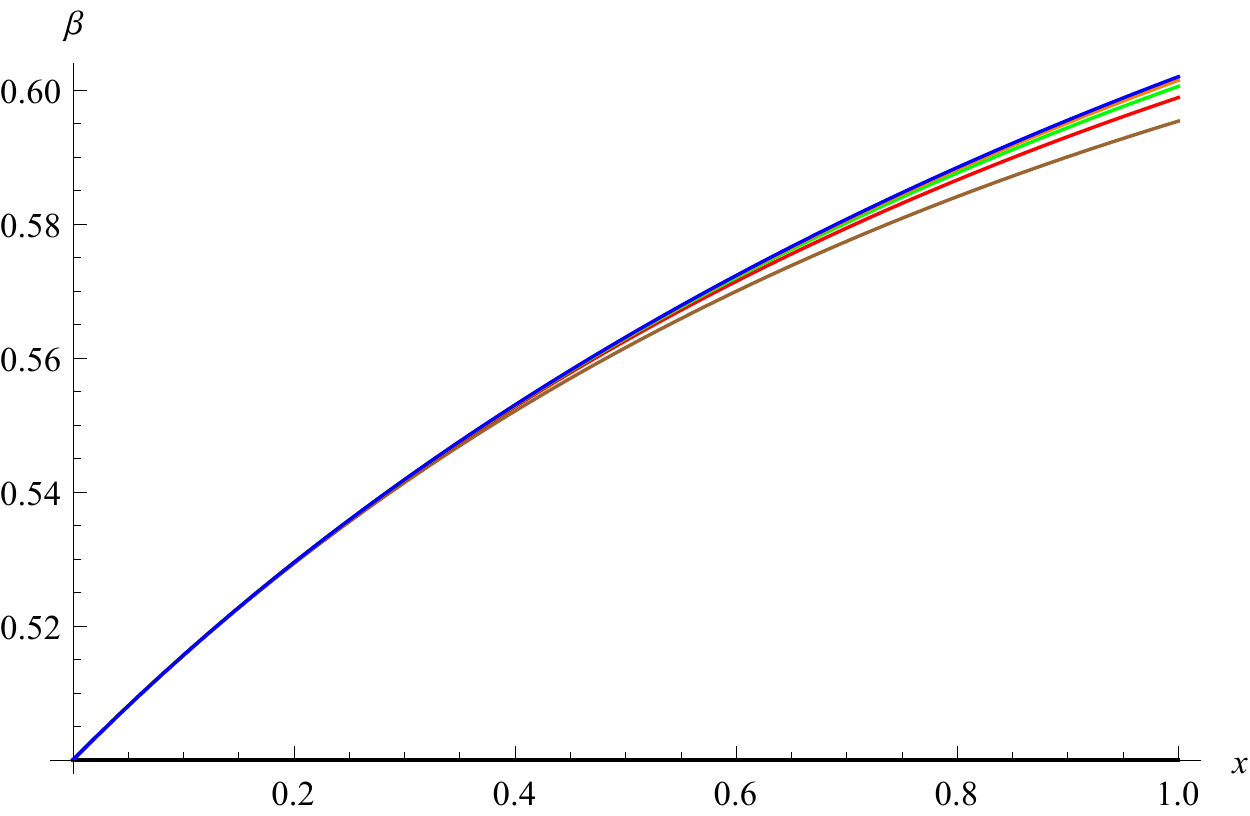}
\end{array}
$$
\caption{(color figures online) Plots for the hypervirial NFW model near the origin. Top-left panel: $\Delta$ versus $x$. Top-right panel: the radial velocity dispersion versus $x$. Bottom-left panel: the tangential velocity dispersion versus $x$. Bottom-right panel: the anisotropy parameter $\beta$ versus $x$. For all panels: the black, brown, red, green, orange and blue curves correspond to the distribution function with 1, 2, 3, 4, 5 and 6 terms, respectively.}
\label{fig4dot5}
\end{figure*}
The plots of the velocity dispersion, anisotropy parameter and $\Delta$ are presented in Fig. \ref{fig4dot5}. We see that the velocity structure remains virtually identical even when additional terms are incorporated which is suggestive of convergence. In Fig. \ref{NFWDFplots}, the distribution function has been plotted for several (fixed) values of $L$ as a function of the dimensionless binding energy. We note that the same line of reasoning presented in our discussion of the small $r$ isochrone distribution functions, as plotted in Fig. \ref{fig3dot1}, is also applicable here.
\begin{figure*}
$$
\begin{array}{ccc}
 \includegraphics[width=6.8cm]{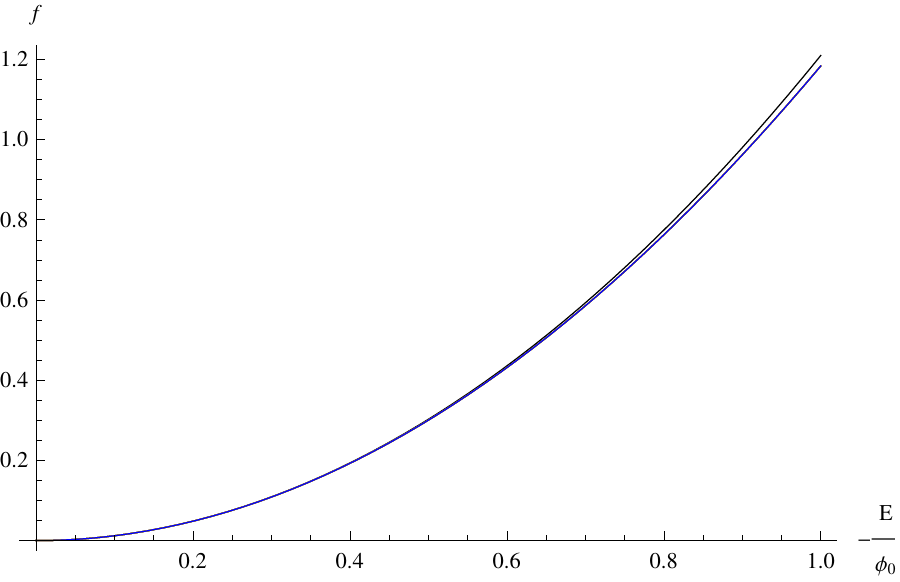} & \includegraphics[width=6.8cm]{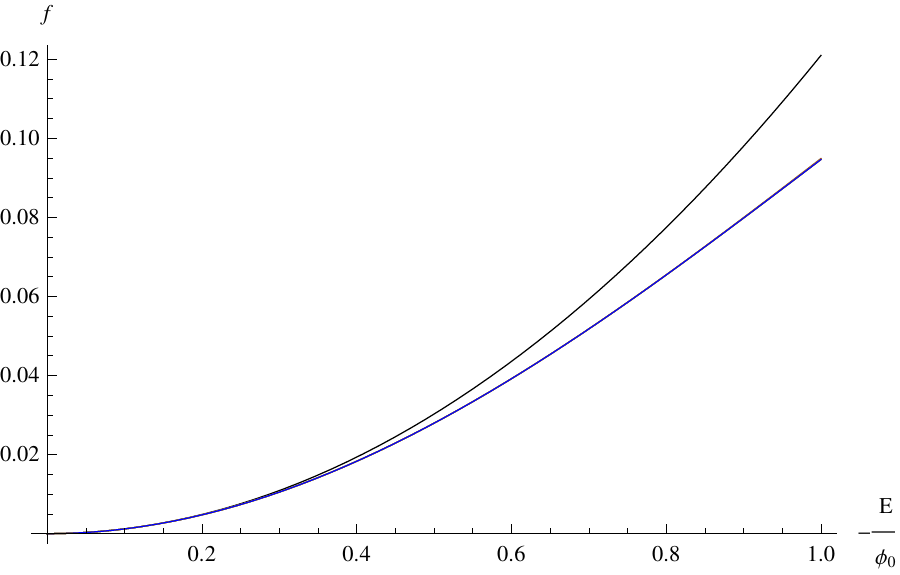} \\
 \includegraphics[width=6.8cm]{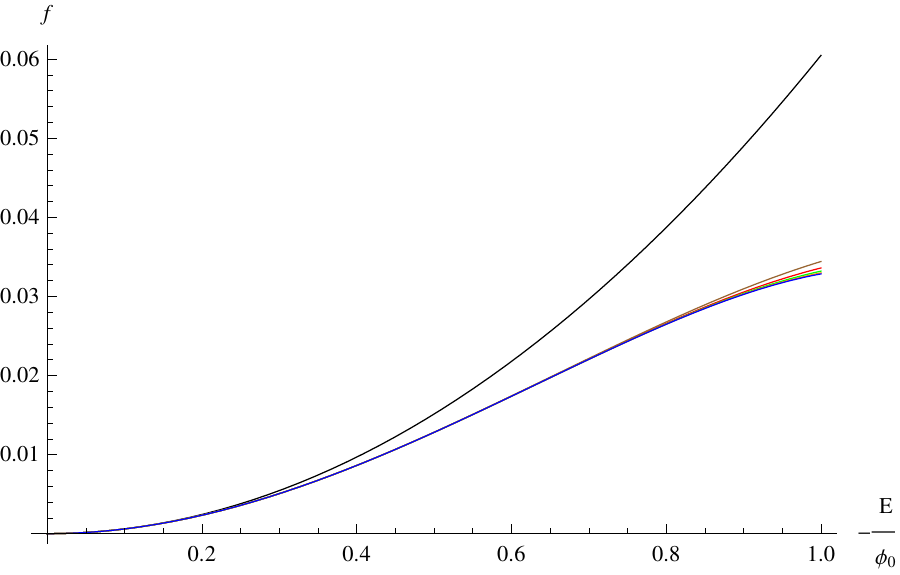} & \includegraphics[width=6.8cm]{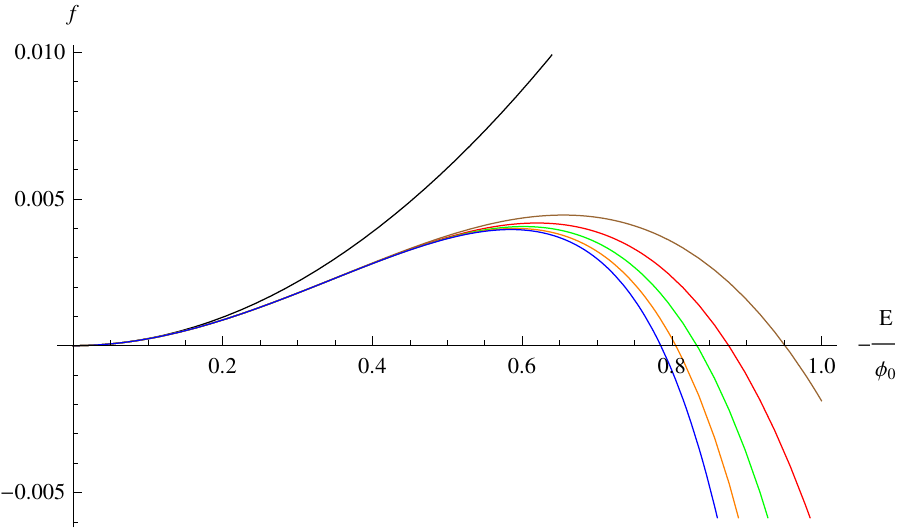}
\end{array}
$$
\caption{(color figures online) Plots of the distribution function as a function of the energy, for different fixed values of angular momentum, for the hypervirial NFW model near the origin. Top-left panel: $L=0.01a\sqrt{\Phi_0}$. Top-right panel: $L=0.1 a\sqrt{\Phi_0}$. Bottom-left panel: $L=0.2 a\sqrt{\Phi_0}$. Bottom-right panel: $L=0.5 a\sqrt{\Phi_0}$. For all panels: the black, brown, red, green, orange and blue curves correspond to the distribution function with 1, 2, 3, 4, 5 and 6 terms, respectively.}
\label{NFWDFplots}
\end{figure*}

\subsection{The hypervirial Jaffe model}
\begin{figure*}
$$
\begin{array}{ccc}
 \includegraphics[width=6.8cm]{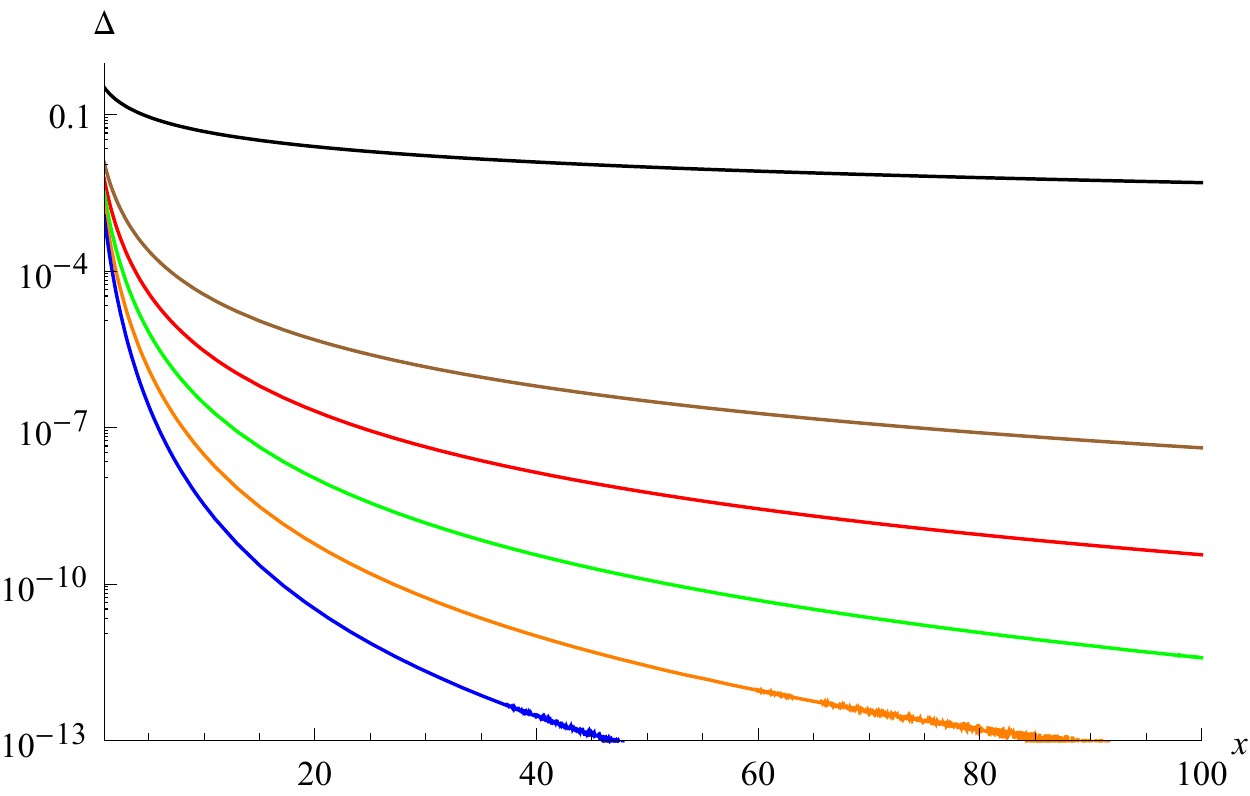} & \includegraphics[width=6.8cm]{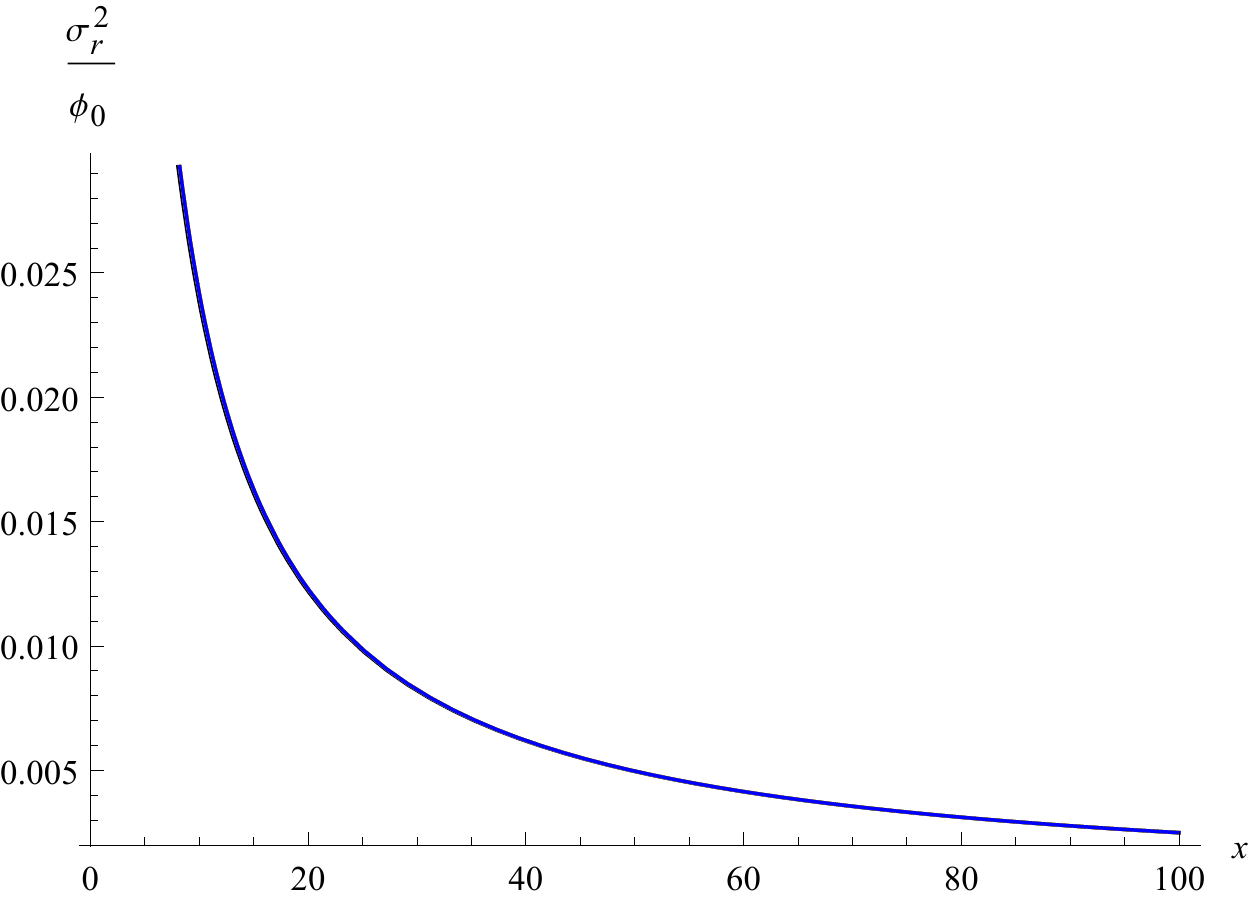} \\
 \includegraphics[width=6.8cm]{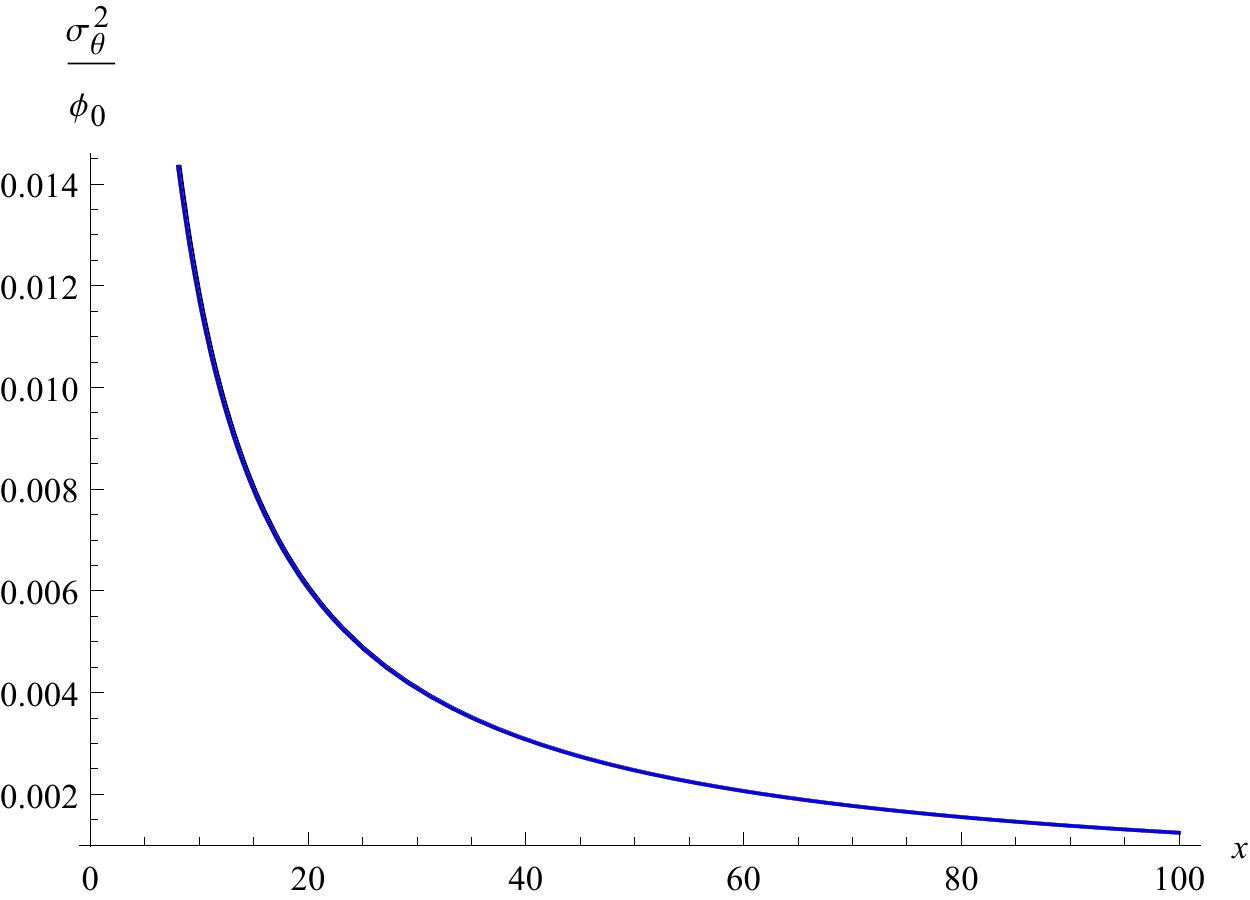} & \includegraphics[width=6.8cm]{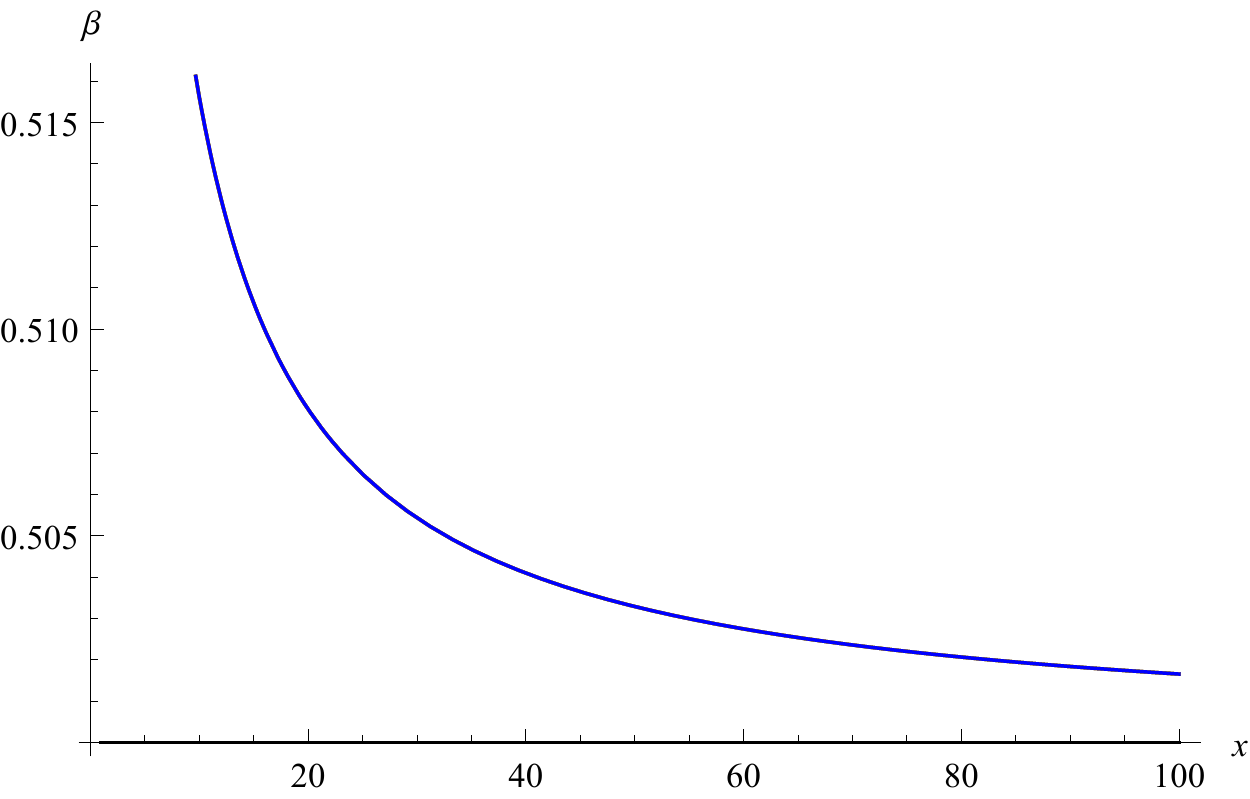}
\end{array}
$$
\caption{(color figures online) Plots for the hypervirial Jaffe model near infinity. Top-left panel: $\Delta$ versus $x$. Top-right panel: the radial velocity dispersion versus $x$. Bottom-left panel: the tangential velocity dispersion versus $x$. Bottom-right panel: the anisotropy parameter $\beta$ versus $x$. For all panels: the black, brown, red, green, orange and blue curves correspond to the distribution function with 1, 2, 3, 4, 5 and 6 terms, respectively.}
\label{fig5}
\end{figure*}

\begin{figure*}
$$
\begin{array}{ccc}
 \includegraphics[width=6.8cm]{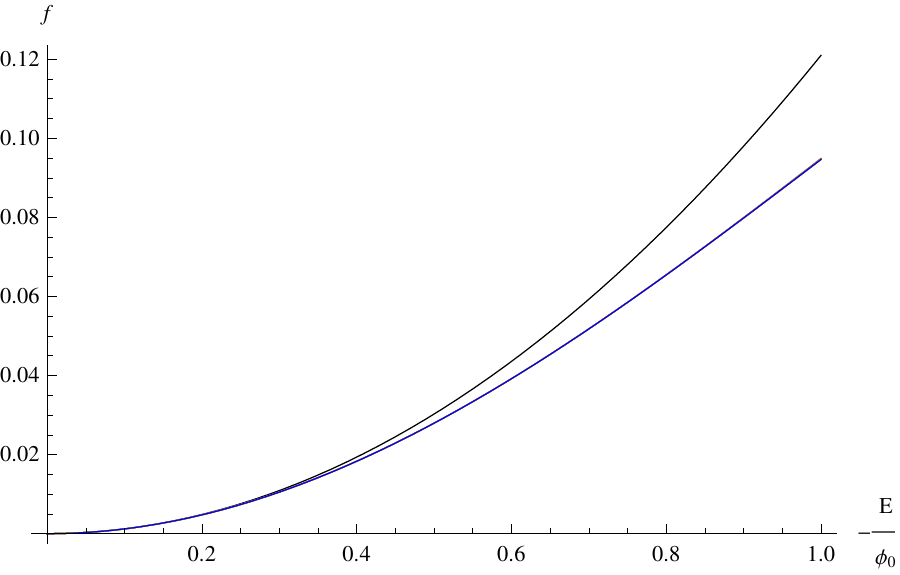} & \includegraphics[width=6.8cm]{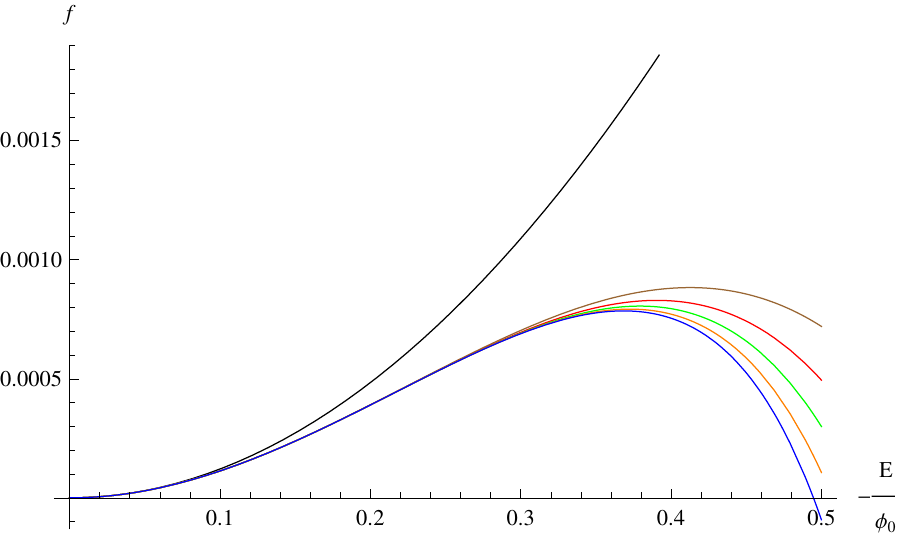} \\
 \includegraphics[width=6.8cm]{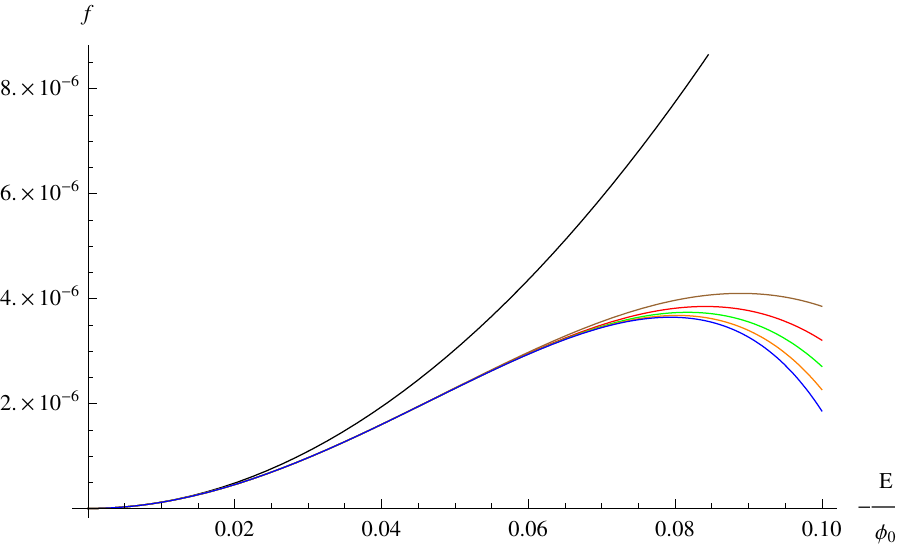} & \includegraphics[width=6.8cm]{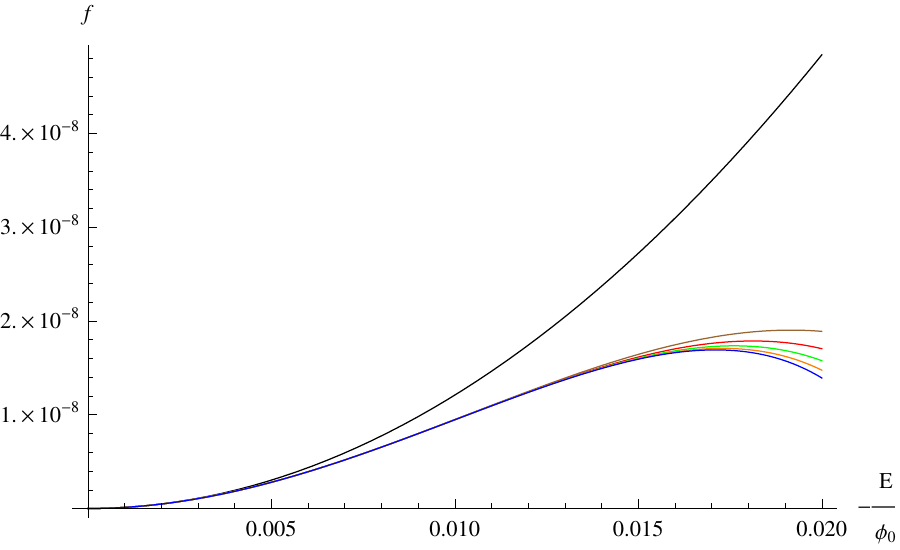}
\end{array}
$$
\caption{(color figures online) Plots of the distribution function as a function of the energy, for different fixed values of angular momentum, for the hypervirial Jaffe model far from the origin. Top-left panel: $L=0.1 a \sqrt{\Phi_0}$. Top-right panel: $L=a \sqrt{\Phi_0}$. Bottom-left panel: $L=10 a \sqrt{\Phi_0}$. Bottom-right panel: $L=100 a \sqrt{\Phi_0}$. For all panels: the black, brown, red, green, orange and blue curves correspond to the distribution function with 1, 2, 3, 4, 5 and 6 terms, respectively.}
\label{JAffeDF}
\end{figure*}

The potential--density pair for the Jaffe model was first obtained by \citet{ja83} and is given by
\begin{equation}\label{PsiJaffe}
\Psi = \frac{GM}{a}\ln{\left(1+\frac{1}{x}\right)},
\end{equation}
\begin{equation}
\rho = \frac{M}{4\pi a^3} \frac{1}{x^2 (1+x)^2}.
\end{equation}
This time, due to the singular nature of $\Psi$ at the origin, we will perform the expansions around infinity instead.
\begin{equation}
x^{5/2}\rho = \frac{M}{4\pi a^3}\left(x^{-3/2} -2x^{-5/2} + 3x^{-7/2} -4x^{-9/2} + 5x^{-11/2} -6x^{-13/2} \cdots\right),
\end{equation}
\begin{equation}
\sqrt{x} \Psi = \frac{GM}{a}\left(x^{-1/2} - \frac{1}{2} x^{-3/2} + \frac{1}{3} x^{-5/2} - \frac{1}{4} x^{-7/2} + \frac{1}{5} x^{-9/2} - \frac{1}{6} x^{-11/2} \cdots\right).
\end{equation}
Curiously, these expansions are exactly the same as for the NFW model, with the replacement of $x$ by $x^{-1}$. Proceeding as for the NFW model, we find the exact same augmented density (\ref{NFWaugdens}) and distribution function (\ref{NFWDF}), but this time this is supposed to be an approximation for the outer region of the model. Indeed, the outer falloff of both the Jaffe and Hernquist models are the same ($r^{-4}$) but their inner cusps differ ($r^{-2}$ for the Jaffe model and $r^{-1}$ for the Hernquist model). The convergence of the distribution function around infinity for the Jaffe model can be established by following the same reasoning as the one used for the isochrone outer branch. 
One can compute the value of $\Delta$ as defined before, but it must be noted that this must be evaluated for large values of $r$. To leading order, we find that $\Delta$ has the following asymptotic behaviour
\begin{equation}
\Delta = \frac{2237}{40320} x^{-7}.
\end{equation}
This expression is in agreement with equation (\ref{NFWDELTA}) because the two expressions are exactly the same if one replaces $x$ by $x^{-1}$ and vice versa, which also ensures that the augmented densities and the distribution functions for these two models are exactly the same, albeit in different regimes. The plots for the velocity dispersion, anisotropy parameter and $\Delta$ are presented in Fig. \ref{fig5}. The convergence of the velocity structure profiles as one includes higher order terms of the distribution function is also evident. In Fig. \ref{JAffeDF}, we plot the distribution functions for the Jaffe model as a function of the dimensionless binding energy, at different fixed values of $L$. The choices of $L$, and the behaviour seen in these plots can be explained by following a similar line of reasoning to the one employed in discussing the large $r$ isochrone model, plotted in Fig. \ref{fig5dot5}.

\subsection{The hypervirial $\gamma=\frac{3}{2}$ Dehnen model}
\begin{figure*}
$$
\begin{array}{ccc}
 \includegraphics[width=6.8cm]{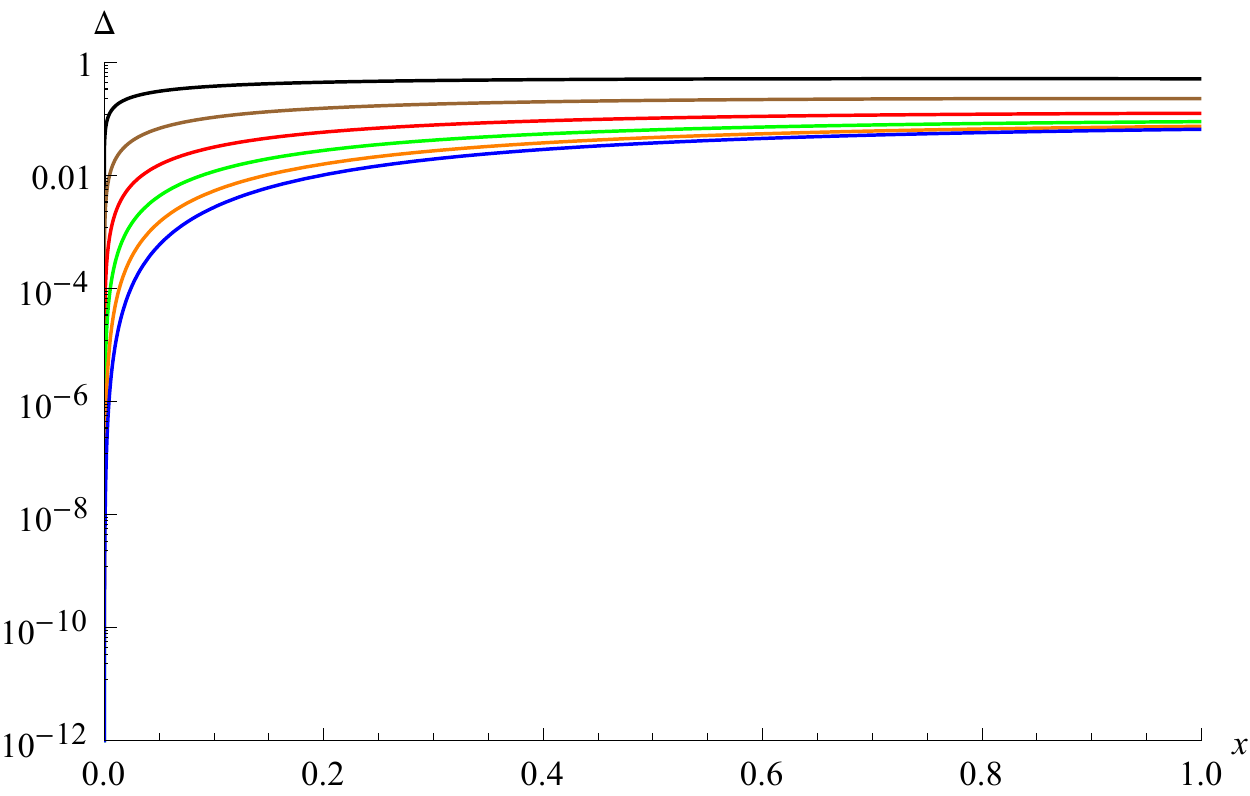} & \includegraphics[width=6.8cm]{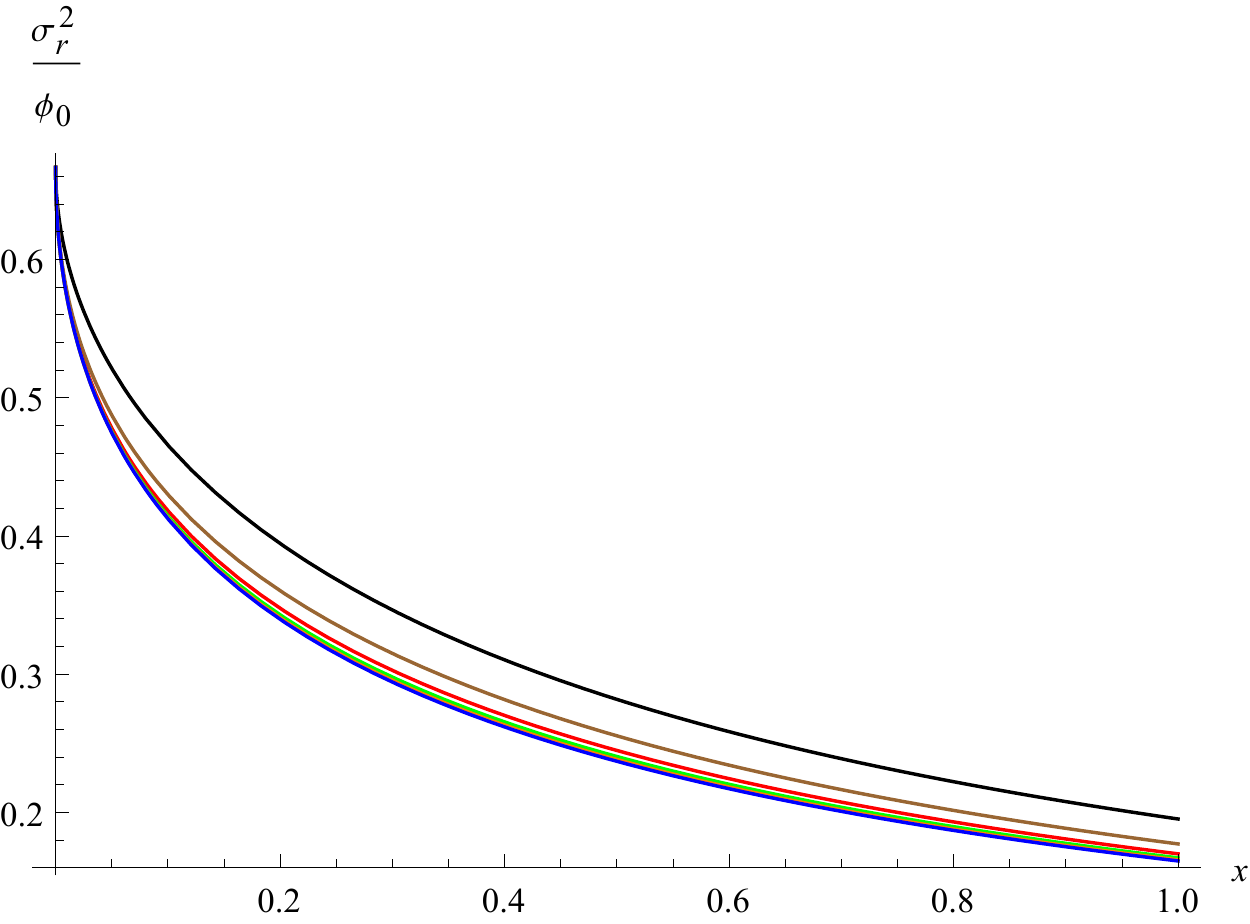} \\
 \includegraphics[width=6.8cm]{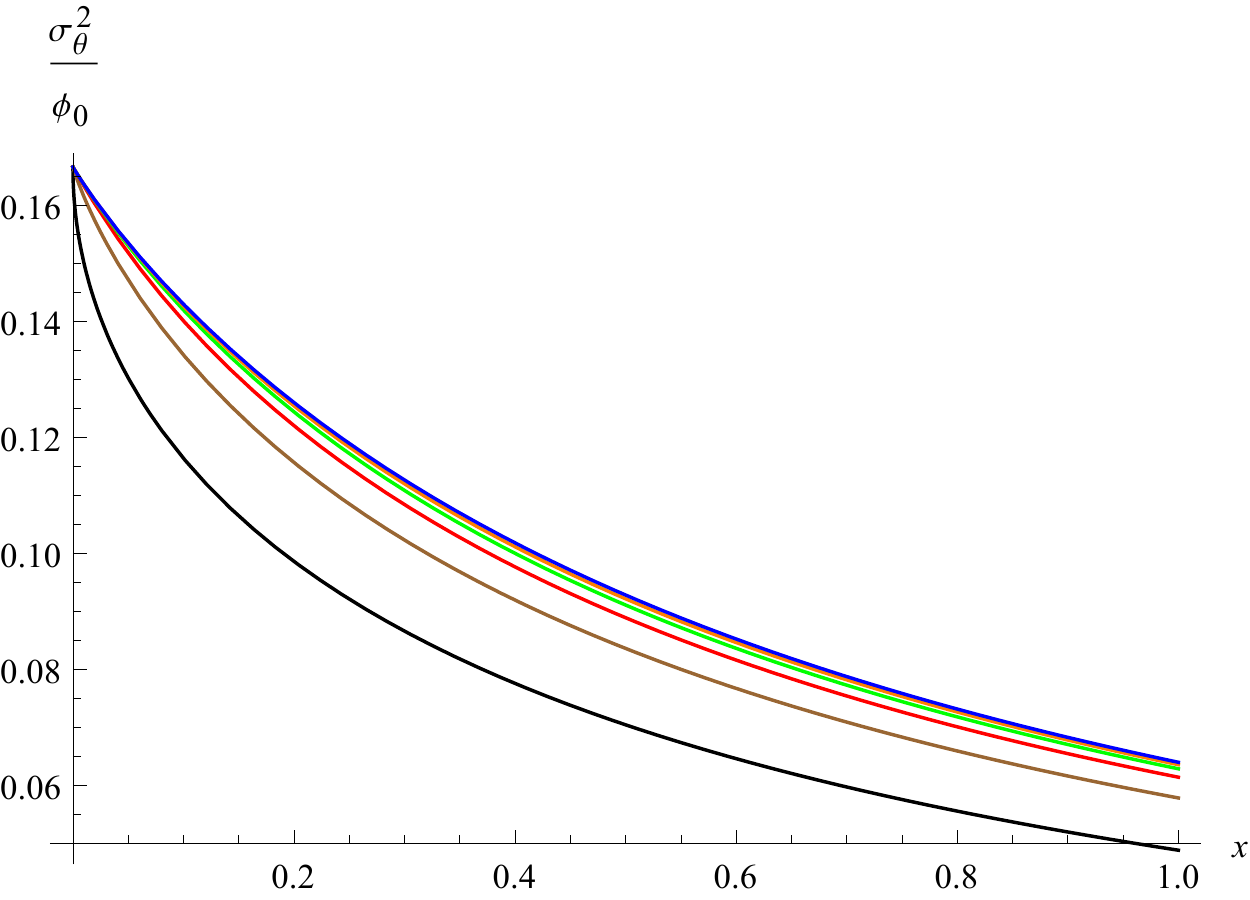} & \includegraphics[width=6.8cm]{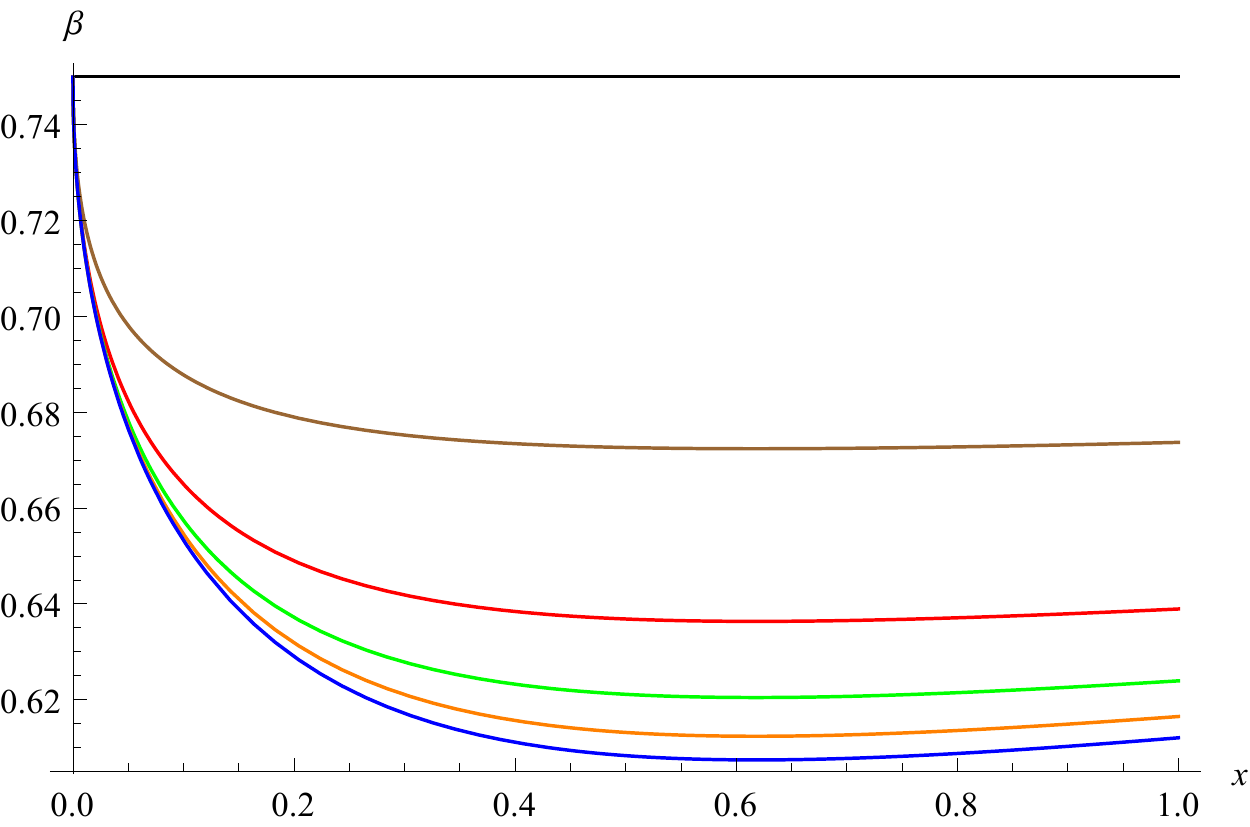}
\end{array}
$$
\caption{(color figures online) Plots for the hypervirial Dehnen model near the origin. Top-left panel: $\Delta$ versus $x$. Top-right panel: the radial velocity dispersion versus $x$. Bottom-left panel: the tangential velocity dispersion versus $x$. Bottom-right panel: the anisotropy parameter $\beta$ versus $x$. For all panels: the black, brown, red, green, orange and blue curves correspond to the distribution function with 1, 2, 3, 4, 5 and 6 terms, respectively.}
\label{fig6}
\end{figure*}

\begin{figure*}
$$
\begin{array}{ccc}
 \includegraphics[width=6.8cm]{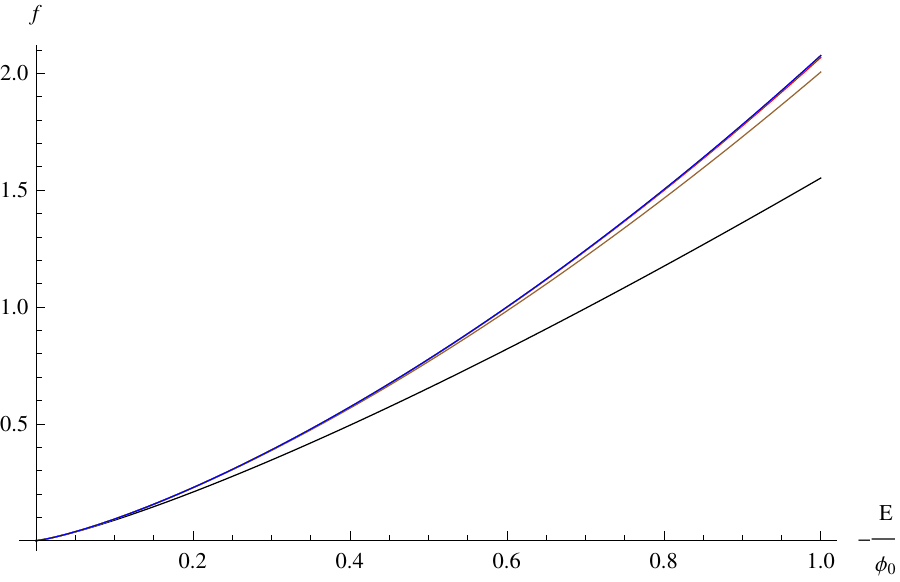} & \includegraphics[width=6.8cm]{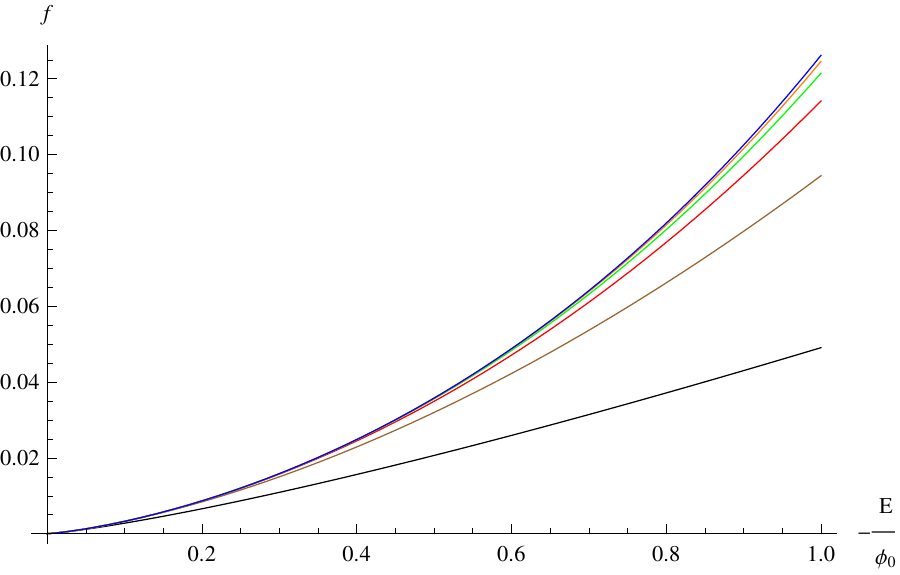} \\
 \includegraphics[width=6.8cm]{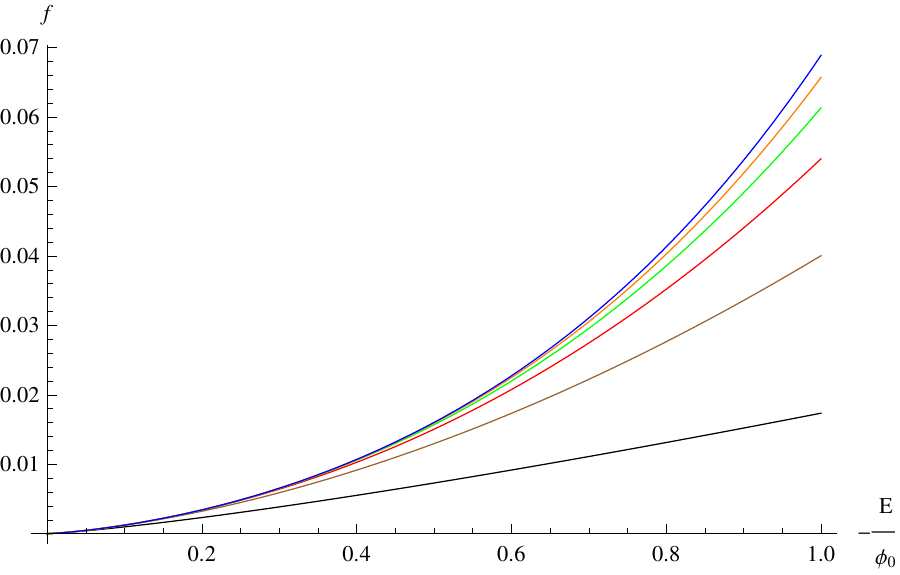} & \includegraphics[width=6.8cm]{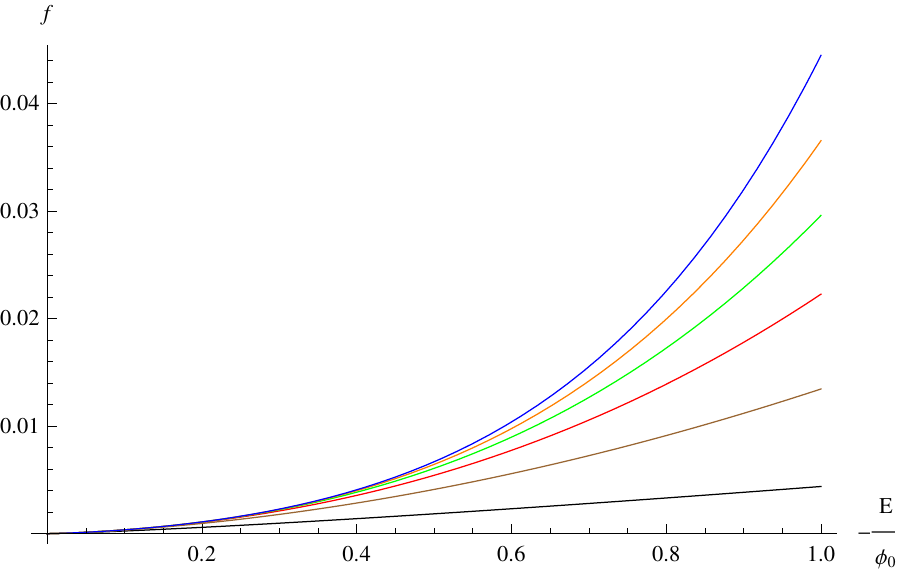}
\end{array}
$$
\caption{(color figures online) Plots of the distribution function as a function of the energy, for different fixed values of angular momentum, for the hypervirial Dehnen model near the origin. Top-left panel: $L=0.01a\sqrt{\Phi_0}$. Top-right panel: $L=0.1 a\sqrt{\Phi_0}$. Bottom left panel: $L=0.2 a\sqrt{\Phi_0}$. Bottom-right panel: $L=0.5 a\sqrt{\Phi_0}$. For all panels: the black, brown, red, green, orange and blue curves correspond to the distribution function with 1, 2, 3, 4, 5 and 6 terms, respectively.}
\label{DehnenDFsmallrplots}
\end{figure*}

Our final example of a hypervirial distribution function is a specific case of the Dehnen models \citep{deh93}. Anisotropic distribution functions of the Dehnen models have been computed, for example, in \citet{bu07}. The Dehnen family of models have the following potential--density pair:
\begin{equation}
\Psi = \frac{GM}{a} \frac{1}{2-\gamma} \left[1-\left(\frac{x}{x+1}\right)^{2-\gamma}\right],
\end{equation}
\begin{equation}
\rho = \frac{M (3-\gamma)}{4\pi a^3} x^{-\gamma} \left(1+x\right)^{\gamma-4}.
\end{equation}
We will only consider the model with $\gamma=\frac{3}{2}$ since this model is known to closely approximate de Vaucouleurs' law $R^{1/4}$ law \citep{deh93,mo10}. We repeat the procedure outlined in the above sections and expand $x^{5/2} \rho$ and $\sqrt{x} \Psi$ about $x=0$. Upon doing so, we obtain the augmented density
\begin{eqnarray}\label{Dehnen}
\rho &=& \frac{3 a^{1/2}}{32 \pi G^2 M} r^{-3/2} \Psi^2  + \frac{3 a}{32 \pi G^3 M^2} r^{-1} \Psi^3 + \frac{15 a^{3/2} }{256 \pi G^4 M^3} r^{-1/2} \Psi^4 + \frac{9 a^2}{256 \pi G^5 M^4} \Psi^5 \\ \nonumber
&+& \frac{105 a^{5/2}}{4096 \pi G^6 M^5} r^{1/2} \Psi^6 + \frac{3 a^3}{128 \pi G^7 M^6} r \Psi^7.
\end{eqnarray}
And the corresponding distribution function is
\begin{eqnarray}
f({\mathcal{E}},L) &=& \frac{3 a^{1/2}}{16 (2^{3/4}) \Gamma(\frac{1}{4}) \Gamma(\frac{9}{4}) \pi^{5/2} G^2 M} L^{-3/2} {\mathcal{E}}^{5/4}  + \frac{9 a}{64 \pi^{3}G^3 M^2} L^{-1} {\mathcal{E}}^2 + \frac{45 a^{3/2} }{64 (2^{1/4}) \pi^{5/2} \Gamma(\frac{3}{4}) \Gamma(\frac{15}{4}) G^4 M^3} L^{-1/2} {\mathcal{E}}^{11/4} \nonumber \\
&+& \frac{9 a^2}{28\sqrt{2} \pi^{3} G^5 M^4} {\mathcal{E}}^{7/2} + \frac{4725 a^{5/2}}{512 (2^{3/4}) \pi^{5/2} \Gamma(\frac{5}{4}) \Gamma(\frac{21}{4}) G^6 M^5} L^{1/2} {\mathcal{E}}^{17/4} + \frac{63 a^3}{128 \pi^{3}G^7 M^6} L {\mathcal{E}}^5.
\end{eqnarray}
That the series above converge near the origin can be seen following a reasoning completely analogous to the previous models. It is interesting to note that the leading term is the Veltmann NFW-like model (with $m=1/2$) discussed in the previous section, which has the same inner cusp as the Dehnen model ($r^{-3/2}$) but a slightly different outer falloff -- $r^{-4}$ for the Dehnen model, and $r^{-7/2}$ for the hypervirial approximation. Note also that all the subleading terms do not contribute to the outer falloff of the hypervirial approximation. The value of $\Delta$ for small values of $x$ is found to be
\begin{equation}
\Delta = \frac{273}{16} x^{3}.
\end{equation}
The velocity dispersions for the above augmented density are found to be
\begin{equation}
\sigma_r^2 = \frac{\Psi}{3} \frac{128+96(\frac{\Psi}{\Phi_0}) x^{1/2}+48(\frac{\Psi}{\Phi_0})^2 x + 24 (\frac{\Psi}{\Phi_0})^3 x^{3/2} + 15 (\frac{\Psi}{\Phi_0})^4 x^2 + 12 (\frac{\Psi}{\Phi_0})^5 x^{5/2} }{128 + 128 (\frac{\Psi}{\Phi_0}) x^{1/2} + 80 (\frac{\Psi}{\Phi_0})^2 x + 48 (\frac{\Psi}{\Phi_0})^3 x^{3/2} + 35 (\frac{\Psi}{\Phi_0})^4 x^2 + 32 (\frac{\Psi}{\Phi_0})^5 x^{5/2}},
\end{equation}
\begin{equation}
\sigma_\theta^2 = \frac{\Psi}{12} \frac{128+192(\frac{\Psi}{\Phi_0}) x^{1/2}+144(\frac{\Psi}{\Phi_0})^2 x + 96 (\frac{\Psi}{\Phi_0})^3 x^{3/2} + 75 (\frac{\Psi}{\Phi_0})^4 x^2 + 72 (\frac{\Psi}{\Phi_0})^5 x^{5/2} }{128 + 128 (\frac{\Psi}{\Phi_0}) x^{1/2} + 80 (\frac{\Psi}{\Phi_0})^2 x + 48 (\frac{\Psi}{\Phi_0})^3 x^{3/2} + 35 (\frac{\Psi}{\Phi_0})^4 x^2 + 32 (\frac{\Psi}{\Phi_0})^5 x^{5/2}},
\end{equation}
and the anisotropy parameter is 
\begin{equation}
\beta = \frac{96+48(\frac{\Psi}{\Phi_0}) x^{1/2}+12(\frac{\Psi}{\Phi_0})^2 x  - \frac{15}{4}(\frac{\Psi}{\Phi_0})^4 x^2 -6 (\frac{\Psi}{\Phi_0})^5 x^{5/2} }{128 + 128 (\frac{\Psi}{\Phi_0}) x^{1/2} + 80 (\frac{\Psi}{\Phi_0})^2 x + 48 (\frac{\Psi}{\Phi_0})^3 x^{3/2} + 35 (\frac{\Psi}{\Phi_0})^4 x^2 + 32 (\frac{\Psi}{\Phi_0})^5 x^{5/2}}.
\end{equation}
The resulting dispersions and anisotropy parameter $\beta$ are plotted in Fig. \ref{fig6} along with the value of $\Delta$. The Dehnen model grows slowly, i.e. one finds that including successive terms does play a major role in increasing the accuracy of $\Delta$ but do not significantly alter the structure of the velocity dispersion profiles. In Fig. \ref{DehnenDFsmallrplots}, we plot the small $r$ distributions functions for the $\gamma=3/2$ Dehnen model, holding $L$ fixed. The insights that can be gleaned from this figure are similar to the ones derivable from Fig. \ref{fig3dot1} that deals with the small $r$ isochrone distribution functions and the discussion for the latter can be found in the subsection on the isochrone models.

\begin{figure*}
$$
\begin{array}{ccc}
 \includegraphics[width=6.8cm]{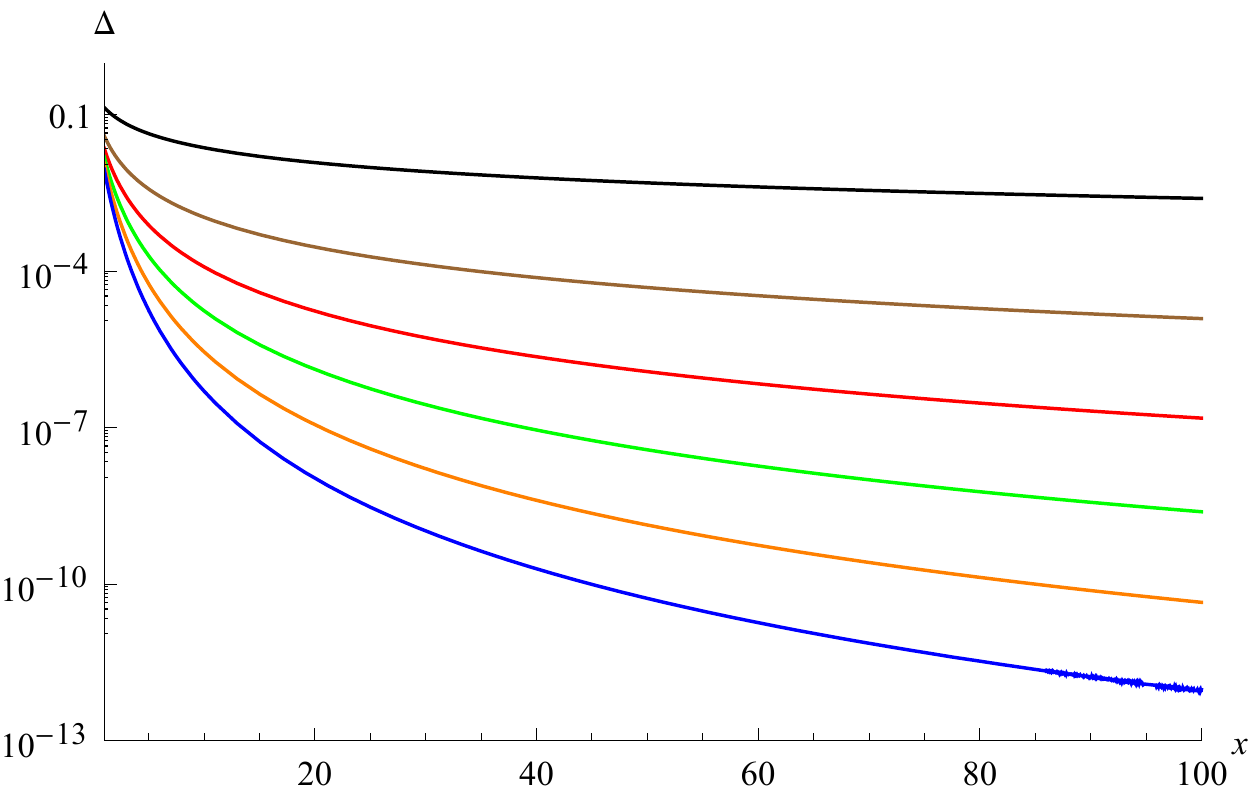} & \includegraphics[width=6.8cm]{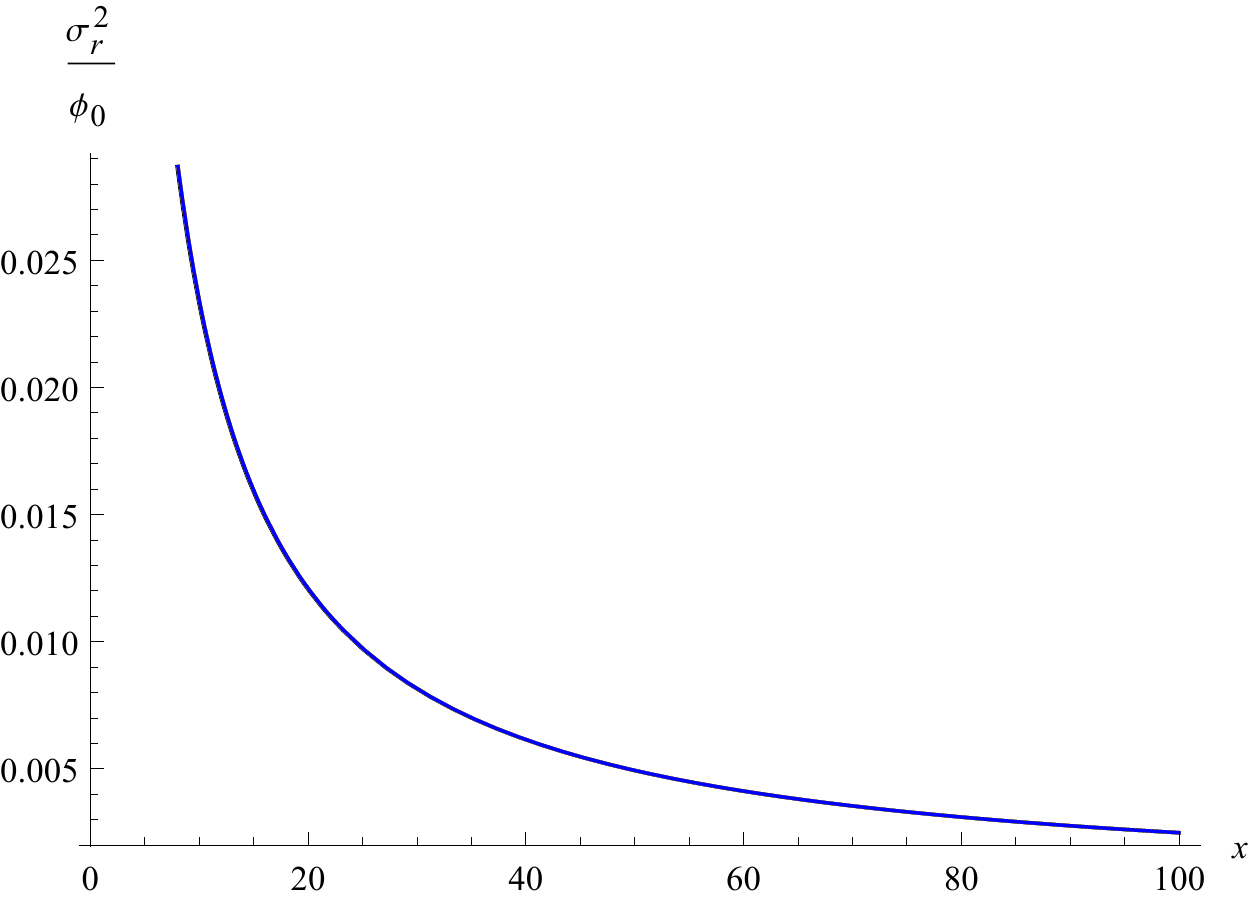} \\
 \includegraphics[width=6.8cm]{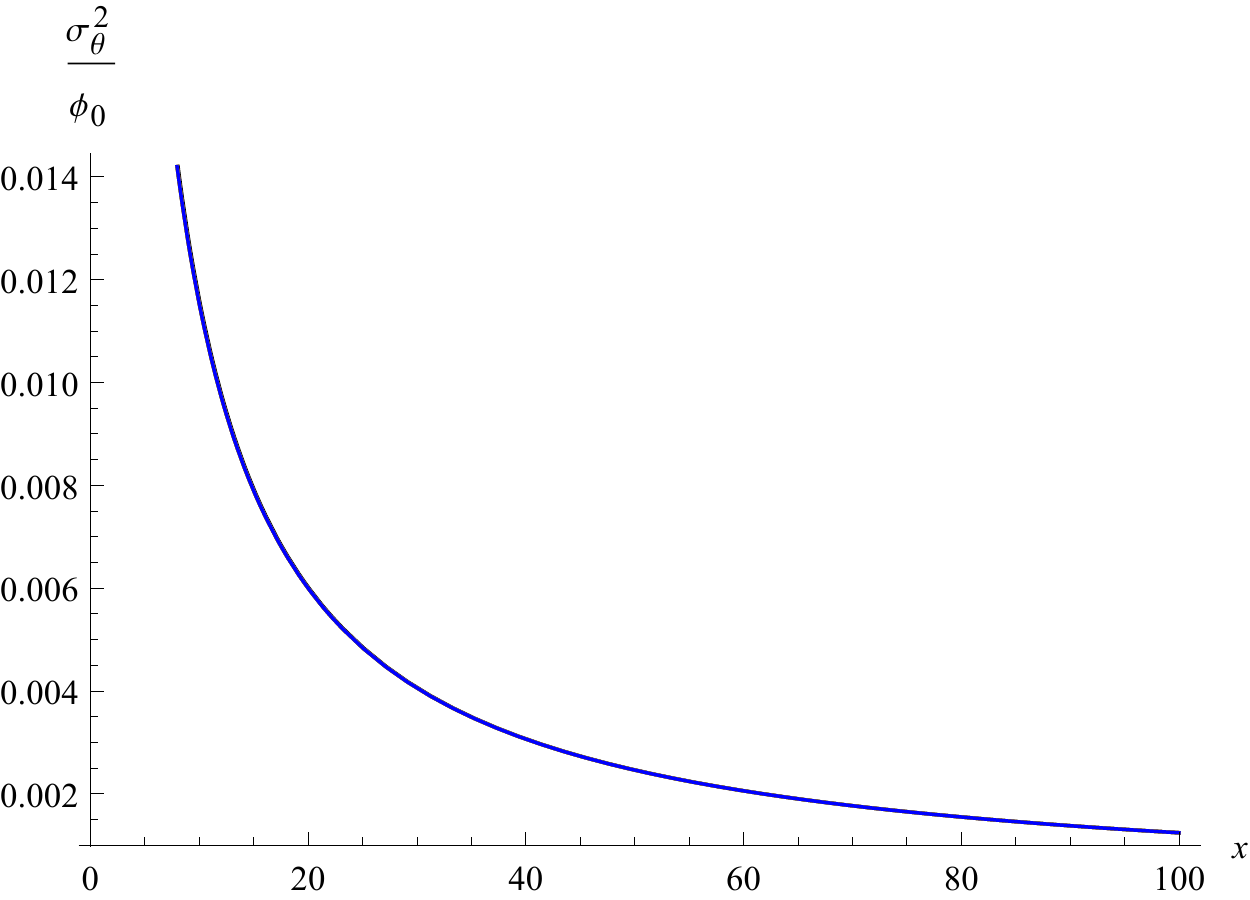} & \includegraphics[width=6.8cm]{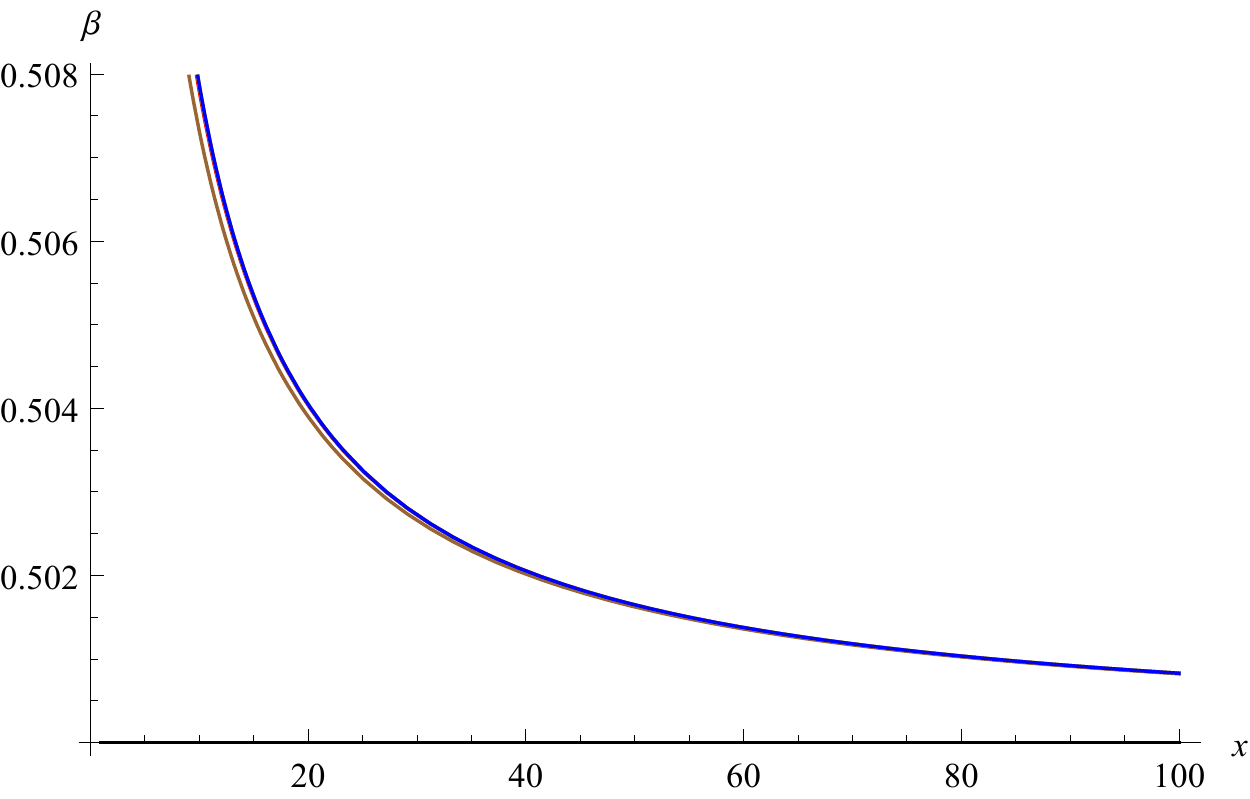}
\end{array}
$$
\caption{(color figures online) Plots for the hypervirial Dehnen model near infinity. Top-left panel: $\Delta$ versus $x$. Top-right panel: the radial velocity dispersion versus $x$. Bottom-left panel: the tangential velocity dispersion versus $x$. Bottom-right panel: the anisotropy parameter $\beta$ versus $x$. For all panels: the black, brown, red, green, orange and blue curves correspond to the distribution function with 1, 2, 3, 4, 5 and 6 terms, respectively.}
\label{fig7}
\end{figure*}

\begin{figure*}
$$
\begin{array}{ccc}
 \includegraphics[width=6.8cm]{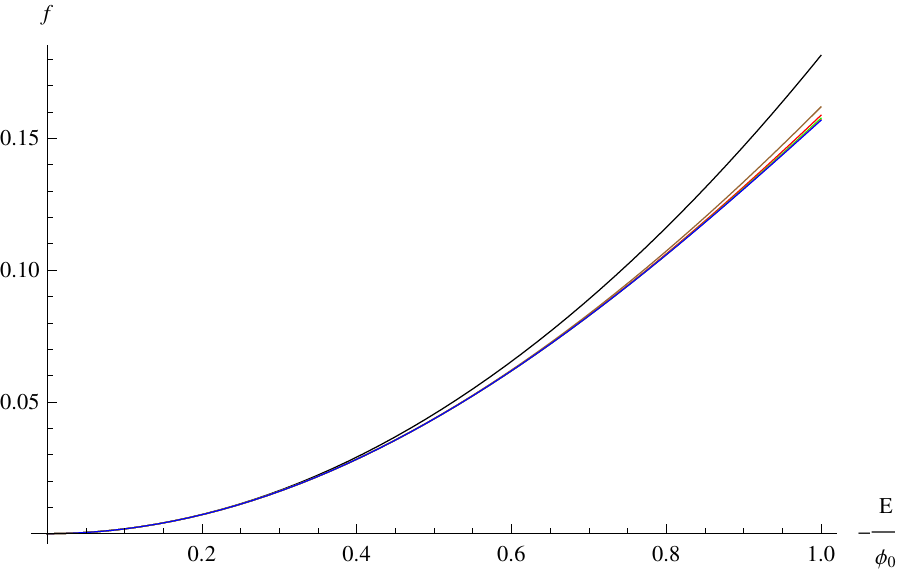} & \includegraphics[width=6.8cm]{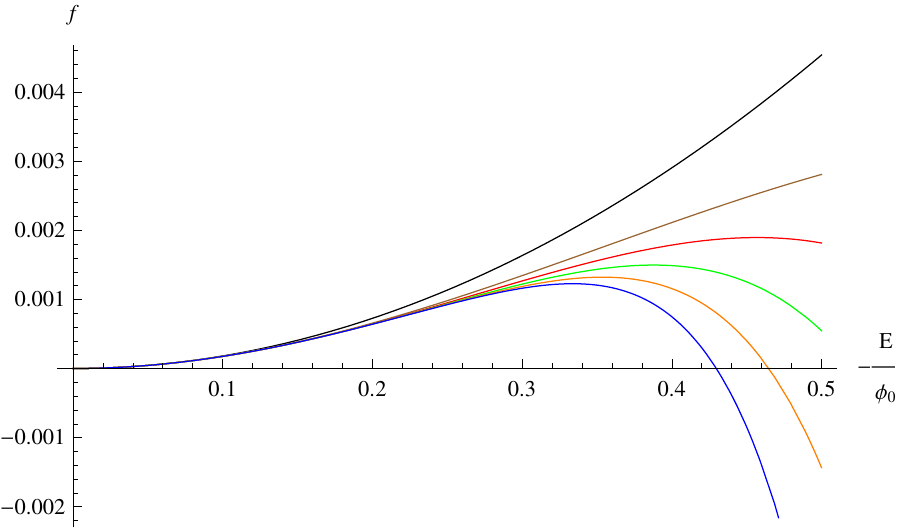} \\
 \includegraphics[width=6.8cm]{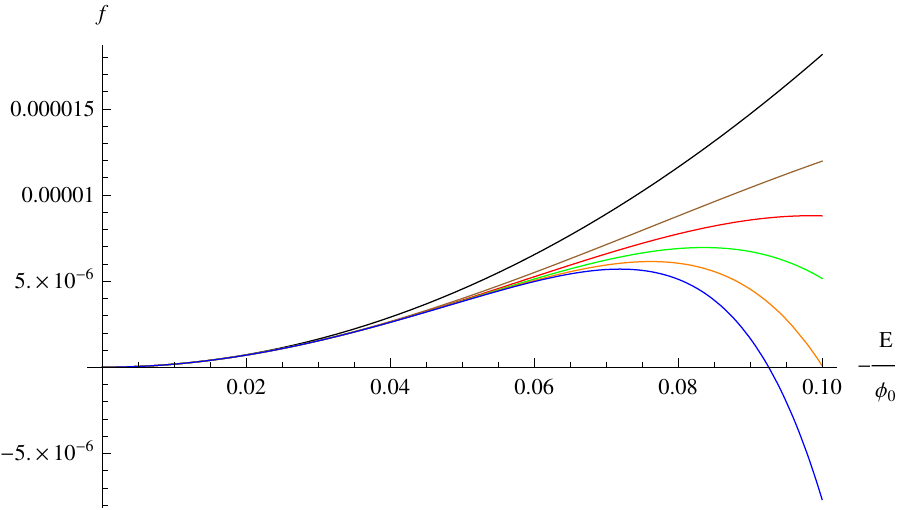} & \includegraphics[width=6.8cm]{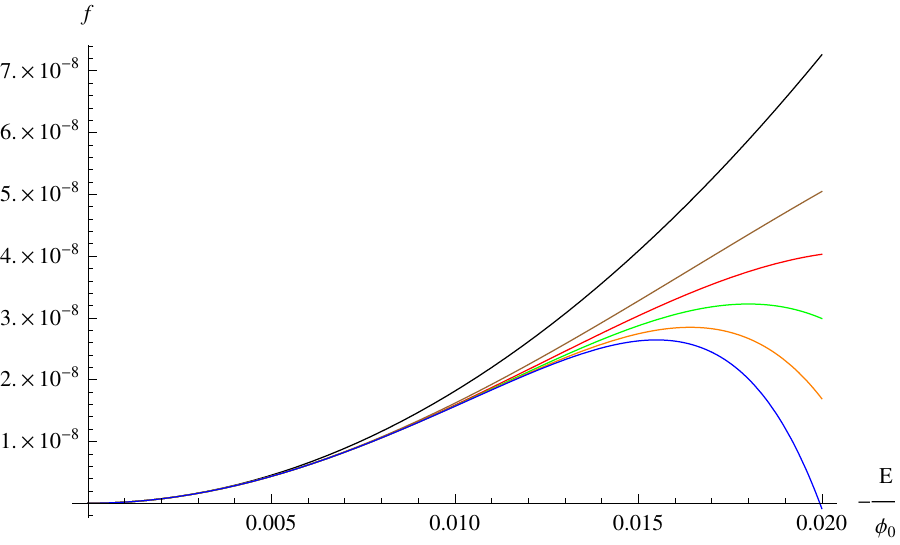}
\end{array}
$$
\caption{(color figures online) Plots of the distribution function as a function of the energy, for different fixed values of angular momentum, for the hypervirial Dehnen model far from the origin. Top-left panel: $L=0.1 a \sqrt{\Phi_0}$. Top-right panel: $L=a \sqrt{\Phi_0}$. Bottom-left panel: $L=10 a \sqrt{\Phi_0}$. Bottom-right panel: $L=100 a \sqrt{\Phi_0}$. For all panels: the black, brown, red, green, orange and blue curves correspond to the distribution function with 1, 2, 3, 4, 5 and 6 terms, respectively.}
\label{DehnenDFlarger}
\end{figure*}

Expanding around infinity, we find the augmented density and distribution function:
\begin{equation}
\rho = \frac{3a}{8\pi G^{3}M^{2}}r^{-1}\Psi^{3}-\frac{3a^2}{32\pi G^{5}M^{4}}\Psi^{5}-\frac{3a^3}{64\pi G^{7}M^{6}}r \Psi^{7}-\frac{15a^4}{256\pi G^{9}M^{8}}r^{2}\Psi^{9}-\frac{195a^5}{2048\pi G^{11}M^{10}}r^{3}\Psi^{11}-\frac{1449a^6}{8192\pi G^{13}M^{12}}r^{4}\Psi^{13},
\end{equation}
\begin{eqnarray}
f{(\mathcal{E},L)} &=& \frac{9a}{16\pi^{3} G^{3}M^{2}} \frac{\mathcal{E}^{2}}{L} - \frac{3 \sqrt{2} a^2}{7 \pi^{3} G^{5}M^{4}}{\mathcal{E}}^{7/2} - \frac{63 a^3}{64 \pi^{3} G^{7}M^{6}} {\mathcal{E}}^5 L \\ \nonumber
&-& \frac{360 \sqrt{2} a^4}{143 \pi^{3} G^{9}M^{8}} {\mathcal{E}}^{13/2} L^2 - \frac{32175 a^5}{2048 \pi^{3} G^{11}M^{10}} {\mathcal{E}}^8 L^3 - \frac{17388 \sqrt{2} a^6}{323 \pi^{3} G^{13}M^{12}} {\mathcal{E}}^{19/2} L^4.
\end{eqnarray}
Again, the convergence of the series around infinity hinges on the fact that $r\Psi^{2} \rightarrow r^{-1}$ as $r \rightarrow \infty$. The value for $\Delta$ is found to be:
\begin{equation}
\Delta = \frac{973}{1024}x^{-6}.
\end{equation}
The six-term distribution has the following velocity dispersion profiles
\begin{equation}
\sigma_r^2 = \frac{\Psi}{6} \frac{-1536+256x(\Psi/\Phi_0)^2+96x^2(\Psi/\Phi_0)^4+96x^3(\Psi/\Phi_0)^6+130x^4(\Psi/\Phi_0)^8+207x^5(\Psi/\Phi_0)^{10}}{-1024+256x(\Psi/\Phi_0)^2+128x^2(\Psi/\Phi_0)^4+160x^3(\Psi/\Phi_0)^6+260x^4(\Psi/\Phi_0)^8+483x^5(\Psi/\Phi_0)^{10}},
\end{equation}
\begin{equation}
\sigma_\theta^2 = \frac{\Psi}{6} \frac{-768+256x(\Psi/\Phi_0)^2+144x^2(\Psi/\Phi_0)^4+192x^3(\Psi/\Phi_0)^6+325x^4(\Psi/\Phi_0)^8+621x^5(\Psi/\Phi_0)^{10}}{-1024+256x(\Psi/\Phi_0)^2+128x^2(\Psi/\Phi_0)^4+160x^3(\Psi/\Phi_0)^6+260x^4(\Psi/\Phi_0)^8+483x^5(\Psi/\Phi_0)^{10}},
\end{equation}
and the corresponding anisotropy parameter is
\begin{equation}
\beta = \frac{-3\left(256+16x^2(\Psi/\Phi_0)^4+32x^3(\Psi/\Phi_0)^6+65x^4(\Psi/\Phi_0)^8+138x^5(\Psi/\Phi_0)^{10}\right)}{-1536+256x(\Psi/\Phi_0)^2+96x^2(\Psi/\Phi_0)^4+96x^3(\Psi/\Phi_0)^6+130x^4(\Psi/\Phi_0)^8+207x^5(\Psi/\Phi_0)^{10}}.
\end{equation}
The velocity dispersion, anisotropy parameter and $\Delta$ are plotted in Fig. \ref{fig7}. As before, the convergence of the multi term velocity dispersion profiles is excellent. In Fig. \ref{DehnenDFlarger}, we plot the large $r$ Dehnen models as a function of the dimensionless binding energy, while holding the angular momentum $L$ fixed. Many of the features that are evident from this graph, such as the existence of a finite radius of convergence that decreases as one increases $L$, can be explained by using the same methodology as the one employed for the large $r$ isochrone models. In particular, we refer to Fig. \ref{fig5dot5} and its corresponding discussion in the subsection on the hypervirial isochrone models.

\section{Asymptotic properties of the hypervirial augmented density ansatz}\label{SectIII}
In this section, we investigate the velocity structure of a generic hypervirial model, in particular the asymptotic values of the anisotropy parameter as $r \rightarrow 0$. From the augmented density (\ref{hypervirialrho}) and the statement of hyperviriality (\ref{hyperviriality}), it follows that
\begin{equation} \label{betahypervirialallmodels}
\beta=\frac{3}{2}-\frac{1}{2}\left(\frac{\sum_{k}C_{k}\Psi^{2p_{k}+1}r^{p_{k}-2}}{\sum_{k}\frac{C_{k}}{p_{k}+1}\Psi^{2p_{k}+1}r^{p_{k}-2}}\right).
\end{equation}
Before we proceed to investigate the behaviour of $\beta$ in the two asymptotic regimes, two important assumptions are made. First, we assume that the potential is finite at the origin, and secondly, we assume that it goes to zero at infinity. It must be emphasized that there exist models such as the isothermal sphere or models with central black holes whose $\Psi$ does not satisfy these properties. But, the majority of the models used in the literature such as the Dehnen, H\'enon isochrone, Veltmann and NFW models do possess this property.
\subsection{The $r\rightarrow0$ limit}
To investigate the behaviour of the potential--density pair in the limit $r \rightarrow 0$, we shall use the method of dominant balance from asymptotic analysis \citep{be99,wh10}. We begin with the Poisson equation
\begin{equation}
\nabla^{2}\Psi + 4\pi G \rho(\Psi,r) =  \nabla^{2}\Psi + 4 \pi G \sum_k C_{k}\Psi^{2p_{k}+1}r^{p_k-2} = 0.
\end{equation}
Now, we investigate the behaviour of the Poisson equation in the limit $r \rightarrow 0$. We will suppose that the terms in the augmented density that have higher powers of $r$ drop out, and that only the term in the augmented density which dominates is the one with the lowest power of $r$. If we make this assumption, the Poisson equation reduces to
\begin{equation}
\nabla^{2}\Psi+ 4\pi G\, C_{\kappa}\Psi^{2p_{\kappa}+1}r^{p_{\kappa}-2} = 0,
\end{equation}
where $p_{\kappa}$ is the smallest value of $p_{k}$, and $C_{\kappa}$ is the corresponding value of $C$. This differential equation can be solved exactly, and one can show that the solution with the boundary conditions on $\Psi$ discussed above has been derived by \citet{ea05,ng13}. It is given by

\begin{equation} \label{pot0}
\Psi\propto\left(1+r^{p_{\kappa}}\right)^{-1/p_{\kappa}},
\end{equation}
and the Laplacian scales as follows
\begin{equation} \label{density0}
\nabla^{2}\Psi\propto r^{p_{\kappa}-2}\left[\left(1+r^{p_{\kappa}}\right)^{-1/p_{\kappa}}\right]^{2p_{\kappa}+1}.
\end{equation}

Now, we need to establish that the assumption we made, namely that the higher order terms in $r$ that appear in the augmented density go to zero, is indeed valid. In short, we need to show that the following equations hold true.

\begin{equation} \label{condition1r0}
\lim_{r\rightarrow0}\frac{C_{k}\Psi^{2p_{k}+1}r^{p_{k}-2}}{C_{\kappa}\Psi^{2p_{\kappa}+1}r^{p_{\kappa}-2}}=0;\quad\quad k\neq \kappa.
\end{equation}
\begin{equation} \label{condition2r0}
\lim_{r\rightarrow0}{4\pi G}\frac{C_{k}\Psi^{2p_{k}+1}r^{p_{k}-2}}{-\nabla^{2}\Psi}=0.
\end{equation}

By using equations (\ref{pot0}) and (\ref{density0}) as well as the relation $p_{k}-p_{\kappa}>0$ for all values of $k$, it is immediately evident that equations (\ref{condition1r0}) and (\ref{condition2r0}) are indeed satisfied. Since we have an asymptotic expression for $\Psi(r)$ in the limit $r\rightarrow0$, we can substitute it in equation (\ref{betahypervirialallmodels}) to find the inner anisotropy parameter. After some algebra, it is found that
\begin{equation} \label{betar0}
\beta_{0}=1-\frac{p_{\kappa}}{2}.
\end{equation}
By using equation (\ref{density0}), it is evident that the inner slope is given by $\gamma_{0}=2-p_{\kappa}$. On combining this relation with the above expression, one obtains the following equality for these models:
\begin{equation} \label{slopeanisotropy1}
\gamma_{0}=2\beta_{0},
\end{equation}
which happens to be exactly the limiting case of the cusp slope--central anisotropy theorem \citep{ae06}. The cusp slope--central anisotropy states that the central slope ($\gamma_0$) and the central anisotropy parameter ($\beta_0$) are related through the inequality $\gamma_0 \geq 2 \beta_0$, and we observe that the hypervirial augmented density ansatz is found to saturate this inequality regardless of the values chosen for the parameters.

\subsection{Discussion}
In the previous subsections, we derived an anisotropy equality for this family of models in the regime $r\rightarrow0$ which relates the density slope in that region to the value of the anisotropy parameter in the same region. It was found that all these models obey equation (\ref{slopeanisotropy1}). The most striking aspect of this relation is that they are completely independent of the choice of the $C_k$ and the $p_k$ and the number of terms $N$ in the hypervirial augmented density ansatz, which proves their generality.

Furthermore, one discovers a second remarkable result by considering equation (\ref{betahypervirialallmodels}). Suppose that we restrict ourselves to those models which have finite total mass. As previously discussed, the potential falls off as $r^{-1}$ in the $r \rightarrow \infty$ limit. By using this property in conjunction with equation (\ref{betahypervirialallmodels}), it can be shown that $\beta_\infty = 1-\frac{p_{\kappa}}{2}$. This might lead us to conclude that $\beta_0 = \beta_\infty$, but this is \emph{not} the case. The reason is simple, and stems from the non-monotonic nature of $\sqrt{r} \Psi(r)$ which was used to construct the augmented density. Because of this property, one ends up with two different branches, and consequently, the augmented density in the region $r \rightarrow 0$ is not the same as the augmented density in the region $r \rightarrow 0$. Hence, the value of $p_\kappa$ for these models is also not identical which in turn implies that $\beta_0 \neq \beta_\infty$ in general. But, if one does have models where the augmented density is the same for all values of $x$, one can indeed see that the equality is preserved.  From the expression for the anisotropy parameter, given by equation (\ref{betahypervirialallmodels}), it is evident that $\beta$ is a function of $r$, except in the special case where $N=1$. If $N=1$, the anisotropy parameter is independent of $r$, and one does indeed have $\beta_0 = \beta_\infty$ for these models. These models also happen to be the one-term hypervirial models discussed in the previous sections, where the two branches coincide. Another set of models for whom this equality is preserved are the two-term hypervirial models studied by \citet{an05} where the augmented density is applicable for all values of $r$.

\section{Conclusion}\label{SectCon}
This paper presents a hybrid approach that combines the knowledge of the potential--density pair with a distribution function comprising of double-power terms. A major advantage of this approach is that one can bypass integral transforms, except for equation (\ref{centralintegral}), and one can instead sum the series to obtain closed form expressions. It must be emphasized  ,however, that not all potential--density pairs lend themselves to such an approach, but a large number of the physically relevant models are amenable to such a treatment. It is also important that the free parameters be chosen such that they ensure that the distribution functions thus generated satisfy the phase-space consistency requirements \citep{an11a,an11b,an12}. The second advantage of this approach lies in the fact that it presents a means of generating a reasonably diverse range of distribution functions for several models. The double-power distribution functions are themselves generated from a double-power augmented density ansatz, which has additional advantages of its own. In particular, one can construct an appropriate double-power augmented density ansatz that is always hypervirial, irrespective of the choice of the potential--density pair, and use it to generate the potential--density pairs of known models to arbitrary accuracy.

In Section \ref{SectI}, we illustrate the method of constructing a more complex augmented density from a basic one by performing simple algebraic operations on the simpler augmented density. The usefulness of this approach is demonstrated by applying it to a concrete example, namely the $m=1/2$ Veltmann model. This model is of relevance because of its similarities to the NFW profile, and for the reason that the basic augmented density is hypervirial. In this section, we explicitly compute some of the simpler distribution functions and their velocity dispersions for this model. In an associated appendix (Appendix \ref{AppA}), we also list some of the corresponding expressions for the Plummer and Hernquist models, which can also be generated using alternative methodologies currently existing in the literature (see, for example, \citet{dj87,ba02,bvh07}).

However, the primary aim of the paper is to generate hypervirial distribution functions for a wide array of models. In order to do so, we start with an appropriate double-power augmented density ansatz in Section \ref{SectII} that satisfies the hypervirial property for any given potential--density pair. The next step involves expanding the potential--density pair as a power series, and matching coefficients with the expression for the augmented density. By doing so, we present a method that gives rise to hypervirial distribution functions that can model a given potential--density pair to an arbitrarily high degree of accuracy. The non-monotonicity of $\sqrt{r}\Psi$ requires two different distribution functions in the two regimes $r\rightarrow0$ and $r\rightarrow\infty$.  This method is employed to construct hypervirial distribution functions for commonly used potential--density pairs in the literature including the H\'enon isochrone, NFW, Jaffe and Dehnen models. After we construct the distribution functions, we also compute the velocity dispersions and the anisotropy parameter, as well as the relative error $\Delta$ which measures the deviation of the augmented density from the actual density. 

In Section \ref{SectIII}, we take the generic hypervirial augmented density that was used to generate the distribution functions for the above models, and investigate its behaviour as $r\rightarrow0$ and $r\rightarrow\infty$. We demonstrate the existence of a universal relation between the anisotropy parameter and the density slope in the asymptotic regime $r \rightarrow 0$. In other words, we show that all the hypervirial distribution functions thus constructed (from the hypervirial augmented density ansatz) saturate the cusp slope--central anisotropy theorem derived by \citet{ae06}. Furthermore, it is found that the asymptotic values of the anisotropy parameter, $\beta_0$ and $\beta_\infty$, are equal for all models that possess a finite total mass and can be described by a single augmented density in all regions. For such models, we show that the asymptotic values of the anisotropy parameter are functionally very simple.

Owing to the existence of these universal properties for the hypervirial augmented densities, it seems reasonable to suppose that observations which establish any of the above properties may be strongly indicative of an underlying hypervirial distribution function. However, it must be emphasized that any potential observational evidence that satisfies any of the above properties will $not$ necessarily indicate hyperviriality because these properties hold true for the hypervirial ansatz but the converse is not necessarily valid. Nonetheless, we believe that further investigations along these lines constitute a promising line of enquiry.

\section{Acknowledgements}
We thank Richard Matzner and Philip Morrison for their support and guidance. We are deeply grateful to Jin An for his insightful suggestions, detailed comments and for pointing out new avenues that we had hitherto been unaware of. This material is based on work supported by the Department of Physics and the Texas Cosmology Center, which is supported by the College of Natural Sciences and the Department of Astronomy at the University of Texas at Austin and the McDonald Observatory.

\appendix
\section{Distribution functions for the Plummer and Hernquist models} \label{AppA}
\subsection{Distribution functions for the Plummer model}
In this appendix, we apply the method developed for the $m=1/2$ Veltmann model to write down a few simple distribution functions for the Plummer and Hernquist models. The augmented density with one-component is
\begin{equation} \label{Plum1}
\rho(x,\Phi) = \frac{3a^{2}}{4\pi G^{5}M^{4}} (-\Phi)^5,
\end{equation}
and the corresponding distribution function is nothing but the isotropic distribution function. Next, we look for two-component distribution functions. The only such model is
\begin{equation} \label{Plum2}
\rho(x,\Psi) =  \frac{3a^{4}}{4\pi G^{7}M^{6}} \Psi^7 \left(1+x^2\right).
\end{equation}
\begin{equation}
f({\mathcal{E}},L) = \frac{64\sqrt{2}a^4}{11G^7M^6{\pi}^3} {\mathcal{E}}^{11/2} + \frac{16\sqrt{2}a^2}{G^7M^6{\pi}^3} L^{2} {\mathcal{E}}^{9/2}.
\end{equation}
This is a particular case of the one-parameter family presented in \citet{ji07c}. Finally, we proceed to derive three-component distribution functions for the Plummer sphere. This can actually be done in three different ways. The first possible choice is
\begin{equation} \label{Plum3a}
\rho(x,\Psi) = \left(\frac{3a^6}{4\pi G^{9}M^{8}}\right)\Psi^{9}\left(1+2x^2+x^4\right),
\end{equation}
\begin{equation}
f({\mathcal{E}},L)=\frac{6144 \sqrt{2} a^6}{715 \pi^{3} G^9 M^8}{\mathcal{E}}^{15/2}+\frac{4608 \sqrt{2} a^4}{143 \pi^{3} G^9 M^8}L^{2} {\mathcal{E}}^{13/2}+\frac{576 \sqrt{2} a^2}{11 \pi^{3} G^9 M^8}L^{4} {\mathcal{E}}^{11/2}.
\end{equation}
This distribution function is yet another special case of the family presented in \citet{ji07c}. The second choice of augmented density and distribution function is
\begin{equation}\label{Plum3b}
\rho(x,\Psi)=\frac{3a^{4}}{4\pi G^{7}M^{6}}\Psi^{7}\left[\left(\frac{a}{GM}\right)^2\Psi^{2}\left(1+x^2\right)+x^2\right].
\end{equation}
\begin{equation}
f({\mathcal{E}},L)=\frac{6144 \sqrt{2} a^6}{715 \pi^{3} G^9 M^8}{\mathcal{E}}^{15/2}+\frac{4608 \sqrt{2} a^4}{143 \pi^{3} G^9 M^8}L^{2} {\mathcal{E}}^{13/2}+\frac{16 \sqrt{2} a^2}{\pi^{3} G^7 M^6}L^{2} {\mathcal{E}}^{9/2}.
\end{equation}
The third possibility is
\begin{equation}\label{Plum3c}
\rho(x,\Psi)=\frac{3a^{4}}{4\pi G^{7}M^{6}}\Psi^{7}\left[1+\left(\frac{a}{GM}\right)^2\Psi^{2}x^2\left(1+x^2\right)\right].
\end{equation}
\begin{equation}
f({\mathcal{E}},L)=\frac{64 \sqrt{2} a^4}{11 \pi^{3} G^7 M^6} {\mathcal{E}}^{11/2}+\frac{4608 \sqrt{2} a^4}{143 \pi^{3} G^9 M^8}L^{2} {\mathcal{E}}^{13/2}+\frac{576 \sqrt{2} a^2}{11 \pi^{3} G^9 M^8}L^{4} {\mathcal{E}}^{11/2}.
\end{equation}
\begin{figure*}
$$
\begin{array}{ccc}
 \includegraphics[width=6.8cm]{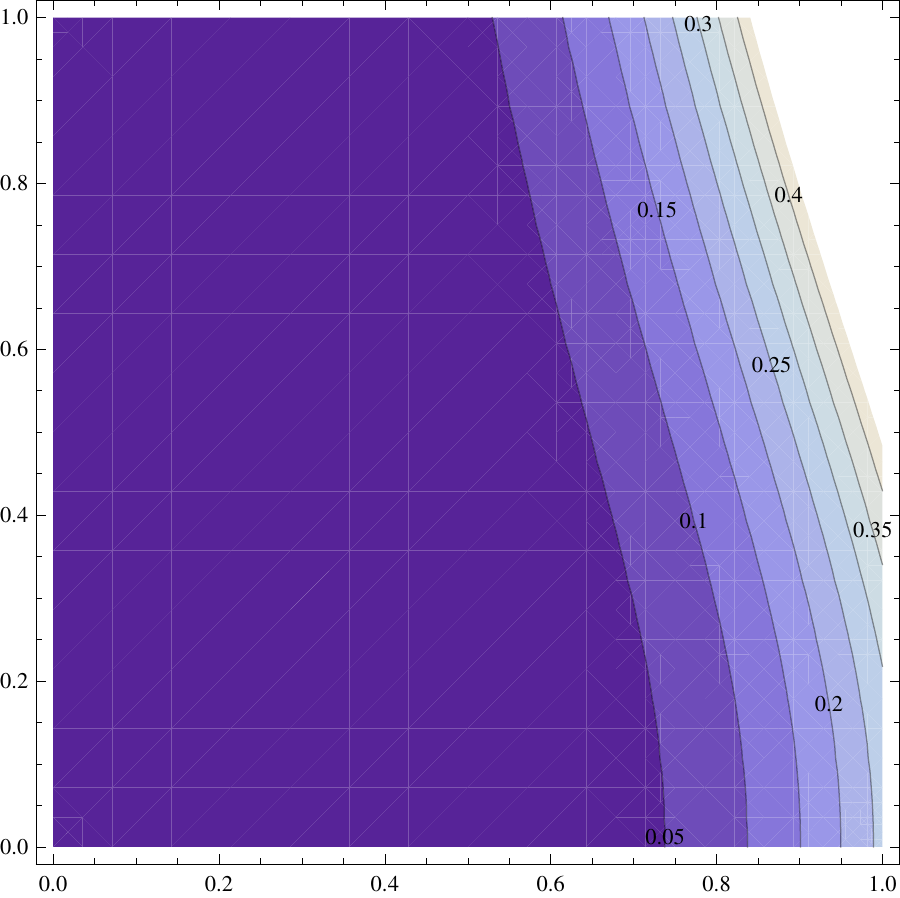} & \includegraphics[width=6.8cm]{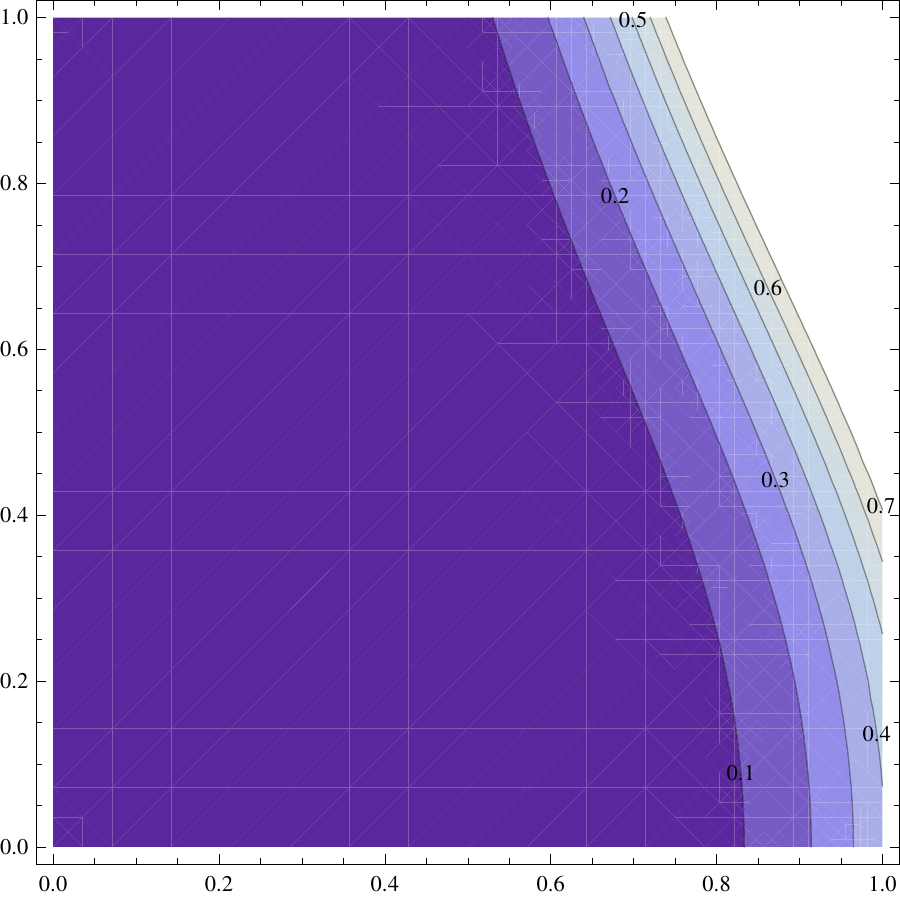} \\
 \includegraphics[width=6.8cm]{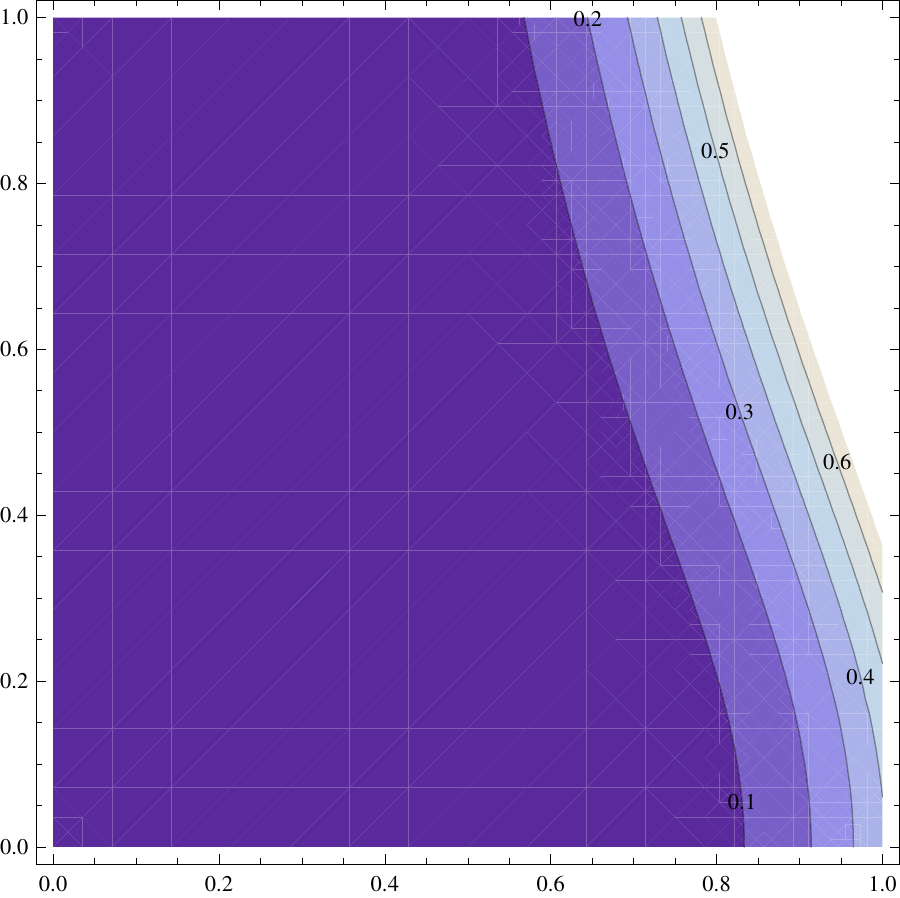} & \includegraphics[width=6.8cm]{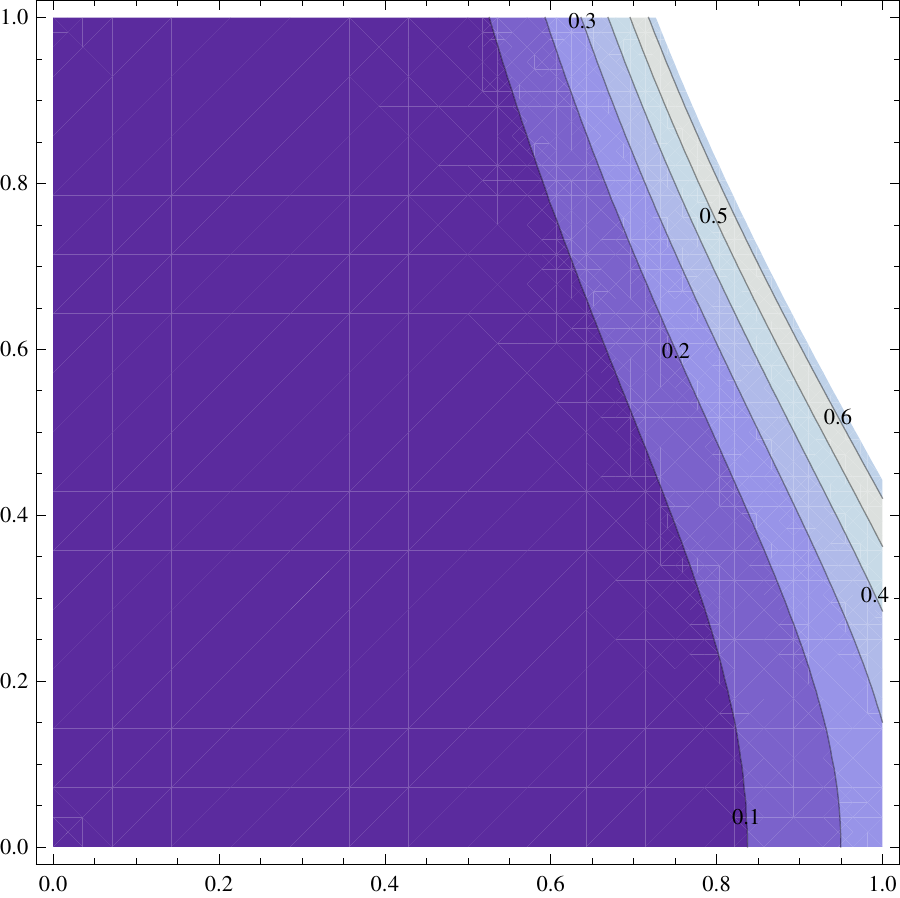}
\end{array}
$$
\caption{(color figures online) Contour plots of the Plummer distribution functions. The horizontal direction is $-E/\Phi_0$ and the vertical direction is $L/(a\sqrt{\Phi_0})$. Top-left panel: the distribution function of equation (\ref{Plum2}). Top-right panel: the distribution function of equation (\ref{Plum3a}). Bottom-left panel: the distribution function of equation (\ref{Plum3b}). Bottom-right panel: the distribution function of equation (\ref{Plum3c}).}
\label{contourPlum}
\end{figure*}

\subsection{The velocity dispersions for the Plummer model distribution functions}\label{AppPlumVel}
\begin{figure*}
$$
\begin{array}{ccc}
 \includegraphics[width=5.2cm]{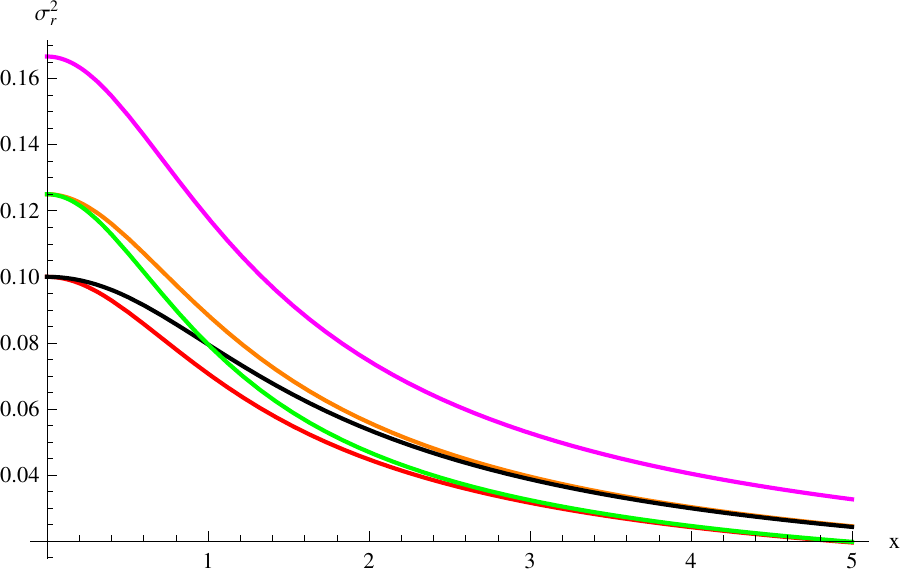} & \includegraphics[width=5.2cm]{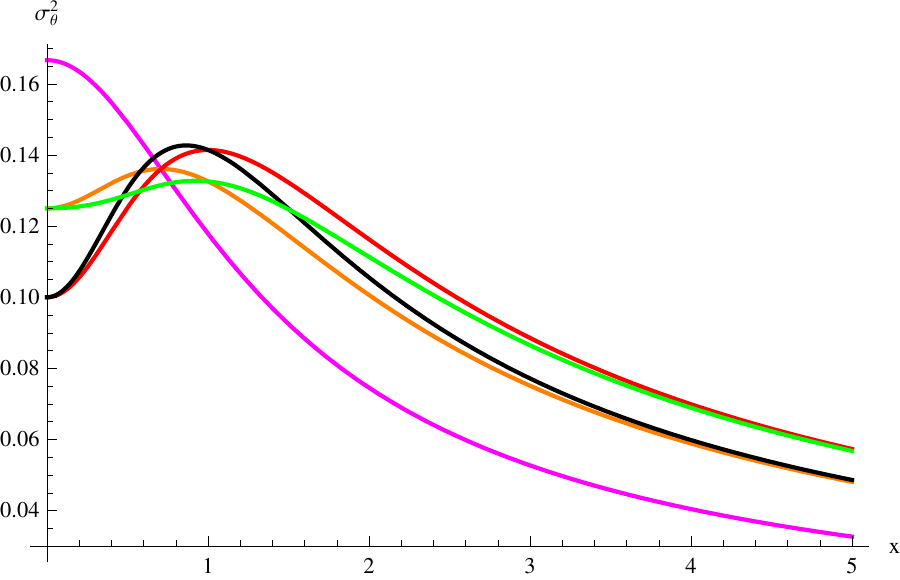} & \includegraphics[width=5.2cm]{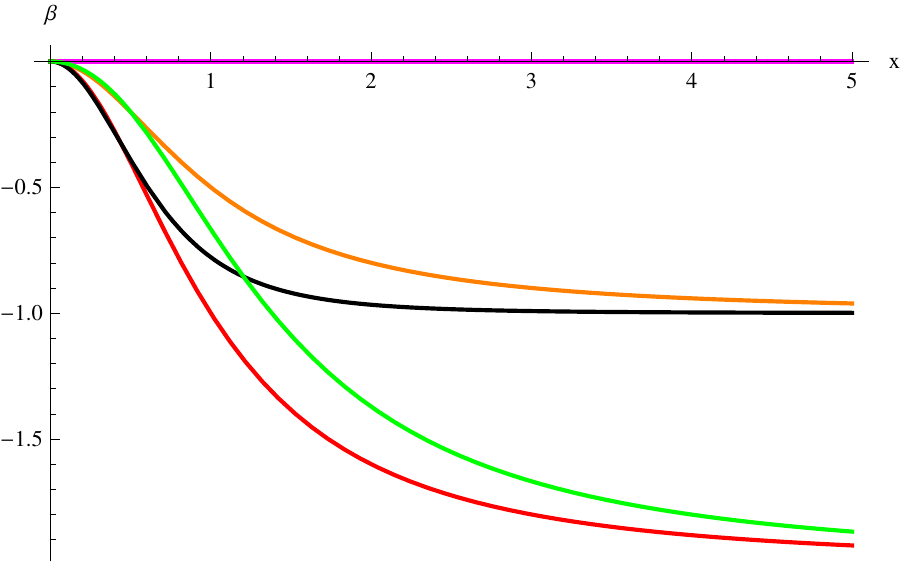}\\
 \quad\quad(a) & \quad\quad(b) & \quad\quad(c)\\
\end{array}
$$
\caption{(color figures online) The velocity dispersion tensor and the anisotropy parameter for multi component Plummer models. Left-hand panel: $\sigma^{2}_{r}$ as a function of $x$. Central panel: $\sigma^{2}_{\theta}$ as a function of $x$. Right-hand panel: $\beta$ as a function of $x$. For all panels: we plotted the one-component model (magenta), the two-component model (orange) and the three-component models (red, black and green, in the order in which they are discussed in Appendix \ref{AppPlumVel}). Note: the velocity dispersions have been converted to dimensionless units before plotting.}
\label{figA1}
\end{figure*}
As usual, we will compute the velocity structure of the models, i.e. the tangential and radial velocity dispersions as well as the anisotropy parameter. For the isotropic Plummer model (\ref{Plum1}), we find
\begin{equation}
\sigma^2_r = \sigma^2_\phi = \frac{1}{6} \frac{\Phi_{0}}{\sqrt{1+x^2}},
\end{equation}
\begin{equation}
\beta = 0.
\end{equation}
For the two-term distribution function, the density is given by equation (\ref{Plum2}). The velocity dispersions and anisotropy parameter are
\begin{equation}
\sigma^2_r = \frac{1}{8} \frac{\Phi_{0}}{\sqrt{1+x^2}},
\end{equation}
\begin{equation}
\sigma^2_\phi = \frac{1}{8}\Phi_{0} \frac{1+2x^2}{(1+x^2)^{3/2}},
\end{equation}
\begin{equation}
\beta = -\frac{x^2}{1+x^2}.
\end{equation}
There are three different three-term distribution functions. The first of these has a density dependence given by equation (\ref{Plum3a}). For this model, the dispersions and anisotropy parameter are
\begin{equation}
\sigma^2_r = \frac{1}{10} \frac{\Phi_{0}}{\sqrt{1+x^2}},
\end{equation}
\begin{equation}
\sigma^2_\phi = \frac{1}{10} \Phi_{0} \frac{1+4x^2+3x^4}{(1+x^2)^{5/2}},
\end{equation}
\begin{equation}
\beta = -2\frac{x^2+x^4}{1+2x^2+x^4}.
\end{equation}
For the second three-term distribution function, the density is given by (\ref{Plum3b}). The velocity dispersions are computed for this model.
\begin{equation}
\sigma^2_r = \frac{\Phi_{0}}{40} \frac{4+5x^2}{(1+x^2)^{3/2}},
\end{equation}
\begin{equation}
\sigma^2_\phi = \frac{\Phi_{0}}{20} \frac{2+9x^2+5x^4}{\left(1+x^2\right)^{5/2}},
\end{equation}
\begin{equation}
\beta = - x^2 \frac{\left(9+5x^2\right)}{\left(1+x^2\right)\left(4+5x^2\right)}.
\end{equation}
For the last Plummer model, which is given by (\ref{Plum3c}), we obtain
\begin{equation}
\sigma^2_r = \frac{\Phi_{0}}{40} \frac{5+4x^2}{\left(1+x^2\right)^{3/2}},
\end{equation}
\begin{equation}
\sigma^2_\phi = \frac{\Phi_{0}}{40}\frac{5+13x^2+12x^4}{\left(1+x^2\right)^{5/2}},
\end{equation}
\begin{equation}
\beta = - 4x^2 \frac{\left(1+2x^2\right)}{\left(1+x^2\right)\left(5+4x^2\right)}.
\end{equation}
The anisotropy parameters for all the Plummer models presented above are either constant or decreasing functions of $r$, and they exhibit tangential anisotropy for all values of $r$. Models with tangential anisotropy have been studied in the context of analysing the substructure of dark matter haloes (see e.g. \citet{an06}). The radial and tangential velocity dispersions, as well as the anisotropy parameter, for the distribution functions for the Plummer model discussed in this section have been plotted in Fig. \ref{figA1}. In Fig. \ref{contourPlum}, the contour plots for some of the Plummer distribution functions are depicted, where the distribution functions depend on both the dimensionless binding energy and the dimensionless angular momentum.

\subsection{Distribution functions for the Hernquist model}
Next, we repeat for the Hernquist model. Again, we start with the one-component distribution function, whose augmented density is
\begin{equation} \label{Her1}
\rho{(x,\Psi)} = \left(\frac{1}{2\pi G^{3}M^{2}}\right) x^{-1}\Psi^{3}.
\end{equation}
This augmented density corresponds to the hypervirial distribution function discussed in Section \ref{SectI}. Next, we proceed to the two-component distribution function for the Hernquist model:
\begin{equation} \label{Her2}
\rho{(x,\Psi)} = \left(\frac{a}{2\pi G^{4}M^{3}}\right)\Psi^{4}\left(1+\frac{1}{x}\right),
\end{equation}
\begin{equation}
f({\mathcal{E}},L)=\frac{16a}{5\sqrt{2}\pi^{3}G^{4}M^{3}} {\mathcal{E}}^{5/2}+\frac{a^{2}}{G^{4}M^{3}\pi^{2}}L^{-1} {\mathcal{E}}^{3}.
\end{equation}
As for the Plummer case, the three-component Hernquist models can be constructed in three distinct ways. The first augmented density and distribution function is
\begin{equation} \label{Her3a}
\rho(x,\Psi) = \left(\frac{a^2}{2\pi G^{5}M^{4}}\right) \Psi^5 x^{-1} (1+x)^2,
\end{equation}
\begin{equation}
f({\mathcal{E}},L)=\frac{5a^{3}}{16\pi^{4}G^{6}M^{4}}L^{-1} {\mathcal{E}}^{4}+\frac{8\sqrt{2}a^{2}}{7\pi^{4}G^{6}M^{4}} {\mathcal{E}}^{7/2}+\frac{5a}{4\pi^{4}G^{6}M^{4}}L {\mathcal{E}}^{3}.
\end{equation}
Alternatively, we have
\begin{equation} \label{Her3b}
\rho{(x,\Psi)} =\frac{a}{2\pi G^{4}M^{3}} \Psi^{4}\left(\frac{\Psi}{\Phi_0}(1+x)+\frac{1}{x}\right),
\end{equation}
\begin{equation}
f({\mathcal{E}},L) = \frac{16 \sqrt{2} a^2}{7 \pi^{3} G^5 M^4} {\mathcal{E}}^{7/2} + \frac{5 a}{\pi^{3} G^5 M^4} L {\mathcal{E}}^{3} + \frac{a^2}{\pi^{3} G^4 M^3}L^{-1} {\mathcal{E}}^{3}.
\end{equation}
Lastly, we also have
\begin{equation} \label{Her3c}
\rho{(x,\Psi)} = \frac{a}{2\pi G^{4}M^{3}} \Psi^{4}\left(1+\frac{\Psi}{\Phi_0}\frac{1+x}{x}\right),
\end{equation}
\begin{equation}
f({\mathcal{E}},L) =  \frac{8\sqrt{2}a}{5 \pi^{3} G^4 M^3} {\mathcal{E}}^{5/2} + \frac{5 a^3}{4 \pi^{3} G^5 M^4} L^{-1} {\mathcal{E}}^{4} + \frac{16 \sqrt{2} a^2}{7 \pi^{3} G^5 M^4} {\mathcal{E}}^{7/2}.
\end{equation}
\begin{figure*}
$$
\begin{array}{ccc}
 \includegraphics[width=6.8cm]{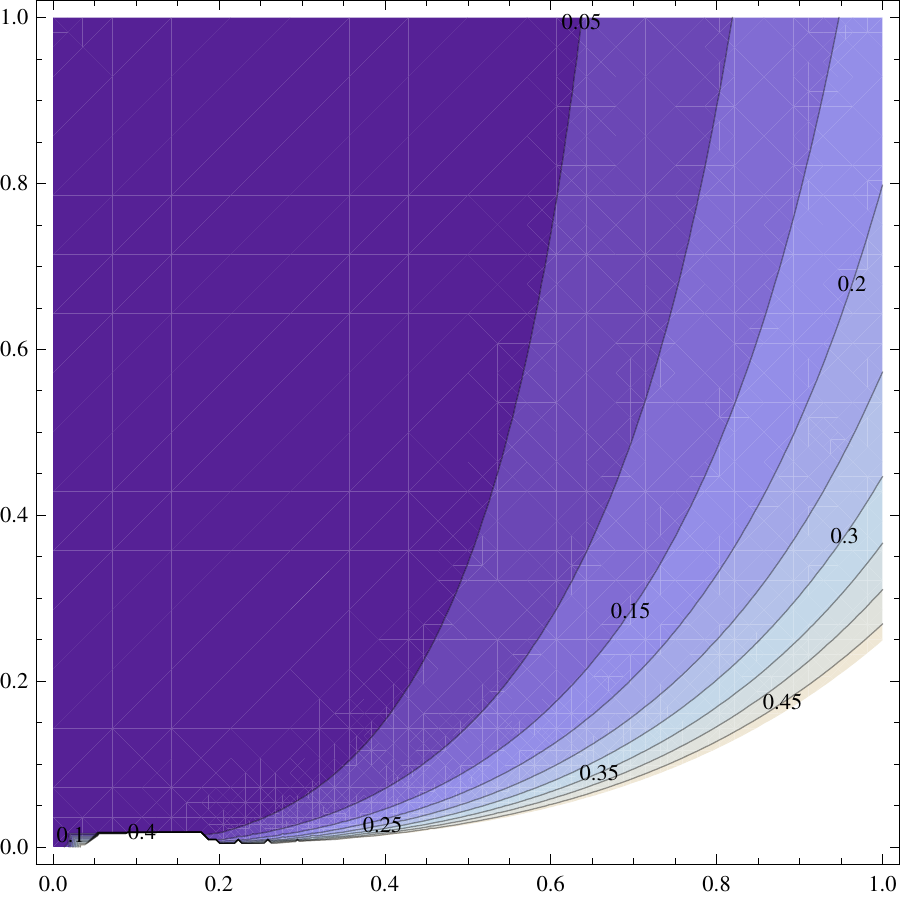} & \includegraphics[width=6.8cm]{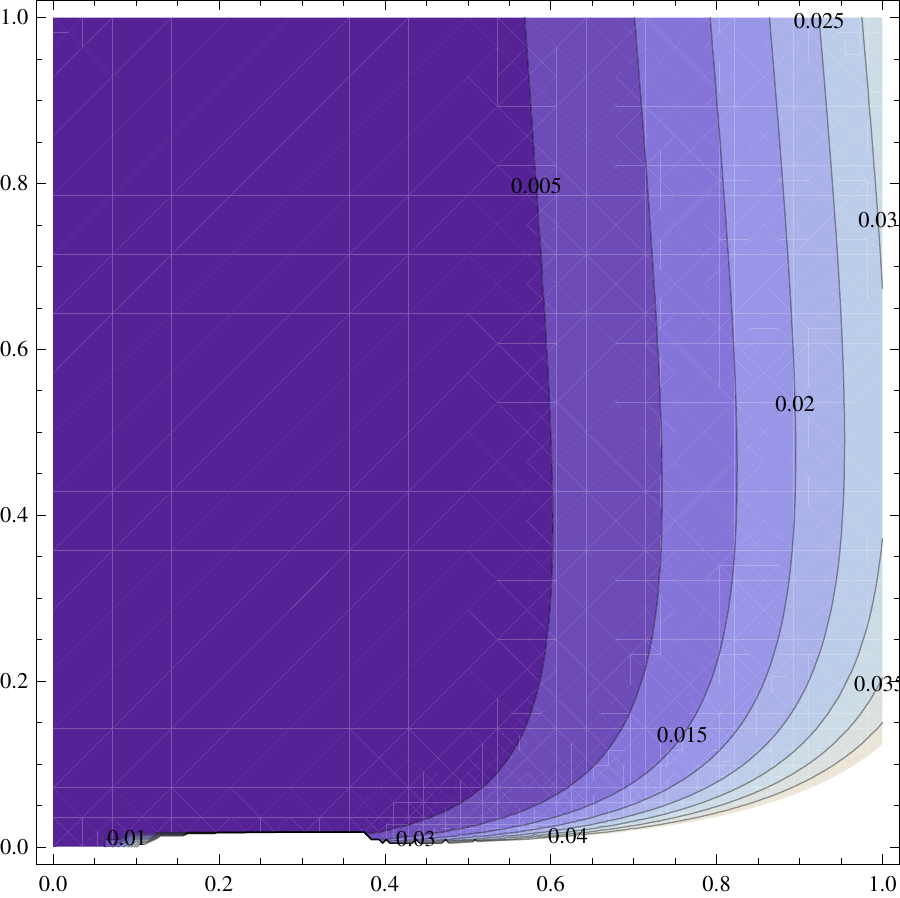} \\
 \includegraphics[width=6.8cm]{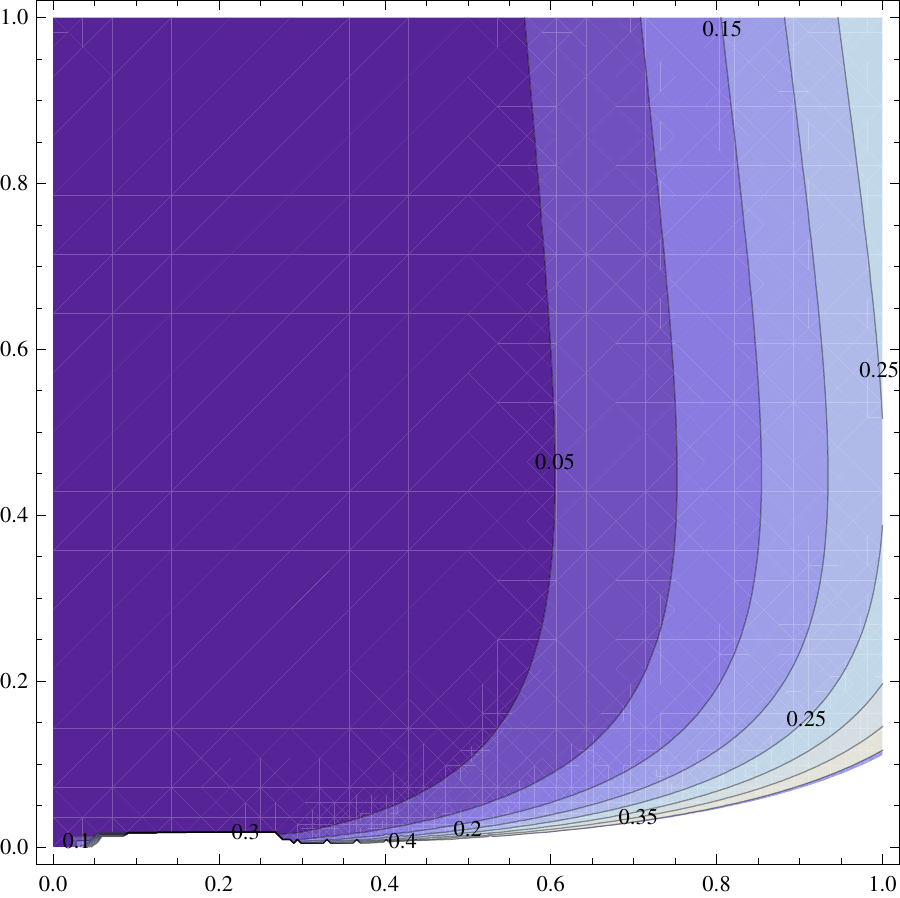} & \includegraphics[width=6.8cm]{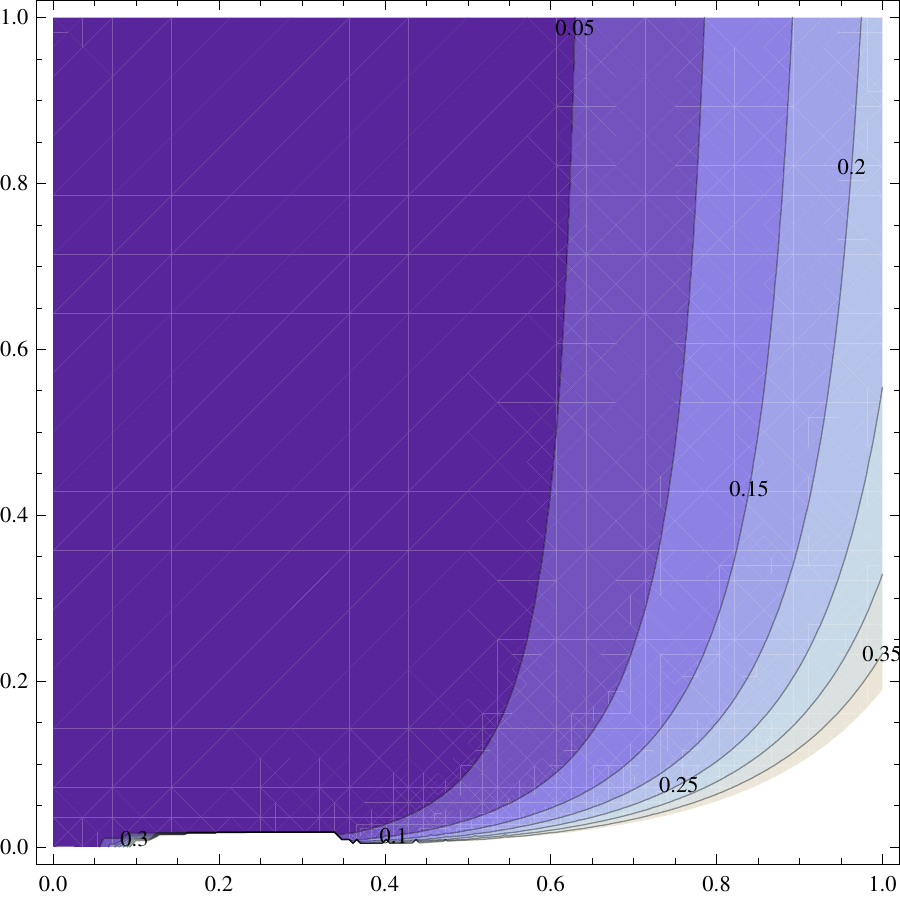}
\end{array}
$$
\caption{(color figures online) Contour plots of the Hernquist distribution functions. The horizontal direction is $-E/\Phi_0$ and the vertical direction is $L/(a\sqrt{\Phi_0})$. Top-left panel: the distribution function of equation (\ref{Her2}). Top-right panel: the distribution function of equation (\ref{Her3a}). Bottom-left panel: the distribution function of equation (\ref{Her3b}). Bottom-right panel: the distribution function of equation (\ref{Her3c}).}
\label{contourHer}
\end{figure*}
\subsection{The velocity dispersions for the Hernquist model distribution functions}\label{AppHernVel}
\begin{figure*}
$$
\begin{array}{ccc}
 \includegraphics[width=5.2cm]{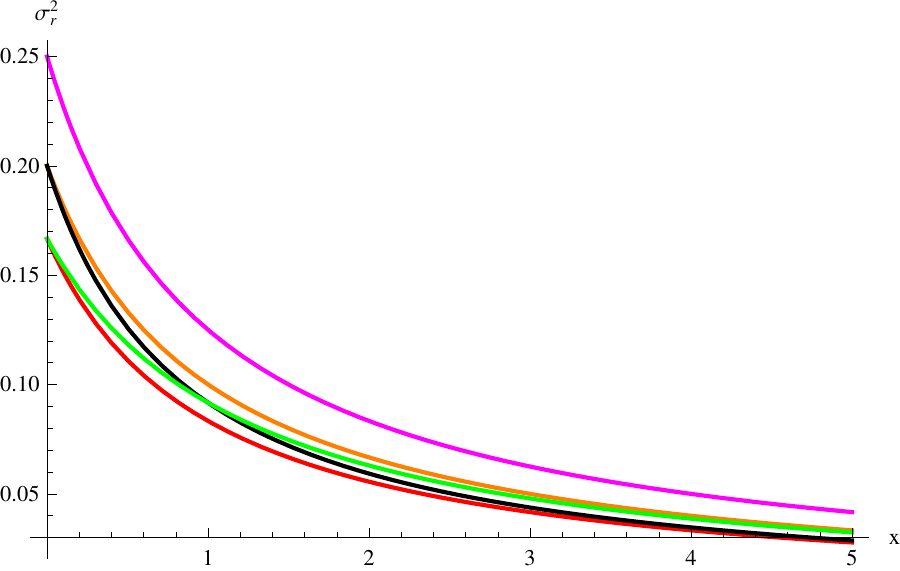} & \includegraphics[width=5.2cm]{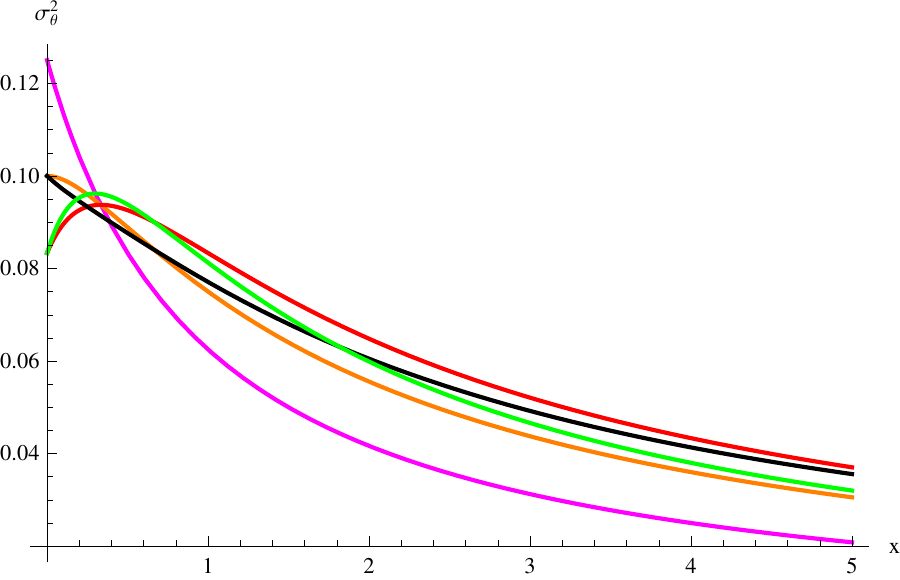} & \includegraphics[width=5.2cm]{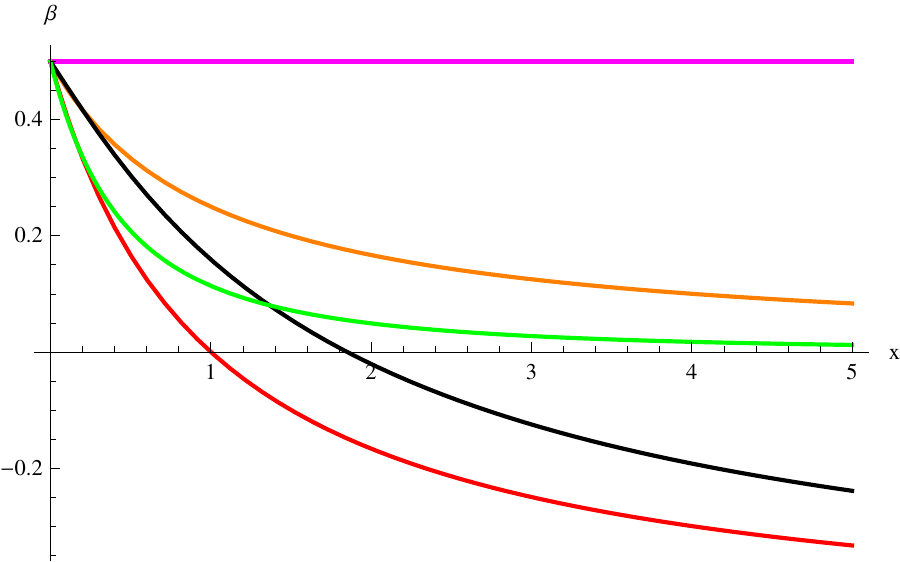}\\
 \quad\quad(a) & \quad\quad(b) & \quad\quad(c)\\
\end{array}
$$
\caption{(color figures online) The velocity dispersion tensor and the anisotropy parameter for multi component Hernquist models. Left-hand panel: $\sigma^{2}_{r}$ as a function of $x$. Central panel: $\sigma^{2}_{\theta}$ as a function of $x$. Right-hand panel: $\beta$ as a function of $x$. For all panels: we plotted the one-component model (magenta), the one-component model (orange), and the one-component models (red, black and green, in the order in which they are discussed in Appendix \ref{AppHernVel}). Note: the velocity dispersions have been converted to dimensionless units before plotting.}
\label{figA2}
\end{figure*}
The simplest model for the Hernquist profile has a density profile that takes on the functional dependence given by equation (\ref{Her1}). For this model, the velocity dispersions and anisotropy parameter are the same as the ones obtained for the hypervirial family.
\begin{equation}
\sigma^2_r = 2 \sigma^2_\phi = \frac{1}{4} \frac{\Phi_{0}}{1+x},
\end{equation}
\begin{equation}
\beta = \frac{1}{2}.
\end{equation}
The second simplest model is described by equation (\ref{Her2}). One can proceed in a manner similar to the Plummer models, and we obtain
\begin{equation}
\sigma^2_r = \frac{1}{5} \frac{\Phi_{0}}{1+x},
\end{equation}
\begin{equation}
\sigma^2_\phi = \frac{\Phi_{0}}{5} \frac{x+\frac{1}{2}}{(1+x)^2},
\end{equation}
\begin{equation}
\beta = \frac{1}{2} \frac{1}{1+x}.
\end{equation}
Now, we move on to the three term distribution functions. The first of these has the simplest functional dependence of the density and is given by equation (\ref{Her3a}). For this model, the quantities are found to be
\begin{equation}
\sigma^2_r = \frac{1}{6} \frac{\Phi_{0}}{1+x},
\end{equation}
\begin{equation}
\sigma^2_\phi = \frac{\Phi_{0}}{12} \frac{3x^2+4x+1}{(1+x)^3},
\end{equation}
\begin{equation}
\beta = \frac{1}{2} \frac{1-x}{1+x}.
\end{equation}
The second three-term distribution function has a density given by equation (\ref{Her3b}). The dispersions and the anisotropy parameters are
\begin{equation}
\sigma^2_r = \frac{\Phi_0}{30} \frac{6+5x}{\left(1+x\right)^2},
\end{equation}
\begin{equation}
\sigma^2_\phi = \frac{\Phi_{0}}{60}\frac{6+16x+15x^2}{\left(1+x\right)^3},
\end{equation}
\begin{equation}
\beta = \frac{1}{2} \frac{6+6x-5x^2}{\left(1+x\right)\left(6+5x\right)}.
\end{equation}
The third three-term distribution function is given by the relation (\ref{Her3c}). The velocity dispersions and the anisotropy parameters are
\begin{equation}
\sigma^2_r = \frac{\Phi_{0}}{30} \frac{5+6x}{\left(1+x\right)^2},
\end{equation}
\begin{equation}
\sigma^2_t = \frac{\Phi_{0}}{60}\frac{5+22x+12x^2}{\left(1+x\right)^3},
\end{equation}
\begin{equation}
\beta = \frac{1}{2} \frac{5}{\left(1+x\right)\left(5+6x\right)}.
\end{equation}
The radial and tangential velocity dispersions, as well as the anisotropy parameter, for the distribution functions for the Hernquist model discussed in this section have been plotted in Fig. \ref{figA2}. We also present contour plots for some of these distribution functions in Fig. \ref{contourHer} and it must be noted that they are functions of both the dimensionless binding energy and angular momentum.

\subsection{Miscellaneous distribution functions for the Veltmann models}
In this final subsection, we present another wide class of distribution functions for the Veltmann models. We start by multiplying and dividing the augmented density (\ref{rho1term}) by a factor of $(1+cx^{m})(\Psi/\Phi_{0})^{m}$ for some constant $c$, then rewrite the denominator using the relation
\begin{equation}
1 - \left(\frac{\Psi}{\Phi_{0}}\right)^{m} = x^{m}\left(\frac{\Psi}{\Phi_{0}}\right)^{m},
\end{equation}
to obtain the new augmented density
\begin{equation}
\rho(\Psi,x) = \frac{\Phi_{0}(1+m)}{4\pi G a^{2}} \frac{x^{m-2}(\Psi/\Phi_{0})^{3m+1}(1+cx^{m})}{(\Psi/\Phi_{0})^{m}+c(1-(\Psi/\Phi_{0})^{m})}.
\end{equation}
Expanding, the augmented density becomes
\begin{equation}
\rho(\Psi,x) = \frac{\Phi_{0}(1+m)}{4\pi c Ga^{2}}x^{m-2}\left(1+cx^m\right) \left(\frac{\Psi}{\Phi_{0}}\right)^{3m+1} \sum_{l=0}^{\infty} \left(1-\frac{1}{c}\right)^l \left(\frac{\Psi}{\Phi_{0}}\right)^{ml}.
\end{equation}
The double-power distribution function that corresponds to the above augmented density is
\begin{eqnarray}
f\left(\mathcal{E},L\right)&=&\frac{\Phi_{0}^{-1/2}(1+m)}{4\pi cGa^{2}\left(2\pi\right)^{3/2}}\Bigg[\left(\frac{L}{\sqrt{2\Phi_{0}a^{2}}}\right)^{m-2}\sum_{l=0}^{\infty}\left(1-\frac{1}{c}\right)^{l}\frac{\Gamma\left(3m+ml+2\right)}{\Gamma\left(\frac{5m}{2}+ml+\frac{3}{2}\right)\Gamma\left(\frac{m}{2}\right)}\left(\frac{\mathcal{E}}{\Phi_{0}}\right)^{ml+(5m+1)/2} \\ \nonumber
&+& c\left(\frac{L}{\sqrt{2\Phi_{0}a^{2}}}\right)^{2m-2}\sum_{l=0}^{\infty}\left(1-\frac{1}{c}\right)^{l}\frac{\Gamma\left(3m+ml+2\right)}{\Gamma\left(2m+ml+\frac{3}{2}\right)\Gamma\left(m\right)}\left(\frac{\mathcal{E}}{\Phi_{0}}\right)^{2m+ml+1/2}\Bigg].
\end{eqnarray}
The anisotropy parameter can be computed from the augmented density.
\begin{equation}
\beta=1-\frac{m}{2}-\frac{m}{2}\frac{cx^{m}}{1+cx^{m}}.
\end{equation}
It is seen that the value of $\beta$ decreases as one moves radially outwards. 

\end{document}